\documentclass[aps, prd, floats, floatfix, twocolumn, superscriptaddress, nofootinbib, showpacs, letterpaper, 10pt, preprintnumbers]{revtex4-1}

\usepackage{graphicx}
\usepackage{amsmath}
\usepackage{amsfonts}
\usepackage{amssymb}
\usepackage{amstext}
\usepackage{tensor}
\usepackage[english]{babel}
\usepackage{blindtext}
\usepackage{helvet}
\usepackage{microtype}
\usepackage{aas_macros}
\usepackage[pdftex,
  pdftitle={Formalism for Primordial Black Hole Formation in Spherical Symmetry},
  pdfauthor={Author Here},
  bookmarks,bookmarksopen=false,
  pdfstartview={FitH},
  linktocpage=true
]{hyperref}

\newcommand{\cn}{{\cal N}}

\begin{document}

\title{Formalism for Primordial Black Hole Formation in Spherical Symmetry}

\author{Jolyon Bloomfield}
\email{jolyon@mit.edu}
\author{Daniel Bulhosa}
\email{dbulhosa@mit.edu}
\author{Stephen Face}
\email{face@mit.edu}
\affiliation{Center for Theoretical Physics, Laboratory for Nuclear Science, and Department of Physics, Massachusetts Institute of Technology, Cambridge, MA 02139, USA}

\date{\today}

\preprint{MIT-CTP-4662}

\begin{abstract}
  We present a comprehensive formalism for the description of primordial black hole formation in spherical symmetry based on the formalisms of Misner, Sharp, and Hernandez, which can be used to predict whether or not a black hole will form, and extract the resulting black hole mass when formation does occur. Rigorous derivations of all aspects of the formalism are provided, including a thorough investigation of appropriate initial and boundary conditions. We connect our formalism with numerous other approaches in the literature. Some implementation details for numerical code are provided. We include animations of simulated primordial black hole formation as supplemental material.
\end{abstract}


\maketitle

\section{Introduction}

The very early universe is understood to consist of a generally isotropic and homogeneous plasma with small perturbations. These perturbations, thought to originate in quantum fluctuations driven outside the horizon during inflation \cite{Guth1980, Starobinsky1980}, form the seeds of structure that give rise to the cosmic microwave background and ultimately the universe as we know it today.

If the primordial perturbations are sufficiently large, it is possible that they might collapse under their own gravity, forming primordial black holes \cite{Yokoyama1998, Yokoyama1999}. This requires perturbations to grow to nonlinear scales. The standard inflationary tale predicts Gaussian scale invariant perturbations with amplitudes $\sim 10^{-5}$, meaning that the probability of forming a nonlinear perturbation is negligible; indeed, primordial black hole production rates are predicted to be vanishingly small in this picture \cite{Young2014}. It is conceivable, however, that there exist features in the inflationary power spectrum at small wavelengths, well below cosmic microwave background radiation scales, that may allow perturbations to become sufficiently nonlinear as to grow under their own gravity and undergo gravitational collapse \cite{GarciaBellido1996, Thompson2005, Bugaev2011} (alternatively, large non-Gaussianities or isocurvature may also provide the necessary conditions).

The idea of primordial black holes, whilst currently strictly hypothetical, is intriguing for a number of reasons. Astrophysically, an appropriate spectrum of primordial black holes may provide the seeds for the supermassive black holes found in the centers of galaxies \cite{Duechting2004}. An entirely different spectrum consisting primarily of lunar mass black holes may be able to explain cold dark matter \cite{Capela2013}. Smaller black holes are also of theoretical interest as a window into the process of black hole evaporation through Hawking radiation \cite{Hawking1974}. Finally, should primordial black holes be identified, this would give insight into the small-scale structure of the very early universe, well beyond what the CMB and large scale structure can provide.

The concept of primordial black holes was first introduced by Zel'dovich and Novikov \cite{Zeldovich1967} in 1966, and again by Hawking in 1971 \cite{Hawking1971}. Theoretical work in the 1970's identified rough criteria for primordial black holes to form \cite{Carr1974, Carr1975}, and early numerical hydrodynamic simulations were undertaken later that decade \cite{Nadezhin1978, Bicknell1979, Novikov1980}. Later work investigated the possibility of primordial black hole formation through exotic processes such as bubble collisions \cite{Crawford1982}, cosmic strings \cite{Hawking1987, Polnarev1988} and phase transitions \cite{Jedamzik1999}.

A large amount of work has gone into computing the number density and mass spectrum of primordial black holes from inflationary models \cite{Green2004, Young2014, Thompson2005, Bugaev2011}, and there exists an extensive literature constraining the mass spectrum through astronomical, astrophysical and cosmological measurements \cite{Green1997, Carr2009, Josan2009, Byrnes2012, Griest2013, Carr2014} (see also \cite{Carr2005, Khlopov2008, Green2014} for reviews). The absence of primordial black holes has in turn been used to place constraints on other cosmological processes.

In this paper, we are concerned with the formation of primordial black holes through inflationary perturbations in the primordial plasma. Most work in the literature deals with spherically symmetric perturbations, and we follow this trend. This simplification is justified by the result from peaks theory \cite{BBKS1986} that the largest peaks resulting from an appropriate probability distribution have a very strong tendency to be (almost) spherical.

In order to understand the primordial black hole number density and mass spectrum, one needs to know the threshold under which perturbations will form primordial black holes. Early work by Carr \cite{Carr1974, Carr1975} considered a collapsing region described by a closed Friedmann-Robertson-Walker (FRW) spacetime surrounded by a flat FRW universe. In the radiation-dominated era, this was found to lead to a threshold value for the perturbation amplitude $\delta_c \sim 1/3$, where $\delta$ is defined to be the fractional mass excess inside the cosmological horizon when the overdense region enters the horizon.

Early hydrodynamical simulations of primordial black hole formation \cite{Nadezhin1978, Novikov1980} were based on the Misner-Sharp formalism \cite{Misner1964}, which describes the gravitational collapse of a thermodynamic fluid under spherical symmetry. This formalism cannot evolve past the formation of a black hole, however. To our knowledge, the first numerical evolution capable of extracting the final state black hole mass was Niemayer and Jedamzik \cite{Niemeyer1999}, who modified stellar collapse code \cite{Baumgarte1995} based upon the Hernandez-Misner formalism \cite{Hernandez1966}, which uses a null slicing condition to avoid singularity formation. Alternative formalisms have also been proposed \cite{Shibata1999, Wands2000, Tanaka2006}. A number of works have used these formalisms to investigate critical phenomena and shock formation near the threshold of black hole formation \cite{Niemeyer1997, Hawke2002, Musco2008, Musco2012}.

Niemayer and Jedamzik found that the critical mass overdensity $\delta_c \sim 0.7$, much larger than had been previously thought. However, Sasaki and Shibata \cite{Shibata1999} pointed out that this was due to their initial conditions containing an unphysical decaying mode. Musco \textit{et al.} \cite{Musco2004} investigated this further, arriving at a more modest $\delta_c \sim 0.4$. This issue gave rise to investigations of how to construct appropriate initial conditions for numerical simulations. A particularly nice method has been presented by Polnarev \textit{et al.} \cite{Polnarev2006, Polnarev2012}.

While the formation condition for a primordial black hole is known to be roughly $\delta_c \sim 0.4$, the precise value is dependent upon the density profile of the perturbation \cite{Hidalgo2008}. Polnarev \textit{et al.} \cite{Nakama2013, Nakama2014} have recently constructed a new criteria which aims to capture the effect of the perturbation profile dependence on the black hole formation condition.

The purpose of this paper is to present a comprehensive formalism for the numerical evolution of spherically symmetric perturbations in the early universe under the influence of their own gravity. We employ the Misner-Sharp \cite{Misner1964} and Hernandez-Misner \cite{Hernandez1966} formalisms, building upon the ideas of Niemayer and Jedamzik \cite{Niemeyer1999} and Polnarev \textit{et al.} \cite{Polnarev2006, Polnarev2012}. Our primary goal is to establish a clean and precise formalism from a somewhat murky literature.

Our numerical formalism, while building on previous results, is entirely new. We correct errors in the literature and present new techniques to improve accuracy in simulations. We analyze the regime of validity of various approximations, and provide a concrete description for how to connect inflation to initial conditions for numerical evolution. We have taken pains to be as rigorous and complete as possible, presenting our derivations in a pedagogical manner. We do not give details of a particular numerical implementation, but point to issues that can lead to numerical troubles where appropriate.

This paper is structured as follows. We begin by deriving the Misner-Sharp formalism in detail in Section \ref{sec:MS}. In Section \ref{sec:cosmology}, we apply the Misner-Sharp formalism to cosmology, and find variables more appropriate for this evolution. We pay particular attention to the black hole formation condition and appropriate cosmological boundary conditions. Section \ref{sec:IC} addresses the subject of initial conditions in detail. We then derive the Hernandez-Misner formalism in Section \ref{sec:hmformalism}, and cast it into variables more suited for cosmological evolution in Section \ref{sec:hmcosmology}. We discuss various coding issues in Section \ref{sec:code}. Those who are only interested in seeing the supplemental animations may like to skip to Section \ref{sec:plots}, where we demonstrate and explain simulations in our formalism. Finally, we conclude in Section \ref{sec:conclusions}. A number of technical appendices are included.

\section{Misner-Sharp Formalism} \label{sec:MS}

We begin by investigating Einstein's equations for a spacetime containing a perfect fluid under the assumption of spherical symmetry. We follow the formalism first laid out by Misner and Sharp \cite{Misner1964}, and aim to present a detailed yet succinct derivation of the equations of motion.

A spacetime is spherically symmetric if it possesses an $SO(3)$ isometry (3D rotation) group, where the action of this group on any given point (the `orbits' of the group) are two-dimensional spheres \cite{WaldBook}. The spacetime metric induces a two-dimensional metric on each such sphere, which must be proportional to the metric of a unit two-sphere,
\begin{align}
  d\Omega^2 = d \theta^2 + \sin^2 \theta d \phi^2
\end{align}
for two coordinates $\theta$ and $\phi$. This two-dimensional metric possesses three Killing vectors, associated with rotations around three axes. Demanding that the spacetime metric also possess these Killing vectors constrains it to take the form
\begin{align}
  ds^2 = g_{AB} dx^A dx^B + R^2(x^A) d\Omega^2 \label{eq:smetric}
\end{align}
where $x^0$ and $x^1$ are arbitrary coordinates and the two-dimensional metric $g_{AB}$ depends only on $x^0$ and $x^1$.

The remaining gauge freedom in the metric \eqref{eq:smetric} lies in redefinitions of $x^0$ and $x^1$. Common gauge choices are to choose coordinates such that $R^2 = (x^1)^2$, known as the radial gauge, or to fix $(x^1)^2 g_{11} = R^2$, known as isotropic radial coordinates. The gauge choice we employ is to fix $g_{01} = 0$, which makes the metric diagonal. We choose coordinates $t$ and $A$, and write the metric as
\begin{align}
  ds^2 = - e^{2 \phi} dt^2  + e^{\lambda} dA^2 + R^2 d \Omega^2 \,. \label{eq:MSmetric}
\end{align}
Here $\phi$, $\lambda$, and $R$ are functions of $A$ and $t$ alone, and we take $R$ to be a monotonically increasing function of $A$. The remaining gauge freedom lies in transformations of the form $t = t(\tilde{t})$ and $A = A(\tilde{A})$.

We describe the matter content of the spacetime as a perfect fluid with stress-energy tensor
\begin{align}
  T_{\mu \nu} = (\rho + P) u_\mu u_\nu + P g_{\mu \nu} \label{eq:perfluid}
\end{align}
where $\rho$ is the energy density, $P$ is the pressure, and $u^\mu$ is the four-velocity of the fluid. Note that because this tensor is diagonal in the local rest frame of the fluid, it cannot describe the energy flow associated with radiation \cite{Misner1964}. In particular, this means that free-streaming neutrinos cannot be described in this formalism. We discuss the use of a perfect-fluid description of the early universe photon bath in Appendix \ref{app:stressenergytensor}. It is convenient to use coordinates such that the coordinates move with the fluid, known as comoving or Lagrangian coordinates. The fluid four-velocity is then given by
\begin{align}
  u^t = e^{-\phi}, \qquad u^i = 0; \; i = r, \theta, \phi \,.
\end{align}

The preliminaries complete, we now compute the equations of motion. We begin with the Einstein equation $G_{\mu \nu} = 8 \pi T_{\mu \nu}$, where we work with $G = c = 1$. The time-time, radius-radius, and time-radius components of the equation are as follows, where we use primes and dots to denote derivatives with respect to $A$ and $t$ respectively.
\begin{subequations}
\begin{align}
  8 \pi e^{2 \phi} \rho ={}&{} \frac{1}{R^2} \left(e^{2 \phi} + \dot{R}^2 - e^{- \lambda + 2 \phi} (R^\prime)^2 \right)
\label{eq:Einsteintt} \\
  &+ \frac{e^{-\lambda}}{R} \left(e^\lambda \dot{R} \dot{\lambda} + e^{2 \phi} R^\prime \lambda^\prime - 2 e^{2 \phi} R^{\prime \prime}\right)
\nonumber \\
  8 \pi e^\lambda P ={}&{} \frac{1}{R^2} \Big[ (R^\prime)^2 + 2 R R^\prime \phi^\prime
\label{eq:Einsteinrr} \\
  &- e^{\lambda -2 \phi} (e^{2 \phi} + \dot{R}^2 - 2 R \dot{R} \dot{\phi} + 2 R \ddot{R}) \Big]
\nonumber \\
  0 ={}&{} \frac{\dot{\lambda} R^\prime + 2 \dot{R} \phi^\prime - 2 \dot{R}^\prime}{R} \label{eq:Einsteintr}
\end{align}
\end{subequations}
Except for the $\theta \theta$ and $\phi \phi$ components, all other components of the Einstein equation vanish identically. The $\theta \theta$ and $\phi \phi$ components are related by symmetry, and as the Einstein equation and the conservation of stress-energy equation are related by the Bianchi identity, these angular components can be neglected in favor of the simpler equations arising from conservation of stress-energy.

We now use conservation of stress-energy, $\nabla_{\mu} T^{\mu \nu} = 0$. The $t$ and $r$ components are
\begin{subequations}
\begin{align}
  0 &= 2 R \dot{\rho} + (P + \rho)(4 \dot{R} + \dot{\lambda} R ) \label{eq:stresst} \\
  0 &= P' + (\rho + P) \phi' \label{eq:stressr}
\end{align}
\end{subequations}
while the $\theta$ and $\phi$ components vanish identically.

The following definitions allow us to bring the equations of motion into a more convenient form.
\begin{subequations}
\begin{align}
  U &= e^{- \phi} \dot{R} \label{eq:Udef}
\\
  \Gamma &= e^{- \lambda/2} R^\prime \label{eq:Gammadef}
\\
  m &= 4 \pi \int_0^A \rho R^2 R^\prime dA \label{eq:mdef}
\end{align}
\end{subequations}
$U$ is the coordinate velocity of a fluid element, as the fluid element at comoving radius $A$ is positioned at radius $R$. Note that this is only a coordinate velocity, not a physical velocity, and therefore can exceed unity. The quantity $\Gamma$ is simply an alternative variable to $\lambda$, and must be positive due to the monotonic nature of $R$. The quantity $m$ is the Misner-Sharp mass, a measure of the gravitational mass contained within the radius $A$. It must also be monotonic in $A$. The following related expressions are useful.
\begin{subequations}
\begin{align}
  e^{- \phi} \ddot{R} &= \dot{U} + U \dot{\phi} \label{eq:rdotdot}
\\
  e^{- \lambda/2} R^{\prime \prime} &= \Gamma^\prime + \frac{1}{2} \Gamma \lambda^\prime \label{eq:rprimeprime}
\\
  m' &= 4 \pi R^2 R' \rho \label{eq:MSmprime}
\end{align}
\end{subequations}

We now write the five equations of motion using these definitions. Begin by solving Eq. \eqref{eq:stressr} for $\phi'$.
\begin{align}
  \phi' = - \frac{P'}{\rho+P} \label{eq:phiprime}
\end{align}
Next, write Eq. \eqref{eq:Einsteintr} as
\begin{subequations}
\begin{align}
  \dot{\lambda} R' &= 2 (\dot{R}^\prime - \dot{R} \phi^\prime) \,, \label{eq:lambdadot}
\\
  \dot{\lambda} &= \frac{2 e^\phi U^\prime}{R^\prime} \label{eq:lambdadot2} \,.
\end{align}
\end{subequations}
Take Eq. \eqref{eq:Einsteintt} and substitute in Eqs. \eqref{eq:Udef}, \eqref{eq:Gammadef}, \eqref{eq:rprimeprime} and \eqref{eq:lambdadot2} to obtain
\begin{align}
  8 \pi \rho R^2 R^\prime &= \frac{d}{dA} \left((1 + U^2 - \Gamma^2) R\right) \,.
\end{align}
Integrating this from $0$ to $A$, imposing the boundary condition $R(t,0)=0$, and using Eq. \eqref{eq:mdef} we find
\begin{align}
  \Gamma^2 = 1 + U^2 - \frac{2m}{R} \,. \label{eq:MSGamma}
\end{align}
Now, multiply Eq. \eqref{eq:stresst} by $2 \pi R R'$ and substitute Eq. \eqref{eq:lambdadot}. Some rearrangement leads to
\begin{align}
  0 = \frac{d}{dt} \left(4 \pi \rho R^2 R'\right) + 4 \pi \frac{d}{dA} \left(P R^2 \dot{R}\right) \,. \label{eq:stresst2}
\end{align}
Integrate this over the radial coordinate from $0$ to $A$, then make use of Eqs. \eqref{eq:Udef} and \eqref{eq:mdef} to obtain
\begin{align}
  \dot{m} = - e^\phi 4 \pi R^2 P U \label{eq:MSmdot}
\end{align}
where we once again employ the boundary condition $R(t,0) = 0$. The last equation is \eqref{eq:Einsteinrr}. Substituting Eqs. \eqref{eq:Udef}, \eqref{eq:Gammadef} and \eqref{eq:rdotdot} yields the following.
\begin{align}
  2 e^{-\phi} R \dot{U} = \Gamma^2 - 1 - U^2 + 2 \Gamma^2 R \frac{\phi^\prime}{R'} - 8 \pi P R^2
\end{align}
This can be further simplified using Eqs. \eqref{eq:phiprime} and \eqref{eq:MSGamma}.
\begin{align}
\dot{U} = - e^\phi \left(\Gamma^2 \frac{P'}{R'(\rho+P)} + \frac{m + 4 \pi P R^3}{R^2}\right) \label{eq:MSUdot}
\end{align}

We now have the six equations that constitute the Misner-Sharp formalism \cite{Misner1964}. Three equations are evolution equations for $R$, $m$ and $U$ (Eqs. \eqref{eq:Udef}, \eqref{eq:MSmdot} and \eqref{eq:MSUdot}), and three equations are constraint equations for $\rho$, $\phi$ and $\Gamma$ (Eqs. \eqref{eq:MSmprime}, \eqref{eq:phiprime}, and \eqref{eq:MSGamma}). Together, they express the content of Einstein's equations in a convenient manner for numerical implementation.

This system of equations must be supplemented by an equation of state $P$. In the original Misner-Sharp formalism, the equation of state was taken to depend on energy density and specific energy (kinetic energy density), and so thermodynamic relations were also required to close the system of equations. For early universe cosmology purposes, we can take pressure to be a function of energy density only, and so we only need the equation of state\footnote{A number of papers in the primordial black hole literature employ thermodynamic relations in their system of equations despite this. To our knowledge, this is unnecessary.}.

Because it describes fluid flows, the Misner-Sharp formalism has the capacity to develop shocks. A typical way to combat this is to introduce an artificial viscosity term into the equations in order to smooth out any shocks \cite{Neumann1950}. The effect of adding a a viscous term is to let pressure $P = P_0 + Q$, where $P_0$ is the physical pressure, and $Q$ is the artificial viscosity. For the moment we simply include it as an unknown artificial viscosity in the equations of motion, and leave a detailed discussion for Section \ref{sec:code}. For the physical pressure, we will use $P_0 = w \rho$, where $w$ is the (possibly time-dependent) equation of state. For radiation, $w = 1/3$, as we detail in Appendix \ref{app:stressenergytensor}.

We complete the formalism by specifying boundary conditions. We utilized $R(t, 0) = 0$ in the derivation of the equations. By the definitions of $U$ and $m$, we then have $U(t, 0) = m(t, 0) = 0$. By spherical symmetry, we have $\rho'(t, 0) = P'(t, 0) = 0$ and thus $\phi'(t, 0) = 0$ also. Using L'H\^{o}pital's Rule we find that
\begin{align}
\lim\limits_{A \rightarrow 0} \Gamma^2 = 1 - \lim\limits_{A \rightarrow 0} \frac{8 \pi \rho R^2 R'}{R'} = 1 \,.
\end{align}
The only remaining boundary condition is to specify $\phi$ at one of the boundaries. This is simply a gauge condition on the time coordinate, and is usually set by matching to asymptotic behavior.

Initial conditions must be set for each of the evolution variables, namely $m(t_0, A)$, $U(t_0, A)$ and $R(t_0, A)$. The first two of these correspond to the local energy density and fluid velocity, while the third of these is the initial gauge condition. The fluid provides the only dynamical degree of freedom; the gauge condition simply evolves with time as dictated by the fluid.

We summarize the Misner-Sharp equations for convenience.
\begin{subequations}
\begin{align}
  \dot{R} &= U e^{\phi} \label{eq:rdot}
\\
  \dot{m} &= - e^\phi 4 \pi R^2 P U \label{eq:mdot}
\\
  \dot{U} &= - e^{\phi} \left(\frac{\Gamma^2 P^\prime}{R^\prime (\rho+ P)} + \frac{m}{R^2} + 4 \pi R P \right) \label{eq:udot}
\\
  \rho &= \frac{m^\prime}{4 \pi R^2 R^\prime}  \label{eq:mprime}
\\
  \phi' &= - \frac{P'}{\rho + P} \label{eq:MSphi}
\\
  \Gamma^2 &= 1 + U^2 - \frac{2m}{R}
\end{align}
\end{subequations}
The boundary conditions are $R(t, 0) = 0$, $m(t, 0) = 0$, $U(t, 0) = 0$ and $\Gamma(t, 0) = 1$, with an arbitrary (nonzero) boundary condition $\rho_{ref}(t)$. For later convenience, we also include the following expression.
\begin{align} \label{eq:rhodot}
  \dot{\rho} &= - (\rho + P) e^\phi \left(2 \frac{U}{R} + \frac{U^\prime}{R^\prime} \right)
\end{align}

When artificial viscosity is turned off and $w$ is constant, we can integrate Eq. \eqref{eq:MSphi} to obtain
\begin{align}
  e^\phi &= \left(\frac{\rho_{ref}}{\rho}\right)^{w/(1+w)} \label{eq:MSphi2}
\end{align}
where $\rho_{ref}$ is some reference density which is simply related to the boundary condition on $\phi$. This formula is useful for analytic analysis.

\section{Cosmological Application of the Misner-Sharp Formalism} \label{sec:cosmology}

The Misner-Sharp formalism was originally designed to study stellar collapse, where pressure vanishes after some radius, and the metric smoothly connects to the Schwarzschild metric. In cosmology, neither pressure nor energy density vanish at large radii, and spacetime is not asymptotically flat. Thus, particular care must be given to the boundary conditions in the Misner-Sharp formalism in this case. In this section, we use the Misner-Sharp formalism to describe cosmological evolution. We begin by investigating the Friedmann-Robertson-Walker (FRW) limit of the equations, before looking beyond the background evolution to find variables more suited to cosmological evolution.

\subsection{Background Evolution}
We begin by exploring the background cosmological evolution. This requires that the energy density is constant in space, yielding $\rho_b' = 0$, where the subscript $b$ indicates the background FRW value. This in turn leads to $\phi_b' = 0$ from Eq. \eqref{eq:MSphi} where we ignore artificial viscosity. By choosing our boundary condition appropriately, we can set $\phi_b = 0$. This gives us $\rho_{ref} = \rho_b$. Equation \eqref{eq:mprime} can now be integrated to obtain
\begin{align}
  m_b = \frac{4 \pi}{3} \rho_b R_b^3 \,.
\end{align}
Using these results, the evolution equations \eqref{eq:rdot}, \eqref{eq:mdot} and \eqref{eq:udot} become the following.
\begin{align}
  \dot{R}_b &= U_b
\\
  \dot{\rho}_b &= - 3 \frac{\dot{R}_b}{R_b} \rho_b (1 + w) \,.
\\
  \dot{U}_b &= - \frac{4 \pi}{3} \rho_b R_b (1 + 3w) \,.
\end{align}
In particular, note that there are no spatial derivatives here, and so the evolution of each function will be independent of $A$. In particular, this allows us to write $R_b(A, t) = a(t) A$ where we exploit reparameterization invariance to write $R_b(A, t_0) = A$ for our initial data, selecting $a(t_0) = 1$. The function $a(t)$ is the usual scale factor of FRW cosmology. Substituting this description of $R$ into the equations of motion then yields the following.
\begin{align}
  \dot{\rho}_b &= - 3 \frac{\dot{a}}{a} \rho_b (1 + w)
\\
  U_b &= A \dot{a}
\\
  \dot{U}_b &= - \frac{4 \pi}{3} \rho_b a A (1 + 3w)
\end{align}
The first of these is recognizable as the continuity equation in cosmology. The last two of these combine to yield
\begin{align}
  \frac{\ddot{a}}{a} &= - \frac{4 \pi}{3} \rho_b (1 + 3w)
\end{align}
which is the acceleration equation. Combined with the continuity equation, this can be used to obtain
\begin{align}
  \partial_t (\dot{a}^2) &= \frac{8 \pi}{3} \partial_t (\rho_b a^2)
\end{align}
which integrates to give
\begin{align}
  \left(\frac{\dot{a}}{a}\right)^2 \equiv H^2 = \frac{8 \pi}{3} \rho_b - \frac{k}{a^2}
\end{align}
for some constant of integration $k$. This is the Friedmann equation with spatial curvature $k$. We can multiply this by $R_b^2$ and substitute for $U_b$ to obtain
\begin{align}
  U_b^2 = \frac{2m_b}{R_b} - k A^2
\end{align}
and thus
\begin{align}
  \Gamma_b^2 = 1 - kA^2 \,.
\end{align}
The metric then becomes the curved FRW metric,
\begin{align}
  ds^2 = - dt^2 + a(t)^2 \left(\frac{dA^2}{1 - kA^2} + A^2 d\Omega^2\right)
\end{align}
where we used $e^\lambda = R'^2 / \Gamma^2$. Thus, the Misner-Sharp equations yield the FRW cosmological evolution, as expected.

Let us specialize to the case of $k = 0$ (flat FRW) with constant equation of state $w$. As we aim to model cosmological evolution in the radiation dominated era immediately after inflation has suppressed all spatial curvature, this is a reasonable regime to investigate.

We can then solve the background cosmological evolution as follows. The continuity equation is integrated to obtain
\begin{align} \label{eq:rhosol}
  \rho_b = \rho_0 a^{-3(1+w)}
\end{align}
where $\rho_0$ is the FRW energy density at time $t_0$. The Friedmann equation then becomes
\begin{align}
  \dot{a} = \sqrt{\frac{8 \pi \rho_0}{3}} a^{1 - 1 / \alpha}
\end{align}
where we have defined $\alpha = 2 / 3 (1+w)$ for convenience ($\alpha = 1/2$ for $w = 1/3$). This can be integrated to obtain
\begin{align}
  \alpha (a^{1/\alpha} - 1) &= \sqrt{\frac{8 \pi \rho_0}{3}} (t - t_0)
\end{align}
where the boundary condition $a(t_0) = 1$ has been used, and we assume $w \neq -1$. It is convenient to set $t_0$ by
\begin{align}
  t_0 = \alpha \sqrt{\frac{3}{8 \pi \rho_0}} \label{eq:t0def}
\end{align}
such that
\begin{align} \label{eq:asol}
  a &= \left(\frac{t}{t_0}\right)^\alpha \,.
\end{align}
The Hubble parameter is then given by
\begin{align}
  H = \frac{\alpha}{t} \,.
\end{align}

\subsection{Beyond the Background} \label{sec:beyondbackground}

While the full Misner-Sharp equations describe gravitational collapse in cosmology, they are not written in terms of the most efficient quantities, particularly from a numerical perspective. As we can solve the background evolution exactly, it is useful to factor out this aspect of the solution from the variables, in order to enhance accuracy and stability. In this subsection, we transform the Misner-Sharp equations in such a manner. This subsection largely draws upon and improves the ideas of Polnarev, Nakama and Yokoyama \cite{Polnarev2012}.

Let
\begin{subequations} \label{eq:newvariables}
\begin{align}
  R &= a A \tilde{R} = R_b \tilde{R} \label{eq:newR}
\\
  \rho &= \rho_b \tilde{\rho} \label{eq:trhodef}
\\
  m &= \frac{4 \pi}{3} \rho_b R^3 \tilde{m} \label{eq:newvariablesm}
\\
  U &= H R \tilde{U} \label{eq:newvariablesU}
\\
  P &= \rho_b \tilde{P} = \rho_b \tilde{\rho} (w + \tilde{Q})
\end{align}
\end{subequations}
where the tilded quantities are unity for FRW evolution, and are even functions ($\tilde{Q}$, which represents a dimensionless form of the artificial viscosity, is also an even function but vanishes for FRW evolution). The choice to scale $\tilde{m}$ and $\tilde{U}$ by $R$ instead of $R_b$ is made because the expected local quantities depend on physical radius rather than the expected FRW radius. In more pragmatic terms, this choice of quantity simplifies the expressions that follow. The Misner-Sharp equations of motion become the following in this notation.
\begin{subequations}
\begin{align}
  \phi' &= - \frac{\tilde{P}'}{\tilde{\rho} + \tilde{P}}
\\
  \frac{1}{H} \dot{\tilde{R}} &= \tilde{R} \left(\tilde{U} e^{\phi} - 1 \right)
\\
  \tilde{\rho} &= \tilde{m} + \frac{A \tilde{R}}{3(A \tilde{R})'} \tilde{m}'
\\
  \frac{1}{H} \dot{\tilde{m}}
  &= \frac{2}{\alpha} \tilde{m}
  - 3 \tilde{U} e^\phi \left( \tilde{P}
  + \tilde{m}\right)
\\
  \Gamma^2 &= 1 + H^2 R^2 \left(\tilde{U}^2 - \tilde{m}\right) \label{eq:gammanew}
\\
  \frac{\dot{\tilde{U}}}{H}
  &= \frac{\tilde{U}}{\alpha} - e^\phi \bigg[
  \Gamma^2 \frac{\tilde{P}'}{H^2 R R' (\tilde{\rho} + \tilde{P})}
\nonumber \\
  & \qquad \qquad \qquad + \frac{1}{2} \left(2 \tilde{U}^2 + \tilde{m} + 3 \tilde{P} \right)
  \bigg]
\end{align}
\end{subequations}
Here, we've made liberal use of the background expressions, the continuity equation and the Friedmann equation.


For the purposes of numerical evolution, the evolution slows down significantly as time increases due to the $1/t$ that appears in the equations of motion (as part of the Hubble parameter). To address this, it is useful to work in logarithmic time. Define
\begin{align}
  \xi = \ln \left( \frac{t}{t_0} \right) = \frac{1}{\alpha} \ln a
\end{align}
such that
\begin{align}
  \partial_t = \frac{H}{\alpha} \partial_\xi \,. \label{eq:dxi}
\end{align}
We can express the Hubble and scale factors in terms of $\xi$ as follows.
\begin{align}
  a &= e^{\alpha \xi} \label{eq:axi}
\\
  H &= \frac{\alpha}{t_0} e^{-\xi} = \frac{1}{R_H} e^{-\xi} \label{eq:Hxi}
\end{align}
Here, we define $R_H = t_0 / \alpha$ as the horizon radius at the beginning of the evolution. For later use, it is convenient to note that
\begin{align}
 H a R_H = e^{(\alpha - 1)\xi} \,,
\end{align}
the horizon scale evolves as
\begin{align}
 R_H(\xi) = \frac{1}{H} = R_H e^\xi \,,
\end{align}
and that the Friedmann equation can be written as
\begin{align}
 H^2 = \frac{8 \pi}{3} \rho_b = \frac{e^{-2\xi}}{R_H^2} \,.
\end{align}
It is also convenient to define $A = \bar{A} R_H$ with $\bar{A}$ dimensionless. Using these variables and the definition \eqref{eq:t0def} for $t_0$, the equations of motion become the following.
\begin{subequations} \label{eq:scaledMS}
\begin{align}
  \phi' &= - \frac{\tilde{P}'}{\tilde{\rho} + \tilde{P}} \label{eq:phifinal}
\\
  \partial_\xi \tilde{R} &= \alpha \tilde{R} \left(\tilde{U} e^{\phi} - 1 \right)
\\
  \tilde{\rho} &= \tilde{m} + \frac{\bar{A} \tilde{R}}{3(\bar{A} \tilde{R})'} \tilde{m}' \label{eq:rhomrel}
\\
  \partial_\xi \tilde{m}
  &= 2 \tilde{m} - 3 \alpha \tilde{U} e^\phi \left( \tilde{P} + \tilde{m}\right)
\\
  \bar{\Gamma}^2 &= \frac{\Gamma^2}{a^2 H^2 R_H^2} = e^{2(1-\alpha) \xi} + \bar{A}^2 \tilde{R}^2 \left(\tilde{U}^2 - \tilde{m}\right) \label{eq:newgamma}
\\
  \partial_\xi \tilde{U}
  &= \tilde{U} - \alpha e^\phi \bigg[
  \bar{\Gamma}^2 \frac{\tilde{P}'}{\bar{A} \tilde{R} (\tilde{R} + \bar{A} \tilde{R}') (\tilde{\rho} + \tilde{P})}
\nonumber \\
  & \qquad \qquad \qquad + \frac{1}{2} \left(2 \tilde{U}^2 + \tilde{m} + 3 \tilde{P} \right)
  \bigg] \label{eq:newu}
\end{align}
\end{subequations}
Here, we repurpose primes to denote derivatives with respect to $\bar{A}$, and found it useful to use $\bar{\Gamma}$ in favor of $\Gamma$. When the artificial viscosity is vanishing, we have
\begin{align}
  e^\phi &= \tilde{\rho}^{-3 \alpha w / 2} \,. \label{eq:MSphinov}
\end{align}
These equations are supplemented by the boundary conditions $\tilde{\rho}' = 0$, $\tilde{R}' = 0$, $\tilde{m}' = 0$ and $\tilde{U}' = 0$ at $\bar{A} = 0$ (i.e., all functions are even functions of $\bar{A}$). A boundary condition on $\phi$ is also required; we suggest using the result \eqref{eq:MSphinov}. Initial data consists of specifying $\tilde{R}$, $\tilde{U}$ and $\tilde{m}$ (or $\tilde{\rho}$) at time $\xi = 0$.

The values of $\tilde{m}$, $\tilde{R}$ and $\tilde{U}$ at the origin are only important for computing derivatives, as the quantities $m$, $R$ and $U$ already vanish at the origin. If some evolution scheme that omits the origin while requiring $\tilde{m}$, $\tilde{R}$ and $\tilde{U}$ to be even functions is used, then the origin can be ignored altogether. If not, then the functions should be evolved subject to the constraint that these functions have a vanishing derivative at the origin. We also point out that the evolution equation for $\tilde{U}$ is not singular at the origin as $\tilde{P}$ is an even function; with artificial viscosity vanishing, $\tilde{P} = w \tilde{\rho}$, and
\begin{align} \label{eq:zeroAlimit}
  \lim_{\bar{A} \rightarrow 0} \frac{\tilde{P}'}{\bar{A}} &= \tilde{\rho}''(0) = \frac{5}{3} \tilde{m}''(0)
\end{align}
where we've used $\tilde{m}' = 0$ at the origin.

The equations \eqref{eq:scaledMS} are the complete dimensionless cosmological Misner-Sharp equations of motion. The only time explicit time dependence is in the expression for $\bar{\Gamma}^2$, and there is no dependence upon $\rho_0$ or $t_0$ at all. This implies that for any initial $\tilde{m}$, $\tilde{R}$ and $\tilde{U}$, the evolution is the same no matter the initial time or density.

\subsection{Black Hole Formation}

We now possess the tools to evolve initial density perturbations to investigate whether they form black holes or not. We thus need a method for detecting the appearance of a black hole. To do this, we search for the creation of an apparent horizon, which if detected, points to the existence of an event horizon\footnote{Note that the lack of an apparent horizon does not necessarily imply that there is no event horizon, but this is typically limited to specific spacetime slicings. For our purposes, we may assume that an apparent horizon will form before the central singularity.}. We point the reader to \S{}7 of \cite{BaumgarteBook} for the following results.

Define $n^a$ to be the timelike unit vector orthogonal to slices of constant $t$, and $s^a$ to be the spacelike unit vector orthogonal to slices of constant $A$. Together, these define an outgoing null vector $k^a = (n^a + s^a)/\sqrt{2}$. In our metric, $k^a = (e^{-\phi}, e^{-\lambda/2}, 0, 0)/\sqrt{2}$.

The expansion of outgoing radial null geodesics is given by $\Theta = m^{ab} \nabla_a k_b$ where $m^{ab} = g^{ab} + n^a n^b - s^a s^b$ is the two-dimensional metric on the hypersurface of constant $t$ and $A$. An apparent horizon forms when $\Theta \le 0$. It is known that
\begin{align}
  \Theta = \frac{1}{4 \pi R^2} k^a \nabla_a (4 \pi R^2)
\end{align}
for spherically symmetric spacetimes. For our metric, we have
\begin{align}
  \Theta = \frac{\sqrt{2}}{R} \left(U + \Gamma\right) \,.
\end{align}
As such, an apparent horizon forms when $U + \Gamma \le 0$. As $\Gamma > 0$, this requires $U$ to be negative. Using Eq. \eqref{eq:MSGamma}, the condition of black hole formation becomes
\begin{align}
  \frac{2m}{R} \ge 1
\end{align}
which is recognizable as the amount of mass inside an circumferential radius $R$ smaller than the Schwarzschild radius, and thus satisfying the hoop conjecture. Using \eqref{eq:newvariablesm} and the Friedmann equation, this yields
\begin{align}
  H^2 R^2 \tilde{m} \ge 1 \,.
\end{align}
Written in terms of our variables, this becomes
\begin{align}
  \tilde{R}^2 \tilde{m} \bar{A}^2 e^{2 (\alpha - 1) \xi} \ge 1 \,. \label{eq:bhcondition}
\end{align}
Thus, one should check for the conditions $\tilde{U} < 0$ and \eqref{eq:bhcondition} as the conditions under which a black hole forms.

After the apparent horizon forms, a singularity at the origin forms soon afterwards. The Misner-Sharp formalism cannot evolve past the formation of this singularity, and so a new technique is needed if more information about the resulting black hole (such as its mass) is desired. This brings us to the Hernandez-Misner formalism, which we introduce in Section \ref{sec:hmformalism}.

\subsection{Outer Boundary Conditions}

Let us now consider a situation where we wish to investigate gravitational collapse near the origin, and desire to smoothly connect to an FRW evolution at some radius. Suppose that our region of interest is $A < A_0$ on the initial time slice. By investigation of the Misner-Sharp equations, we see that it is insufficient for the local quantities ($R$, $U$, $\rho$ and $\phi$) to be equal to their FRW values to ensure FRW evolution for $A \ge A_0$, as an incorrect mass inside the radius $A_0$ will then cause the evolution to differ from FRW. Thus, in order to ensure FRW evolution for $A \ge A_0$, we must also require $m(A_0) = m_b(A_0)$. This matching condition implies that any overdensity must be surrounded by a corresponding underdensity in order to smoothly connect to FRW. It is important to note that this condition is not guaranteed to be preserved under time evolution, as we discuss below.

In terms of the tilde variables, the matching condition is that $\tilde{m}$, $\tilde{U}$ and $\tilde{R}$ become unity at the boundary, and remain unity beyond it. If $m(A_0) = m_b(A_0)$, then
\begin{align}
  0 = 4 \pi \int_0^{A_0} R^2 R' (\rho - \rho_b) dA' \,.
\end{align}
Letting $\tilde{\rho} = 1 + \delta_\rho$, this is equivalent to
\begin{align}
  0 = \int_0^{A_0} R^2 R' \rho_b \delta_\rho dA' \,. \label{eq:deltamcond}
\end{align}
Let $\tilde{m} = 1 + \delta_m$. Then from Eq. \eqref{eq:scaledMS}, we have
\begin{align}
  \delta_\rho &= \frac{1}{3 (\bar{A} \tilde{R})^2 (\bar{A} \tilde{R})'} \left[(\bar{A} \tilde{R})^3 \delta_m\right]' \,.
\end{align}
Inserting this into Eq. \eqref{eq:deltamcond} along with $R = a \bar{A} H_R \tilde{R}$ yields
\begin{align}
  0
  = \int_0^{A_0} \left[(\bar{A} \tilde{R})^3 \delta_m\right]' dA'
  = \left.(\bar{A} \tilde{R})^3 (\tilde{m} - 1) \right|_{\bar{A} = A_0 / R_H} \,.
\end{align}
If instead of matching to FRW at $A_0$, we wish to match to FRW asymptotically, then our condition becomes
\begin{align} \label{eq:limit}
  \lim_{\bar{A} \rightarrow \infty} (\bar{A} \tilde{R})^3 (\tilde{m} - 1) = 0 \,.
\end{align}
This extends the linearized argument given by Polnarev and Musco \cite{Polnarev2006}.

For the purposes of matching to FRW numerically, it is helpful to have the initial data approach the background values at the boundary of the computational domain. However, as the evolution equations depend on derivatives of the local variables, data that matches to FRW at the boundary at one time is not guaranteed to match at a later time. In particular, a central perturbed region tends to gradually expand out, particularly when it doesn't form a black hole.

This means that there is no physical boundary condition at the outer boundary of the computational domain. However, in order to solve the equations, an outer boundary needs to be specified, as there are two characteristics to the equations of motion, one traveling inwards and one traveling outwards. Without a boundary condition, no control over the ingoing characteristic is exercised, and the evolution rapidly becomes unstable. Just like a one-dimensional wave equation on a finite domain, a single boundary condition is required at the outer boundary.

A number of works in the literature have employed the Dirichlet boundary condition $\rho(A_0) = \rho_b$. This boundary condition does an excellent job of maintaining stability at the outer boundary, but is unphysical in that it asserts that what is outside the boundary remains FRW for all time, contrary to the argument given above. As such, it creates a reflecting boundary condition where outgoing waves reflect negatively off the outer boundary. We experimented with a Neumann boundary condition $\tilde{\rho}' = 0$ at the boundary, and as expected, found that waves were reflected positively.

One way in which to ameliorate this issue is to begin the evolution with a sufficiently large domain such that traveling waves do not reach the boundary before it can be determined whether a black hole has formed/will not form. However, this is computationally expensive. One solution to this, advocated by Nakama \textit{et al.} \cite{Nakama2013}, is to begin with a reasonably small domain that matches to FRW on the boundary, and watch for the boundary to deviate from FRW by a small quantity. Then, expand the domain by appending FRW values to the edge of the domain, and watch the new boundary for deviations. This would seem to work well, except for the fact that they required pinning $\rho$ at the boundary to its FRW value, and watching $U$ deviate, whereas in principle, both should be allowed to vary. In a similar vein, Hawke and Stewart \cite{Hawke2002} suggested adding extra gridpointss at the boundary of the domain as the evolution proceeds.

We therefore desire a boundary condition on the outer boundary that allows waves to travel past the outer boundary with minimal reflection. To do so, we turn to the method of characteristics, the detailed application of which can be found in Appendix \ref{app:char}. We extract inward ($u_2$) and outward ($u_3$) moving characteristics from solutions to the linearized equations \eqref{eq:fullm}. The equations of motion \eqref{eq:charwave} that $u_2$ and $u_3$ satisfy are inhomogeneous, with the inward and outward moving characteristics sourcing each other. This complicates the problem of constructing a nonreflecting boundary condition. In particular, we cannot simply set the amplitude of inward waves to be zero at the boundary because they are sourced by outward waves that have passed the boundary. Thus, the condition on $u_2$ for a nonreflecting boundary depends on the entire history at the boundary.

For reasons of practicality, we pursue a local condition which gives very small reflections. Following a suggestion by Kidder \textit{et al.} \cite{Kidder2000}, we try a Robin boundary condition, $\left(\bar{A}^n u_2\right)' = 0$. This condition is an assumption that $u_2$ behaves like $\sim \bar{A}^{-n}$ outside the computational domain. Combining this with Eq. \eqref{eq:inwave} to compute a condition we are able to impose, we find
\begin{align}
\partial_\xi u_2+c_s \frac{n}{\bar{A}}u_2 = \frac{1}{4c_s}u_1 + \frac{2c_s}{\bar{A}}\left(u_2 + u_3\right),
\end{align}
where $c_s = e^{\xi/2}/\sqrt{12}$ is the linearized wave speed, and we have specialized to $w = 1/3$\footnote{The speed of sound isn't $c_s^2 = 1/3$ as one might expect. This is due to the use of dimensionless variables; converting back to $t$ and $A$ yields the usual result.}. Converting this back from the characteristic variables, we obtain
\begin{align}
  \partial_\xi u_2 = \frac{\delta_m'}{4} - \delta_U' - \frac{\partial_\xi \delta_U}{c_S}
\end{align}
and so the following boundary condition for $\partial_\xi{\delta_U}$.
\begin{align}
\partial_\xi\delta_U &= -\frac{1}{4}\delta_m - \frac{2c_s^2}{\bar{A}}\delta_m' \nonumber\\ &\quad+ \frac{nc_s}{2\bar{A}}\left(\partial_\xi{\delta_m}+c_s\delta_m'\right)+\frac{c_s}{4}\delta_m'-c_s\delta_U' \label{eq:outerbc}
\end{align}
From experimentation, we find that this condition works well for $n=3$. From further experimentation, we also found that $n=1$ for the $\partial_\xi \delta_m$ term and $n=3$ for the $c_s \delta_m'$ term worked even better. We thus suggest using
\begin{align}
	\partial_\xi\delta_U &= -\frac{1}{4}\delta_m + \left(\frac{c_s}{4} - \frac{1}{2} \frac{c_s^2}{\bar{A}}\right) \delta_m'
+ \frac{c_s}{2\bar{A}} \partial_\xi{\delta_m} - c_s\delta_U' \label{eq:outerb}
\end{align}
as the outer boundary condition.

\begin{figure*}[t]
    \includegraphics[width=0.9\textwidth]{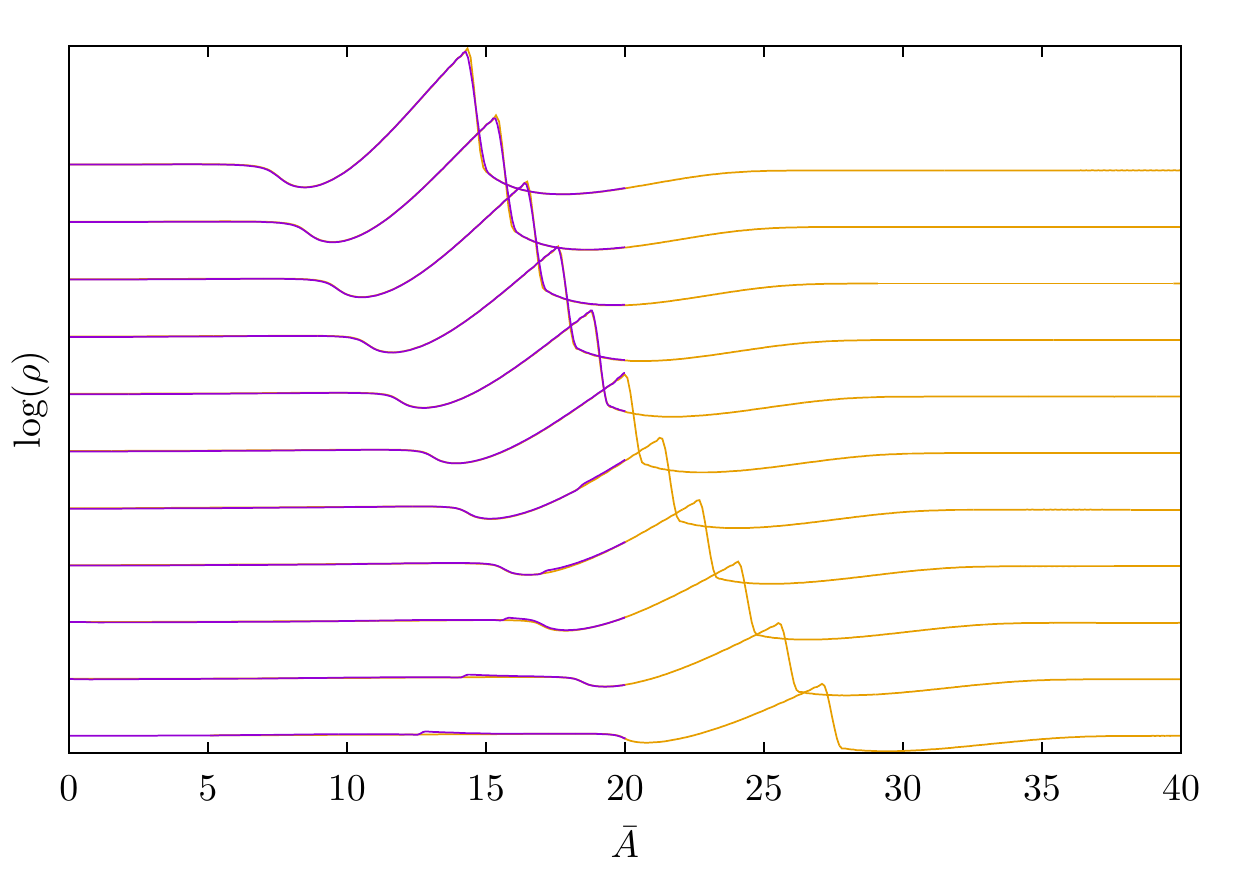}
    \caption{\label{fig:plotboundary}Demonstration of our boundary condition. This plots density as a function of comoving radius at various times. The top curve is the earliest curve in this set, and the bottom curve is the latest. The curve in purple has an outer boundary at $\bar{A} = 20$, while the curve in teal has an outer boundary at $\bar{A} = 40$. The $y$ axis is logarithmic, but the overall scaling is unimportant; it was chosen this way simply to show data at different timesteps clearly.\endgraf\setlength{\parindent}{1em}
    This run did not form a black hole, but instead sent a wave of material propagating outwards. We see that our boundary condition allows the purple curve to track the teal curve remarkably well, despite terminating halfway along the domain. The largest deviations occur in the bottom two curves, which show that there was a small reflected component at the boundary. As the characteristics evolve as $\sim 1/\bar{A}$, even a very small reflected component grows as it moves towards the origin.}
\end{figure*}

In our experimentation, we found that the linear approximation is typically a good approximation at the outer boundary. However, as we know the nonlinear speed of wave propagation for a perfect fluid (as is needed when computing the CFL condition for numerical evolution, see Section \ref{sec:code}), we tried substituting this into our boundary condition instead of the linearized speed of sound. We found that this didn't noticeably decrease the reflected portion of the wave.

Having a minimally reflecting boundary helps with the numerical evolution in a number of ways. It allows waves to escape the computational domain, which in turn allows us to reduce our domain size without affecting the evolution. This significantly increases the computational speed of our evolutions.

There are two caveats to this boundary condition. The first is that it was found by using linear perturbation theory. As such, if a wave with nonlinear amplitude hits the boundary, the condition doesn't work nearly so well (although it works better than the reflecting boundary conditions). The second is that it is physically possible (though unlikely, for appropriate initial conditions) for a wave to exit the boundary and then fall back in again. Our boundary condition does not allow for this occurrence. Regardless, the boundary condition works remarkably well, as we show in Figure \ref{fig:plotboundary}.

\section{Initial Conditions} \label{sec:IC}

The formalism presented above has evolution equations for $\tilde{m}$, $\tilde{R}$ and $\tilde{U}$, and constraint equations for calculating $\tilde{\rho}$, $\bar{\Gamma}$ and $\phi$. As such, initial data for $m$, $R$, and $U$ should be specified on a surface of constant $t$ (although $\tilde{m}$ may be substituted for $\tilde{\rho}$ or $\phi$ if desired). We note that the initial data for $\tilde{R}$ is simply a gauge choice; the physical initial data is $\tilde{m}(R)$ and $\tilde{U}(R)$, describing the fluid energy density and velocity.

It is intended that the initial conditions are initially generated from inflationary perturbations. Small perturbations outside the horizon grow, enter the horizon and become nonlinear before undergoing gravitational collapse. While these perturbations are small, they are well described in the linear regime. Linear fluid perturbations in cosmology are known to have two modes: a growing mode, and a decaying mode. In the time between the end of inflation and the beginning of our numerical evolution, the decaying mode should have had sufficient time to vanish almost completely. As such, the initial data should represent this, and be comprised of the growing mode only\footnote{It is not critical that the initial data consist of only a growing mode if sufficient time is allowed for the decaying mode to decay before any measurements are taken from the simulation. However, it is convenient to remove the decaying mode in order to begin simulations closer to black hole formation, and also to better simulate initial conditions from inflation.}. Disregard for this issue affected early work on this subject \cite{Niemeyer1999}, as pointed out in Ref. \cite{Shibata1999}.

More recently, \cite{Polnarev2006} developed a formulation that selects initial data comprising only a growing mode in the linear regime. This formalism was expanded in \cite{Polnarev2012}, and utilized in \cite{Nakama2013}. In this section, we clarify and further develop this formalism.

\subsection{Linearized Equations of Motion}
We begin by computing the linearized equations of motion. The quantities that we need to linearize are $\tilde{R}$, $\tilde{m}$, and $\tilde{U}$, which are all unity at zeroth order (for simplicity, we take the artificial viscosity $\tilde{Q} = 0$ at linear order; see below). For efficiency, we also linearize $\tilde{\rho}$ and $\phi$. We do not need to linearize $\bar{\Gamma}$, as $\bar{\Gamma}^2$ only appears in the equations of motion as a coefficient of $\tilde{\rho}'$, which is itself first order.

We linearize these quantities as follows.
\begin{subequations}
\begin{align}
  \tilde{R} & = 1 + \epsilon \delta_R
\\
  \tilde{U} & = 1 + \epsilon \delta_U \label{eq:deltaU}
\\
  \tilde{m} & = 1 + \epsilon \delta_m \label{eq:deltam}
\\
  \tilde{\rho} & = 1 + \epsilon \delta_\rho
\\
  e^\phi & = 1 + \epsilon \delta_\phi
\end{align}
\end{subequations}
Here, we use $\epsilon$ as an order counting parameter. It is straightforward to linearize Eqs. \eqref{eq:scaledMS} from these definitions.
\begin{subequations} \label{eq:firstordereqs}
\begin{align}
  \delta_\phi &= - \frac{3 w \alpha}{2} \delta_\rho \label{eq:deltaphieq}
\\
  \partial_\xi \delta_R &= \alpha \left(\delta_U + \delta_\phi\right) \label{eq:deltareq}
\\
  \delta_\rho &= \delta_m + \frac{\bar{A}}{3} \delta_m' \label{eq:deltarhoeq}
\\
  \partial_\xi \delta_m
  &= 3 \alpha w \delta_m - 2 \delta_U - 2 \delta_\phi - 3 \alpha w \delta_\rho
\\
  \partial_\xi \delta_U
  &= (1 - 2 \alpha) \delta_U - \delta_\phi
  - \frac{\alpha}{2} \left(\delta_m + 3 w \delta_\rho \right)
\nonumber \\
  & \qquad - \frac{3w \alpha^2}{2} \frac{\delta_\rho'}{\bar{A}} e^{2 (1 - \alpha) \xi} \label{eq:deltaueqprelim}
\end{align}
\end{subequations}
Note that $\delta_R$ only appears in the evolution equation for $\delta_R$, and is otherwise absent from this system of equations. This justifies the particular choice of variables in Eqs. \eqref{eq:newvariables}. We can thus ignore $\delta_R$ entirely for the moment, and focus on $\delta_m$ and $\delta_U$. Eliminating $\delta_\phi$ and $\delta_\rho$, we obtain the following evolution equations.
\begin{subequations} \label{eq:reducedlinear}
\begin{align}
  \partial_\xi \delta_m &= 3 w \alpha \delta_m - 2 \delta_U \label{eq:deltameq}
\\
  \partial_\xi \delta_U
  &= (1 - 2 \alpha) \delta_U
  - \frac{\alpha}{2} \delta_m
\nonumber\\
  &\qquad
  - \frac{w\alpha^2}{2} e^{2(1 - \alpha) \xi} \left(\frac{4 \delta_m'}{\bar{A}} + \delta_m''\right)
  \label{eq:deltaueq}
\end{align}
\end{subequations}
The first of these can be solved for $\delta_U$, which can then used to eliminate $\delta_U$ in the second equation. This leaves us with a single second-order PDE.
\begin{align}
  & \partial^2_\xi \delta_m
  - (3 - 5 \alpha) \partial_\xi \delta_m
  + \left[(1 - 2 \alpha) 3 w - 1\right] \alpha \delta_m
\nonumber\\
  &\qquad =
  w\alpha^2 e^{2(1 - \alpha) \xi} \left(\frac{4 \delta_m'}{\bar{A}} + \delta_m''\right)
  \label{eq:fullm}
\end{align}
It turns out that this equation is separable, and can be solved using a mode expansion. Unfortunately, all of the spatial modes diverge as $\bar{A} \rightarrow 0$, indicating that the linear approximation breaks down for small $A$. We thus look towards other methods.

We begin by neglecting the RHS, which reduces the equation to a straightforward ODE. The solution will be of the form $\delta_m = C \exp(\beta \xi)$, which yields the characteristic equation
\begin{align}
  \beta^2
  - (3 - 5 \alpha) \beta
  + [(1 - 2 \alpha) 3 w - 1] \alpha
  = 0 \,.
\end{align}
This has solutions
\begin{align}
  \beta_1 = \frac{w - 1}{w + 1} \quad \mathrm{and} \quad \beta_2 = 2 (1 - \alpha) \,. \label{eq:betadefs}
\end{align}
For $-1 < w < 1$, the first solution is a decaying mode, while for $w > -1/3$, the second solution is a growing mode. For our purposes, we will assume $w > -1/3$, and take only the second solution.
\begin{align}
  \delta_m(\bar{A}, \xi) = \delta_{m0} e^{2 (1 - \alpha) \xi} \label{eq:deltamsol}
\end{align}
Here, we let $\delta_{m0} = \delta_m(\bar{A},\xi = 0) = \delta_m(\bar{A}, t=t_0)$. Using this in Eq. \eqref{eq:deltameq}, we can compute the solution for $\delta_U$.
\begin{align} \label{eq:deltausol}
  \delta_U (\bar{A}, \xi) &= - \frac{\alpha}{2} \delta_m (\bar{A}, \xi)
\end{align}
To compute $\delta_R$, we take Eq. \eqref{eq:deltareq} and substitute in the expressions for $\delta_\phi$, $\delta_\rho$, $\delta_U$ and $\delta_m$.
\begin{align}
  \partial_\xi \delta_R &= \alpha(\delta_U + \delta_\phi)
\\
  &= - \frac{\alpha}{2} \left[ 2 (1 - \alpha) \delta_{m0} + \alpha w \bar{A} \delta_{m0}' \right] e^{2 (1 - \alpha) \xi}
\end{align}
The solution to this is
\begin{align}
  \delta_R &= - \frac{\alpha}{2} \left[ \delta_{m0} + \frac{w}{1+3w} \bar{A} \delta_{m0}' \right] e^{2 (1 - \alpha) \xi} \label{eq:deltarsol}
\end{align}
where we have discarded the constant homogeneous solution which is a part of the background evolution.

These expressions for $\delta_m$, $\delta_U$ and $\delta_R$ provide us with exactly what we were looking for: a relationship between the dynamical variables that picks out the growing mode in initial data. Note that the time dependence of all three quantities is the same. These results are equivalent to Eqs. (4.24), (4.25) and (4.28) in Ref. \cite{Polnarev2012}.

\subsection{A Derivative Expansion}

We now come back to the terms we discarded on the RHS of Eq. \eqref{eq:fullm}. The solution we found for $\delta_m$ in Eq. \eqref{eq:deltamsol} is the zeroth order solution $\delta_m^0$ in a derivative expansion solution to Eq. \eqref{eq:fullm}. Here, we flesh out the details of the derivative expansion.

We would like terms with derivatives acting on them to be suppressed compared to terms without derivatives. In order to accomplish this, we formally let $\bar{A} \rightarrow \bar{A} / \sqrt{\varepsilon}$, where $\varepsilon$ is an order counting parameter for our derivative expansion (as distinct from $\epsilon$, which is the order counting parameter for the perturbative expansion). Equation \eqref{eq:fullm} becomes the following under this transformation.
\begin{align}
  & \partial^2_\xi \delta_m
  - (3 - 5 \alpha) \partial_\xi \delta_m
  + \left[(1 - 2 \alpha) 3 w - 1\right] \alpha \delta_m
\nonumber\\
  &\qquad =
  \varepsilon w\alpha^2 e^{2(1 - \alpha) \xi} \left(\frac{4 \delta_m'}{\bar{A}} + \delta_m''\right)
\label{eq:newfullm}
\end{align}
We see that this demotes the terms on the RHS to contribute at first order in the derivative expansion (our solution \eqref{eq:deltamsol} is first order in perturbations, but zeroth order in the derivative expansion).

Let us write our solution to Eq. \eqref{eq:newfullm} as
\begin{align}
  \delta_m(\bar{A}, \xi) = \sum_{n=0}^{\infty} \varepsilon^n \delta_m^n(\bar{A}, \xi)
\end{align} 
where our order counting parameter keeps track of the order of each solution. Our desire is that terms of successively higher order become less and less important.

We can substitute this solution into Eq. \eqref{eq:newfullm}, and demand that the equation of motion be satisfied order by order. At $O(\varepsilon^n)$ ($n > 0$), the equation of motion becomes
\begin{align}
  & \partial^2_\xi \delta_m^n
  - (3 - 5 \alpha) \partial_\xi \delta_m^n
  + \left[(1 - 2 \alpha) 3 w - 1\right] \alpha \delta_m^n
\nonumber\\
  &\qquad =
  w\alpha^2 e^{2(1 - \alpha) \xi} \left(\frac{4 \delta_m^{(n-1) \prime}}{\bar{A}} + \delta_m^{(n-1)\prime\prime}\right) \,.
\end{align}
The LHS here is identical to the LHS of Eq. \eqref{eq:fullm}, and so has a homogeneous solution identical to Eq. \eqref{eq:deltamsol}, which can be absorbed into $\delta_m^0$. Only the particular solution is then of relevance. The RHS of this equation consists only of known functions from the previous order. As such, this is an ODE rather than a PDE: the spatial dependence is completely described by the spatial dependence of the previous order.

Let us then look at the time dependence of this equation of motion. At zeroth order, the solution has time dependence given by $\exp[2 (1-\alpha) \xi]$. The first order equation of motion then becomes
\begin{align}
  & \partial^2_\xi \delta_m^1
  - (3 - 5 \alpha) \partial_\xi \delta_m^1
  + \left[(1 - 2 \alpha) 3 w - 1\right] \alpha \delta_m^1
\nonumber\\
  &\qquad =
  w \alpha^2 \left(e^{2(1 - \alpha) \xi}\right)^2 \left(\frac{4 \delta_{m0}'}{\bar{A}} + \delta_{m0}''\right) \,.
\end{align}
The time dependence of the particular solution must then be given by $\exp[4 (1-\alpha) \xi]$. The full solution is
\begin{align}
  \delta_m^1 = e^{4(1 - \alpha) \xi} \frac{2w}{7 + 30w + 27 w^2} \left(\frac{4 \delta_{m0}'}{\bar{A}} + \delta_{m0}''\right) \,.
\end{align}

Assume that at $n^{\mathrm{th}}$ order, the time dependence of the solution is given by $\exp[2 (n+1) (1-\alpha) \xi]$. Then at $(n+1)^{\mathrm{th}}$ order, the equation of motion is
\begin{align}
  & \partial^2_\xi \delta_m^{n+1}
  - (3 - 5 \alpha) \partial_\xi \delta_m^{n+1}
  + \left[(1 - 2 \alpha) 3 w - 1\right] \alpha \delta_m^{n+1}
\nonumber\\
  &\qquad =
  w\alpha^2 e^{2(1 - \alpha) \xi} \left(\frac{4 \delta_m^{n \prime}}{\bar{A}} + \delta_m^{n\prime\prime}\right)
\end{align}
where in the second line, we substituted the time dependence from the $n^{\mathrm{th}}$ order solution. The particular solution to this ODE will be proportional to $\exp[2 (n+2) (1-\alpha) \xi]$, and hence by induction, the time dependence at $n^{\mathrm{th}}$ order will be $\exp[2 (n+1) (1-\alpha) \xi]$. All spatial dependence at a given order is determined by the spatial dependence at the previous order alone. Thus, a recursion relation can be constructed relating the solution at $n^{\mathrm{th}}$ order to the solution at $(n-1)^{\mathrm{th}}$ order. 

We thus (in principle) have the full solution to Eq. \eqref{eq:fullm} as an infinite sum in a derivative expansion.

\subsection{Beyond Linear Order}
Having solved the equations of motion at first order in perturbations through the use of a derivative expansion, we now look to higher order in perturbation theory. For the moment, let us neglect all spatial derivatives in the equations of motion, and consider the effect of expanding the equations of motion to higher order in perturbations. Let us use the expansion
\begin{align}
  X = 1 + \sum_{n = 1}^\infty \epsilon^n X_n (\bar{A}, \xi)
\end{align}
where $X$ represents $\tilde{m}$, $\tilde{U}$, $\tilde{R}$, $\tilde{\rho}$ and $e^{\phi}$. At $O(\epsilon^n)$, the equations of motion will have homogeneous terms that depend on the functions $X_n$, and source terms that depend on $X_m$ with $m < n$. The homogeneous terms will have equations of motion identical to the first order equations \eqref{eq:firstordereqs}, and thus, identical solutions, which can be absorbed into the first order solutions. We thus look at particular solutions only.

The first order solution has time dependence given by $\exp[2 (1-\alpha) \xi]$, and therefore, the source terms for the second order equations will have time dependence given by $(\exp[2 (1-\alpha) \xi])^2$. The solution to the second order equations will thus have time dependence $\exp[4 (1-\alpha) \xi]$. Assuming that for $O(\epsilon^n)$ and lower, the solution have time dependence $\exp[2 n (1-\alpha) \xi]$, we can again use induction to show that the solutions at all orders have this time dependence.

It is curious that the time-dependence of both the derivative expansion and the perturbative expansion increase with powers of $\exp[2 (1-\alpha) \xi]$. This motivates combining the two expansions into a single expansion wherein each higher order has time dependence given by an increasing power of $\exp[2 (1-\alpha) \xi]$. Indeed, this is achieved by the expansion
\begin{align} \label{eq:seriessol}
  X = 1 + \sum_{n = 1}^\infty \epsilon^n e^{2 n (1-\alpha)\xi} X_n (\bar{A})
\end{align}
where $X$ represents $\tilde{m}$, $\tilde{U}$, $\tilde{R}$, $\tilde{\rho}$ and $e^{\phi}$, and we have let $\varepsilon = \epsilon$. The zeroth order terms are unity, while the first order terms are first order in perturbations and zeroth order in the derivative expansion (there are no terms that are zeroth order in perturbations and first order in the derivative expansion, as unperturbed terms have no spatial dependence). Second order terms are the sum of second order terms in perturbations and terms that are first order in both perturbations and the derivative expansion. 

The equations of motion in this expansion are as in Eqs. \eqref{eq:scaledMS} except for the $\partial_\xi \tilde{U}$ equation, which upon substituting $\bar{A} \rightarrow \bar{A} / \sqrt{\epsilon}$ and $\partial_{\bar{A}} \rightarrow \sqrt{\epsilon} \partial_{\bar{A}}$ multiplies $\rho'$ by $\epsilon$. As previously, $\epsilon$ is only an order counting parameter.

\subsection{Second Order Solutions}
The expansion \eqref{eq:seriessol} is essentially the same as was used in Ref. \cite{Polnarev2012}, where they constructed recursion relations in order to obtain higher order solutions from lower order ones. We do not aim to rederive their recursion relations. Instead, we provide expressions to assist in the construction of initial data, where we specialize to $w = 1/3$.

We provide expressions for $\tilde{m}$, $\tilde{U}$, $\tilde{R}$, $\tilde{\rho}$ and $e^\phi$ to second order in perturbations, based off the function $\delta_{m0}$. We expand these quantities as follows.
\begin{subequations}
\begin{align}
  \tilde{m} (\bar{A}, \xi) &= 1 + e^\xi \delta_m^1(\bar{A}) + e^{2 \xi} \delta_m^2(\bar{A})
\\
  \tilde{U} (\bar{A}, \xi) &= 1 + e^\xi \delta_U^1(\bar{A}) + e^{2 \xi} \delta_U^2(\bar{A})
\\
  \tilde{R} (\bar{A}, \xi) &= 1 + e^\xi \delta_R^1(\bar{A}) + e^{2 \xi} \delta_R^2(\bar{A})
\\
  \tilde{\rho} (\bar{A}, \xi) &= 1 + e^\xi \delta_\rho^1(\bar{A}) + e^{2 \xi} \delta_\rho^2(\bar{A})
\\
  (e^\phi) (\bar{A}, \xi) &= 1 + e^\xi \delta_\phi^1(\bar{A}) + e^{2 \xi} \delta_\phi^2(\bar{A})
\end{align}
\end{subequations}
The first order solutions are
\begin{subequations}
\begin{align}
  \delta_m^1 &= \delta_{m0}
\\
  \delta_U^1 &= - \frac{1}{4} \delta_{m0}
\\
  \delta_R^1 &= - \frac{1}{8} \left( \delta_{m0} + \delta_\rho^1 \right)
\\
  \delta_\rho^1 &= \delta_{m0} + \frac{\bar{A}}{3} \delta_{m0}'
\\
  \delta_\phi^1 &= - \frac{1}{4} \delta_\rho^1
\end{align}
\end{subequations}
where $\delta_{m0} (\bar{A})$ is a free function. These expressions are equivalent to Eqs. \eqref{eq:deltamsol}, \eqref{eq:deltausol}, \eqref{eq:deltarsol}, \eqref{eq:deltaphieq} and \eqref{eq:deltarhoeq}.

The second order solutions are as follows.
\begin{widetext}
\begin{subequations}
\begin{align}
  \delta_m^2
  &=
  \frac{\delta_U^1}{5} \left( 2 \delta_U^1 - 6 \delta_m^1 - \delta_\rho^1 \right)
  + \frac{\delta_\rho^1}{40} \left(10 \delta_m^1 - 3 \delta_\rho^1\right)
  + \frac{\delta_\rho^{1\prime}}{10 \bar{A}}
\\
  \delta_U^2 &= \frac{3}{20} \left[ \delta_U^1 \left(\delta_m^1 + \delta_\rho^1 - 2 \delta_U^1\right) - \frac{(\delta_\rho^1)^2}{4} - \frac{\delta_\rho^{1 \prime}}{2 \bar{A}} \right]
\\
  \delta_R^2 &= \frac{1}{16} \left[4 \delta_R^1 \delta_U^1 + 4 \delta_U^2 - \delta_\rho^2 + \delta_\rho^1 \left( \frac{5}{8} \delta_\rho^1 - \delta_R^1 - \delta_U^1 \right) \right]
\\
  \delta_\rho^2 &= \delta_m^2 + \bar{A} \left[\frac{\delta_m^{2 \prime}}{3} - (\delta_\rho^1 - \delta_m^1) \delta_R^{1\prime}\right]
\\
  \delta_\phi^2 &= \frac{1}{32} \left[5 (\delta_\rho^1)^2 - 8 \delta_\rho^2\right]
\end{align}
\end{subequations}
\end{widetext}
The contributions at second order in perturbations (terms containing $\delta^2$) and first order in derivative expansions (terms containing $\delta'$) are clearly evident in $\delta_m^2$ and $\delta_U^2$. These expressions allow for the construction of initial data to second order that consist solely of growing modes, based on the choice of a single function $\delta_{m0}(\bar{A})$.

\subsection{Constraints on Initial Data}

There are a number of constraints on initial data, arising from both physical and desirable considerations. Here, we construct the various constraints, and then apply them to the solutions found above. We assume that initial data is specified in terms of $\tilde{m}$, $\tilde{R}$ and $\tilde{U}$.

\subsubsection{Physical Constraints}
The most basic physical constraint is that $\tilde{m}$ and $\tilde{R}$ are positive, as neither $m$ nor $R$ can become negative. As $U$ can be negative, so may $\tilde{U}$. By spherical symmetry, we also demand that $\tilde{m}$, $\tilde{R}$ and $\tilde{U}$ are even functions, which specifies that they have zero derivative at the origin. Note that this precludes the possibility of a sharp cusp in the energy density.

Next, we require both $R$ and $m$ to be monotonically increasing with $\bar{A}$. These yield the constraints
\begin{align}
  0 &< \tilde{R} + \bar{A} \tilde{R}' \label{eq:condition1}
\\
  0 &\le \tilde{m} + \frac{\bar{A} \tilde{R}}{3(\tilde{R} + \bar{A} \tilde{R}')} \tilde{m}'
\end{align}
respectively. The second of these is equivalent to $\rho \ge 0$.

From Eq. \eqref{eq:newgamma}, we have
\begin{align}
  \tilde{R}^2 \left(\tilde{m} - \tilde{U}^2\right) < \frac{1}{\bar{A}^2} \label{eq:newgammacond}
\end{align}
by requiring $\bar{\Gamma}^2 > 0$ on the initial slice. This is equivalent to requiring that the metric component $g_{AA}$ is positive, so that we have the correct metric signature.

A similar requirement arises from demanding that the initial data does not already describe a black hole. From Eq. \eqref{eq:bhcondition}, this requires
\begin{align}
  \tilde{R}^2 \tilde{m} < \frac{1}{\bar{A}^2}
\end{align}
whenever $\tilde{U}$ is negative.

\subsubsection{Desirable Constraints}

In addition to the physical constraints described above, there are a few desirable constraints also. The first is a numerical issue: it is recommended that initial data connect smoothly to FRW at the boundary of the computational domain. This improves the accuracy of the evolution by reducing unphysical reflections from the boundary.

The second desirable constraint, as discussed previously, is that the initial data should consist of a growing mode only. This motivates us to use the solutions found in the previous subsection, which have been crafted to consist purely of the growing mode to second order in perturbations. This initial data is written in terms of a single function only, $\delta_{m0}$.

The expansion \eqref{eq:seriessol} has two conditions on its applicability: both the perturbative expansion and the derivative expansion must be valid. The expansion is constructed so as to carry equal weight to increasing the order of perturbations or the order of the derivative expansion. As such, this requires $|\delta_{m0}| \ll 1$ and also $|4 \delta_{m0}'/\bar{A} + \delta_{m0}''| \ll |\delta_{m0}|$ (roughly speaking; of course, when $\delta_{m0} = 0$, this condition is difficult to satisfy, but violating it doesn't necessarily mean that the derivative expansion fails to converge). As the solution \eqref{eq:seriessol} suppresses higher order terms more strongly as $\xi \rightarrow - \infty$, it is sufficient to ensure that these conditions are satisfied for the initial data at $\xi = 0$.

If the expansion fails to satisfy either of these criteria, this does not mean that the initial data is invalid; it may still satisfy the physical constraints above. All that this means is that the initial data no longer necessarily consists solely of the growing mode. It is not imperative that the decaying mode is nonexistent in the initial data; so long as the evolution starts with the perturbation sufficiently far outside the cosmological horizon, the decaying mode will have time to decay before the perturbation enters the horizon, which is when the typical black hole formation threshold is calculated.

\subsubsection{Linear Constraints}
We now write the above constraints in terms of the function $\delta_{m0}$, specialized to $w = 1/3$. For analytic purposes, we work to linear order only; it should be borne in mind that these constraints may be too strict when considering the effect of second order terms. Keep in mind that we are mostly interested in positive $\delta_{m0}$, as this corresponds to an overdensity that can undergo gravitational collapse.

Positivity of $\tilde{R}$ requires
\begin{align}
  \delta_{m0}' < \frac{6(4 - \delta_{m0})}{\bar{A}} \label{eq:constraint1}
\end{align}
while positivity of $\tilde{m}$ requires $\delta_{m0} > -1$.

To ensure that the various functions are even functions with zero derivative at the origin, we require $\delta_{m0}$ to also be an even function. Requiring $R$ to be monotonically increasing yields the constraint
\begin{align}
  8 \delta_{m0}' + \bar{A} \delta_{m0}'' &< \frac{6(4 - \delta_{m0})}{\bar{A}} \,.
\end{align}
For $m$ to be monotonically increasing ($\rho \ge 0$), we have
\begin{align}
  0 \le 1 + \delta_{m0} + \frac{\bar{A}}{3} \delta_{m0}' \,.
\end{align}

The requirement that $\bar{\Gamma}^2 > 0$ gives us the strongest constraint on the amplitude of $\delta_{m0}$.
\begin{align}
  \delta_{m0} < \frac{2}{3 \bar{A}^2} \label{eq:constraint2}
\end{align}
Evidently, we cannot be even modestly nonlinear in $\delta_{m0}$ far enough outside the horizon of the initial data. In particular, for $\tilde{A} > 2.5$ (two and a half horizon scales from the origin in the initial data), $\tilde{m}$ is constrained to be less than 1.1, and so we are well within the linear regime using these variables. The only permitted nonlinear evolution that the initial data cannot capture is thus within the first couple of horizon spans. So long as the evolution begins with features of order the horizon scale or larger, this perturbative approach to initial data should be completely sufficient.

For a black hole to be present in the initial data requires $\tilde{U}$ to be negative, corresponding to $4  < \delta_{m0}$. As this is well outside the linear regime, we do not expect any black holes to exist in the initial data constructed above.

To match smoothly to FRW at large $\bar{A}$ simply requires $\delta_{m0} \rightarrow 0$ as $\bar{A}$ approaches the boundary of the computational domain. Whenever the perturbative expansion \eqref{eq:seriessol} is valid at the level at which is it truncated, the initial data consists of the growing mode only, by construction.

The perturbative expansion requires $|\delta_{m0}| \ll 1$. It is harder to accurately state what the derivative expansion requires; a sufficient condition is $|4 \delta_{m0}'/\bar{A} + \delta_{m0}''| \ll |\delta_{m0}|$. One can also look to the derivative expansion terms in $\delta_m^2$ and require them to be less than $\delta_m^1$, which yields the constraint
\begin{align}
  |\delta_{m0}| \gg \frac{1}{30} \left|\frac{4 \delta_{m0}'}{\bar{A}} + \delta_{m0}'' \right| \,.
\end{align}
To be entirely satisfied that the derivative expansion can be truncated with minimal error requires us to compare the next order term in the derivative expansion. Thankfully, the derivative expansion terms are not difficult to compute. At first order in perturbations and up to second order in the derivative expansion, we have the following terms.
\begin{subequations}
\begin{align}
  \delta_m^1 &= e^{\xi} \delta_{m0}
\\
  \delta_m^2 &= \frac{e^{2 \xi}}{30} \left(\frac{4 \delta_{m0}'}{\bar{A}} + \delta_{m0}''\right)
\\
  \delta_m^3 &= \frac{e^{3\xi}}{2520} \left(
    \delta_{m0}''''
  + \frac{8 \delta_{m0}'''}{\bar{A}}
  + \frac{8 \delta_{m0}''}{\bar{A}^2}
  - \frac{8 \delta_{m0}'}{\bar{A}^3}
  \right)
\end{align}
\end{subequations}
Assuming one can accurately calculate fourth order derivatives, it should be safe to truncate the expansion at second order in perturbation theory when the third order term here is small compared to the others.

\subsection{Relation to the Literature}
We now relate our results to other works in the literature.

\subsubsection{Density and Mass Relationships}

We begin by writing the relationship between $\tilde{\rho}$ and $\tilde{m}$ in various ways. The exact relationship is given by Eq. \eqref{eq:rhomrel}, and can also be written as
\begin{align}
  \tilde{\rho} &= \frac{[(\bar{A} \tilde{R})^3 \tilde{m}]'}{[(\bar{A} \tilde{R})^3]'} \,.
\end{align}
Using the result \eqref{eq:limit}, this can be integrated to give
\begin{align}
  \tilde{m} = 1 + \frac{1}{(\bar{A} \tilde{R})^3} \int_\infty^{\bar{A}} [(\bar{A} \tilde{R})^3]' \delta_\rho d\bar{A}
\end{align}
where $\delta_\rho = \tilde{\rho} - 1$ is just the usual cosmological density contrast. The linearized version of these two relationships are
\begin{align}
  \delta_\rho &= \frac{[\bar{A}^3 \delta_m]'}{3 \bar{A}^2} = \delta_m + \frac{\bar{A}}{3} \delta_m'
  \label{eq:rhoequal}
\\
  \delta_m &= \frac{3}{\bar{A}^3} \int_\infty^{\bar{A}} \bar{A}^2 \delta_\rho d\bar{A}
\end{align}
with $\delta_m = \tilde{m} - 1$.

\subsubsection{Fractional Mass Excess}

A number of papers compute the fractional mass excess inside the cosmological horizon, given the (much-overworked) symbol $\delta$. This quantity has been computed in different ways, with variations based on gauge and a choice on how the background mass is quantified. Carr's original formulation \cite{Carr1974, Carr1975} used a uniform Hubble constant gauge with two shells of unequal density. A more recent approach by Harada \textit{et al.} \cite{Harada2013} revisited Carr's approach. Furthermore, they conveniently represented their result in the gauge we have employed.

The fractional mass excess associated with a perturbation is computed when the perturbation enters the horizon. In our language, the horizon is located at $\bar{A}_H = e^\xi$ under background evolution; this is the comoving radius at which the comparison is performed \footnote{Note that the actual particle horizon will be slightly different due to nonlinear evolution.}. The perturbation is said to enter the horizon when the density at the horizon is equal to the FRW density, followed by a sub-FRW underdensity to compensate for the overdensity inside the horizon. The time of horizon crossing is a gauge-dependent quantity, but within a given gauge, it is well-defined.

The fractional mass excess $\delta$ is then defined by
\begin{align}
  \delta = \frac{m(\bar{A}_H)}{m_b(\bar{A}_H)} - 1
\end{align}
at this time, where $m_b(\bar{A}_H)$ is given by
\begin{align}
  m_b(\bar{A}) = \frac{4 \pi}{3} R_H^3 \bar{A}_H^3 \rho_b \,.
\end{align}
In our language, the fractional mass excess is simply given by
\begin{align}
  \delta = \tilde{m} (\bar{A}_H) - 1 \,,
\end{align}
evaluated at the time of horizon crossing.

Given an initial $\delta_{m0}$, we can compute the expected fractional mass excess under the linear evolution described above. Let $\bar{A}_{pert}$ be the comoving radius at which $\delta_\rho(\bar{A}_{pert}) = 0$ in the initial data. The horizon will grow to this scale at time $\xi = \ln (\bar{A}_{pert})$. Taking into account the growth of $\tilde{m}$ during this time, the fractional mass excess will be approximated by
\begin{align}
  \delta = \delta_{m0} (\bar{A}_{pert}) (\bar{A}_{pert})^{2(1-\alpha)} \,,
\end{align}
although non-linearities typically become important before the perturbation is completely inside the horizon. 

This result mirrors the result derived by Polnarev and Musco \cite{Polnarev2006}, who defined
\begin{align}
  \tilde{\delta} (\bar{A}) = (\tilde{m} (\bar{A}) - 1) e^{2(\alpha - 1) \xi}
\end{align}
in our language, such that under linear evolution, $\tilde{\delta}$ is constant. When evaluated at $\bar{A}_{pert}$, $\tilde{\delta} = \delta$, the fractional mass excess.

Harada \textit{et al.} \cite{Harada2013} calculated that black holes are likely to form for $\delta > \delta_c$ with
\begin{align}
  \delta_c = \frac{3(1+w)}{5 + 3w} \sin^2\left(\frac{\pi \sqrt{w}}{1 + 3w}\right) \,. \label{eq:crit}
\end{align}
This result is given in the gauge we employ here. For $w = 1/3$, this yields $\delta_c \approx 0.4135$.

In order to extract the fractional mass excess from a numerical run, one should monitor the density $\tilde{\rho}$ at the horizon. When it drops beneath one, read off $\delta = \tilde{m} - 1$ at that point in space and time.

Using the fractional mass excess as an indicator of whether or not a black hole will form has some obvious deficiencies: if the density profile is such that the density contrast becomes negative briefly before becoming positive again, this method will not capture the appropriate overdensity. Similarly, a long tail of a positive density contrast can also lead to erroneous results. In general, the dependence of black hole formation on the details of the density profile are not particularly well captured by the estimate in Eq. \eqref{eq:crit}, although it works reasonably well as a rule of thumb. This is to be expected, as Harada \textit{et al.} were more interested in correctly estimating the dependence of $\delta_c$ on the equation of state than the initial profile. In order to overcome some of these deficiencies, Nakama \textit{et al.} \cite{Nakama2013} have begun exploring more appropriate quantities for identifying black hole formation based on the density profile.

\subsubsection{Curvature Perturbation}

An approach to generating appropriate initial conditions based on curvature perturbations has been championed by Polnarev \cite{Polnarev2006, Polnarev2012, Nakama2013}. This approach is based around the fact that
\begin{align}
  \Gamma^2 = 1 - K A^2
\end{align}
for a curved FRW universe. Accordingly, Polnarev and Musco \cite{Polnarev2006} suggested letting
\begin{align}
  \Gamma^2 = 1 - K(A, t) A^2 \label{eq:Kdef}
\end{align}
and describing the dynamics of the curvature perturbation $K(A, t)$. It was claimed that this picks out the growing mode; indeed, the growing mode is the particular solution of an ODE driven by the existence of the curvature perturbation. The decaying mode exists as the homogeneous solution of the equation, which was neglected entirely.

The method utilizes a time-dependent small parameter $\epsilon^2$ where
\begin{align}
  \epsilon = \frac{1}{\dot{a}^2 r_0^2}
\end{align}
for some arbitrary, initially super-horizon comoving scale $r_0$. It is useful to note that
\begin{align}
  \epsilon = \frac{t_0^2}{\alpha^2 r_0^2} e^{2(1-\alpha)\xi} \,.
\end{align}
Eventually, the various quantities $\delta_X^1$ are related to $K(A, t_0)$ at first order in perturbations. The relationship between $K$ and our variables comes from matching Eqs. \eqref{eq:gammanew} and \eqref{eq:Kdef}.
\begin{align}
  K(\bar{A}, \xi) = \frac{1}{R_H^2} e^{2 (\alpha - 1) \xi} \tilde{R}^2 \left(\tilde{m} - \tilde{U}^2\right)
\end{align}
Substituting $\tilde{m}$, $\tilde{U}$ and $\tilde{R}$ to linear order, we obtain
\begin{align}
  K(\bar{A}, \xi) &= \frac{1}{R_H^2} \left(\delta_m^1 - 2 \delta_U^1\right)
\\
  &= \frac{1}{R_H^2} \left(1 + \alpha \right) \delta_{m0} \,.
\end{align}

There are two ways to use $K$. The first is just to use it as a method to construct initial data. However, because of various approximations, when the actual curvature perturbation is computed from initial data constructed using $K$, it doesn't return exactly $K$. The second approach is to use $K$ as an evolution variable to replace $U$. The issue with this approach is that $K$ is related to $U^2$ instead of $U$, and as such is ambiguous whenever $U$ may be negative.

Despite these deficiencies, many aspects of this method were used as inspiration for our derivations, and we mostly arrive at the same results. Two major points that we changed were to construct completely dimensionless equations, and to completely ignore the auxiliary variable $K$. Working directly with $U$, $m$ and $R$ (or their tilded proxies) yields a more precise and clearer description of the system. In particular, our equations of motion are completely absent of any scale (especially the scale $r_0$ which depends on the initial data), some auxiliary equations of motion have been eliminated completely, the growing and decaying modes have been distinctly identified, and all initial data is constructed based around the linear mass perturbation $\delta_{m0}$. If desired, the curvature perturbation can always be computed.

\subsubsection{Bardeen Gauge Invariant Variables}

In order to connect to cosmological perturbation theory, it is useful to find the relation between our variables and the Bardeen gauge invariant variables \cite{Bardeen1980}. As the Bardeen variables are gauge invariant at linear order, here we work to linear order with our variables also. The relationships are derived in Appendix \ref{app:perturbation}, and reproduced here for convenience\footnote{The quantities $\Phi$ and $\Psi$ correspond to the Newtonian gauge metric
\begin{align}
  ds^2 = - (1 + 2 \Phi) dt^2 + a^2 (1 - 2 \Psi) dx^2 \,.
\end{align}
}.
\begin{align}
  \Phi = \Psi = - \frac{1}{2} e^{2 (\alpha - 1)\xi} \int^\infty_{\bar{A}} \bar{A} \, \delta_m d \bar{A}
\end{align}
As $\delta_m$ grows as $\exp(2(1-\alpha) \xi)$ in the linear regime, we see that $\Phi$ and $\Psi$ are constant in this description, as we expect. We also find that the Bardeen variables are equal, as expected for a perfect fluid with no anisotropic shear stress.

Heuristically, we can see that for superhorizon scales, even very linear contributions to $\delta_m$ can cause nonlinear contributions to $\Phi$ and $\Psi$. This suggests that perturbation theory in $\delta_m$ and $\delta_U$ has a broader regime of validity than standard cosmological perturbation theory. We now show this more rigorously.

By using $\delta_\rho = \delta_m + \bar{A} \delta_m'/3$ and integration by parts, we can write
\begin{align}
  \int^\infty_{\bar{A}} \bar{A} \delta_m d \bar{A}
  =
  \bar{A}^2 \delta_m - 3 \int^\infty_{\bar{A}} \bar{A} \delta_\rho d \bar{A}
\end{align}
where the limit $\bar{A}^2 \delta_m \rightarrow 0$ as $\bar{A} \rightarrow \infty$ by Eq. \eqref{eq:limit} and the squeeze theorem. Using this, we have
\begin{align}
  \Phi = \frac{1}{2} e^{2 (\alpha - 1)\xi} \left(- \bar{A}^2 \delta_m + 3 \int^\infty_{\bar{A}} \bar{A} \, \delta_\rho d \bar{A}\right) \,.
\end{align}

Generally speaking, when $\delta_m$ is large and positive, the second term will be subdominant. To see this, consider some $\bar{A}$ beyond which $\delta_\rho \le 0$. Then
\begin{align}
  \bar{A}^2 \delta_m &= \frac{3}{\bar{A}} \int_{\bar{A}}^\infty \bar{A}^2 (-\delta_\rho) d\bar{A} > 3 \int_{\bar{A}}^\infty \bar{A} (-\delta_\rho) d\bar{A}
\end{align}
where the equality comes from integrating Eq. \eqref{eq:rhoequal}. Thus, we can write
\begin{align}
  \Phi, \, \Psi &\simeq - \frac{1}{2} \bar{A}^2 \delta_{m0}
\end{align}
where we use the linear initial conditions. Requiring both $\Phi > -1/2$ and $\Psi < 1/2$ to ensure a non-singular metric, we have
\begin{align}
  \delta_{m0} &< \frac{1}{\bar{A}^2}
\\
  \delta_{m0} &> - \frac{1}{\bar{A}^2}
\end{align}
as conditions on the initial data (at linear order). These are very similar to the condition \eqref{eq:constraint2}. In particular, this shows that the constraint \eqref{eq:constraint2} on $\delta_{m0}$ is not constraining the standard cosmological perturbation theory variables to be incredibly small at large radii.

\subsubsection{Connection to Inflation}

We now address the question of how to relate our evolution variables to a given profile generated by inflation. Our goal is to obtain an expression for $\delta_{m0}$ based on these initial conditions.

The first step is to identify the initial time for our evolution. Consider that at the end of inflation, we know the Hubble scale $H_i$ and the characteristic proper length scale of the perturbation, $R_\ast$. The ratio of the proper length scale of the perturbation to the horizon ratio, as a function of $\xi$, is then
\begin{align}
  \kappa \equiv \frac{R_\ast (\xi)}{R_{horizon} (\xi)} = R_\ast H_i e^{(\alpha - 1)(\xi - \xi_i)}
\end{align}
where $\xi_i$ is the value of $\xi$ at the end of inflation. We would like to start our evolution $(\xi = 0)$ when the perturbation is still well outside the horizon, $\kappa \sim 5 - 10$. This then sets $\xi_i$ to be
\begin{align}
  \xi_i = \frac{1}{1 - \alpha} \ln \left(\frac{\kappa}{R_\ast H_i}\right) \,.
\end{align}
For typical large scale perturbations, $\xi_i$ will be somewhat large and negative. From Eq. \eqref{eq:seriessol}, we see that both the perturbative and derivative expansion approximations become increasingly better for more negative $\xi_i$. As such, it should be sufficient to describe perturbations in this regime by using just the linear expansion. The goal of obtaining initial conditions is then successful once we obtain $\delta_{m0}$.

Let us assume that at the end of inflation, we have the gauge invariant metric perturbation $\Phi$ or $\Psi$, and the gauge invariant velocity perturbation, $V$ (see Appendix \ref{app:perturbation}). From these, we can obtain $\delta_m$ and $\delta_U$ as
\begin{align}
  \delta_m &= \frac{2}{\bar{A}} e^{2 (1 - \alpha)\xi_i} \Phi'
\\
  \delta_U &= \frac{e^{\xi_i} V'}{R_H \bar{A}}
\end{align}
where primes again refer to derivatives with respect to $\bar{A}$\footnote{Note that $\bar{A}$ is the coordinate system defined by $A = R_H \bar{A}$, where $R_H$ is the horizon scale at $\xi = 0$. One should be careful in translating a coordinate system at inflation into the coordinate system at the beginning of this evolution.}. Knowing $\Delta$ provides no further information, as $\Delta = \delta_\rho$, and $\delta_m$ can be converted from one to the other. Note also that we cannot recover any information about $\delta_R$, which is gauge-dependent. This is unimportant; $\delta_R$ has no effect on linear evolution, and rapidly converges to a solution that tracks $\delta_m$ and $\delta_U$ appropriately. We should also point out that these formulas require $\Phi', \, V' \rightarrow 0$ sufficiently rapidly as $\bar{A} \rightarrow 0$ (this is a requirement of spherical symmetry).

Now that we know $\delta_m$ and $\delta_U$ at the end of inflation, we need to write these in combinations that consist of the growing and decaying modes appropriately. From Eq. \eqref{eq:betadefs}, we have
\begin{align}
  \delta_m &= G e^{2 (1-\alpha)\xi_i} + D e^{\xi_i 3 (w-1) \alpha /2}
\\
  \delta_U &= - \frac{\alpha}{2} G e^{2 (1-\alpha)\xi_i} + \frac{1}{2} D e^{\xi_i 3 (w-1) \alpha /2}
\end{align}
where we make use of Eq. \eqref{eq:deltameq}. We are only interested in $G$, which we obtain through
\begin{align}
  G &= \frac{1}{1 + \alpha} (\delta_m - 2 \delta_U) e^{2 (\alpha - 1)\xi_i} \,.
\end{align}
This provides a simple formulation to convert from Bardeen gauge invariant variables at the end of inflation to initial data for numerical evolution.

\section{Hernandez-Misner Formalism} \label{sec:hmformalism}

Though the Misner-Sharp formalism is very practical for the numerical study of spherically symmetric spacetimes, it breaks down if a singularity is formed. When simulating the formation of a primordial black hole, this makes it difficult to calculate its final mass. In order to deal with the formation of a singularity, Hernandez and Misner developed a formalism that uses radially outward traveling light rays to define a time coordinate \cite{Hernandez1966}. The idea is to define a time coordinate based on the time at which an outgoing radial light ray starting from a given event reaches a distant observer. As time runs slower in a deep gravitational well, in this coordinate system, it takes infinite coordinate time for an event horizon to form, and so a singularity never forms. This formalism, which has typically been built on the work of Baumgarte \textit{et al.} \cite{Baumgarte1995}, has been applied to primordial black hole formation in Refs. \cite{Niemeyer1999, Hawke2002, Musco2004, Musco2008, Nakama2013}. Our presentation of the formalism is somewhat simplified compared to previous implementations, as we eliminate the need for various thermodynamic quantities.

Hernandez and Misner define a new coordinate $u$ by the following differential relation.
\begin{align} \label{eq:dudef}
e^{\psi} du = e^{\phi} dt - e^{\lambda \slash 2} dA
\end{align}
Here, $e^\psi$ is an integrating factor used to make $du$ a perfect differential. Note that if $du=0$ then $e^{\phi} dt=e^{\lambda / 2} dA$, so along a path such that $d \theta = d \phi =0$ we have by Eq. \eqref{eq:MSmetric} that $ds^2=0$. Furthermore $du=0$ implies that
\begin{align}
  \frac{dA}{dt} = e^{\phi - \lambda / 2} \label{eq:nullray} \,.
\end{align}
Together, these relations show that $u$ is an outgoing null coordinate. This coordinate system is known as ``observer time coordinates'', as each $u$ corresponds to what an observer at infinity sees looking towards the origin. Substituting Eq. \eqref{eq:nullray} into the metric \eqref{eq:MSmetric}, we find that the metric in this coordinate system is
\begin{align}
  ds^2 = -e^{2 \psi} du^2 -2 e^{\psi + \lambda / 2} du dA + R^2 d \Omega^2 \,. \label{eq:HMmetric}
\end{align}

Given this metric, the equations of motion could be derived from Einstein's equations. However, this is an incredibly tedious computation, and the same results can be arrived at by appropriately transforming quantities from the Misner-Sharp formalism. In order to carry this derivation out, we first define the following invariant operators.
\begin{align}
D_t = u^{\mu} \partial_{\mu}, \qquad D_r = v^{\mu} \partial_{\mu} \label{eq:HMderiv}
\end{align}
Here, $u^{\mu} = (e^{-\phi}, 0, 0, 0)$ is the comoving fluid four-velocity and $v^{\mu} = (0,e^{-\lambda / 2},0,0)$ in the Misner-Sharp coordinate system. By using the metric \eqref{eq:MSmetric} and the form of these operators in the Misner-Sharp coordinate system, we see that $D_t$ is the derivative with respect to proper time at a fixed spatial coordinate, and $D_r$ is the derivative with respect to proper distance at a fixed time and angle.

We can write the Misner-Sharp equations in terms of these operators as follows.
\begin{subequations} \label{eq:MHeqs}
\begin{align}
  D_t R &= U \label{eq:rdotMH}
\\
  D_t {m} &= - 4 \pi R^2 P U \label{eq:mdotMH}
\\
  D_t {U} &= - \left(\frac{\Gamma D_r P}{ \rho + P} + \frac{m + 4 \pi R^3 P}{R^2} \right) \label{eq:UdotMH}
\\
  D_r m &= 4 \pi \rho R^2 D_r R  \label{eq:mprimeMH}
\\
  \Gamma^2 &= 1 + U^2 - \frac{2m}{R}  \label{eq:gamma}
\end{align}
\end{subequations}
Here we have used $\Gamma = e^{-\lambda/2} R' = D_r R$. We have omitted the equation for $\phi'$ since this quantity does not appear in the Hernandez-Misner coordinate system.

We now look at how various quantities transform when changing to the null coordinate. By looking at the transformation properties of the metric, we see that the circumferential radius $R$ is invariant under the coordinate transformation. Both $P$ and $\rho$ are scalar quantities that do not transform. The operators $D_t$ and $D_r$ are invariant by construction. Thus $U$ is also invariant by Eq. \eqref{eq:rdotMH}. Equation \eqref{eq:mprimeMH} can be understood as a definitional equation for $m$; it too is invariant under the coordinate transformation. When combined with the boundary condition $m(A = 0) = 0$, this defines $m$ consistently across reference frames.

Thus, to arrive at dynamical equations specific to the Hernandez-Misner coordinate system we can simply substitute the appropriate expressions for $D_t$ and $D_r$ in Eqs. \eqref{eq:MHeqs}. Through careful use of the chain rule, the derivatives become
\begin{align} \label{eq:invariantderivs}
D_t = e^{- \psi} \partial_u,
\qquad
D_r = e^{- \lambda / 2} \partial_{A} - e^{-\psi} \partial_u \,.
\end{align}
Substituting these into the invariant form of the Misner-Sharp equations \eqref{eq:MHeqs} and simplifying, we arrive at the Hernandez-Misner evolution equations.
\begin{subequations} \label{eq:HM}
\begin{align}
  \dot{R} &= e^{\psi} U \label{eq:HMR}
\\
  \dot{m} &= - e^{ \psi } 4 \pi R^2 P U \label{eq:HMM}
\\
  \dot{U} &= - \frac{e^\psi}{1 - w - \tilde{Q}} \bigg[
\Gamma e^{- \lambda / 2} \frac{P'}{\rho + P}
+ \frac{m + 4 \pi R^3 P}{R^2}
\nonumber\\&\quad
- \frac{\Gamma \rho}{\rho + P} e^{-\psi} \dot{\tilde{Q}}
+ (w + \tilde{Q}) \left(2 \frac{U\Gamma}{R} + e^{-\lambda/2} U' \right)\bigg] \label{eq:HMu}
\\
  \rho &= \frac{e^{-\lambda/2} m'}{4 \pi R^2 (\Gamma - (w + \tilde{Q}) U)}
  = \frac{m'}{4 \pi R^2 R'} \frac{\Gamma + U}{\Gamma - (w + \tilde{Q}) U} \label{eq:HMmprime}
\\
  \Gamma^2 &= 1 + U^2 -\frac{2 m}{R} \label{eq:HMGamma}
\\
  e^{\lambda/2} &= \frac{R'}{\Gamma + U} \label{eq:HMlambda}
\end{align}
\end{subequations}
Here, overdots denote derivative with respect to $u$ and primes denote derivatives with respect to $A$. In the density equation, we wrote $P = \rho (w + \tilde{Q})$, where we anticipate that $\tilde{Q}$ will not depend on density; doing the same in the $U$ equation of motion leads to difficulties (see Section \ref{sec:code}). Something to be cautious of in this coordinate system is that
\begin{align}
  \Gamma = D_r R = e^{- \lambda / 2} R' - U
\end{align}
where the last term did not appear in the Misner-Sharp coordinates. Deriving the equation for $\dot{U}$ is a little tricky; one must be careful with $D_r \rho$. This is most easily calculated by looping through the Misner-Sharp coordinates. Let
\begin{align}
  D_r \rho &= e^{- \lambda / 2} \rho' - e^{-\psi} \partial_u \rho
\end{align}
and then use
\begin{align}
  e^{-\psi} \partial_u \rho = D_t \rho = e^{-\phi} \partial_t \rho \,.
\end{align}
Using the result \eqref{eq:rhodot} followed by $U'/R' = D_r U / \Gamma$ to convert back to the null coordinate system yields the equation of motion for $\partial_u U$.

The system of equations \eqref{eq:HM} is incomplete, as we require an equation for calculating $\psi$. As $du$ is a perfect differential, we may write
\begin{align}
  \partial_t \partial_A u = \partial_A \partial_t u \,.
\end{align}
and use the derivatives from Eq. \eqref{eq:dudef} to obtain
\begin{align} \label{eq:psi1}
  (D_t + D_r) \psi = D_r \phi + \frac{1}{2} D_t \lambda \,.
\end{align}
Next, note that we can rewrite Eq. \ref{eq:stressr} as
\begin{align}
  D_r \phi = - \frac{D_r P}{\rho + P} \label{eq:Drphi}
\end{align}
and Eq. \eqref{eq:lambdadot2} as
\begin{align} \label{eq:Dtlambda}
  D_t \lambda = \frac{2 D_r U}{\Gamma} \,.
\end{align}
The quantity $D_k \equiv D_t + D_r = e^{-\lambda/2} \partial_A$ is a derivative along an outgoing null ray (as the derivative is taken with constant $u$). Combining all these results, we can write Eq. \eqref{eq:psi1} as
\begin{align}
  \Gamma D_k \psi &= D_k U + \frac{m}{R^2} + 4 \pi R P \label{eq:HMpsi}
\end{align}
where we employ Eq. \eqref{eq:UdotMH}. It turns out that this is not necessarily the best equation to evolve, however. Instead, apply $D_k$ to \eqref{eq:HMGamma} we obtain
\begin{align}
  \Gamma D_k \Gamma = U D_k U + \frac{D_k R m}{R^2} - \frac{D_k m}{R} \,,
\end{align}
and substitute $D_k m$ from \eqref{eq:HMmprime}, $m/R^2$ from \eqref{eq:HMpsi}, and $D_k R = U + \Gamma$. After the dust settles, we obtain
\begin{align} \label{eq:HMpsi3}
  e^\psi D_k [e^{-\psi} (\Gamma + U)] = - 4 \pi R (\rho + P)
\end{align}
or alternatively,
\begin{align} \label{eq:HMpsi2}
  \partial_A [e^{-\psi} (\Gamma + U)] = - e^{\lambda/2 - \psi} 4 \pi R (\rho + P) \,.
\end{align}
When combined with an appropriate boundary condition, this closes the Hernandez-Misner system of equations.

A number of boundary conditions from the Misner-Sharp formalism are invariant under the coordinate transformation. In particular, at the origin, $R(0) = 0$, $m(0) = 0$ and $U(0) = 0$. As previously, $\Gamma(0) = 1$. The conditions to match smoothly to FRW also remain the same as previously: all local quantities must be equivalent to their FRW values, and the same mass matching condition must also be met. We discuss the outer boundary condition in the cosmological context below.

Beware that $R$ is not an odd function of $A$ any more, as continuing the function to negative values of $A$ continues to go back further in time. If instead the time coordinate increases again as $A$ becomes negative, then $R$ becomes even, but with a cusp at the origin. The implication of this is that quantities such as $\rho$ which were even functions in the Misner-Sharp formalism are no longer even functions with zero derivative at the origin.

The Hernandez-Misner formalism was originally formulated to describe a collapsing object surrounded by vacuum in an asymptotically flat spacetime. The time coordinate was taken to be the clock time of a static observer at future null infinity. As we do not have an asymptotically flat spacetime, such an observer doesn't exist. Instead, we place an observer at the boundary of the computational domain. At this boundary, we know the quantities $m$, $R$, $U$, $\Gamma$, $\rho$ and $\lambda$, as well as the coordinates $u$ and $A$. Let us also assume that we know the cosmic time coordinate that corresponds to this $u$ and $A$ (this can be found when initial data is constructed, see below). Then by the differential relation \eqref{eq:dudef} with $dA = 0$, we have
\begin{align}
  \frac{du}{dt} = e^{\phi - \psi} \,.
\end{align}
It is a convenient gauge choice to have $u$ for this observer correspond to the observer's cosmic time coordinate $t$ also, which sets $du/dt = 1$ and thus $\psi = \phi$. Given the Misner-Sharp time, we can compute $e^\psi$ at the boundary through
\begin{align}
  e^\psi = e^\phi &= \left(\frac{\rho_b}{\rho}\right)^{3 \alpha w/2}
\end{align}
with
\begin{align}
  \rho_b = \frac{3 \alpha^2}{8 \pi t^2}
\end{align}
where we have used Eqs. \eqref{eq:MSphi2}, \eqref{eq:rhosol}, \eqref{eq:t0def} and \eqref{eq:asol}. When the density matches the appropriate FRW energy density, $e^\psi = 1$.

\begin{figure}[t]
    \includegraphics[width=\columnwidth]{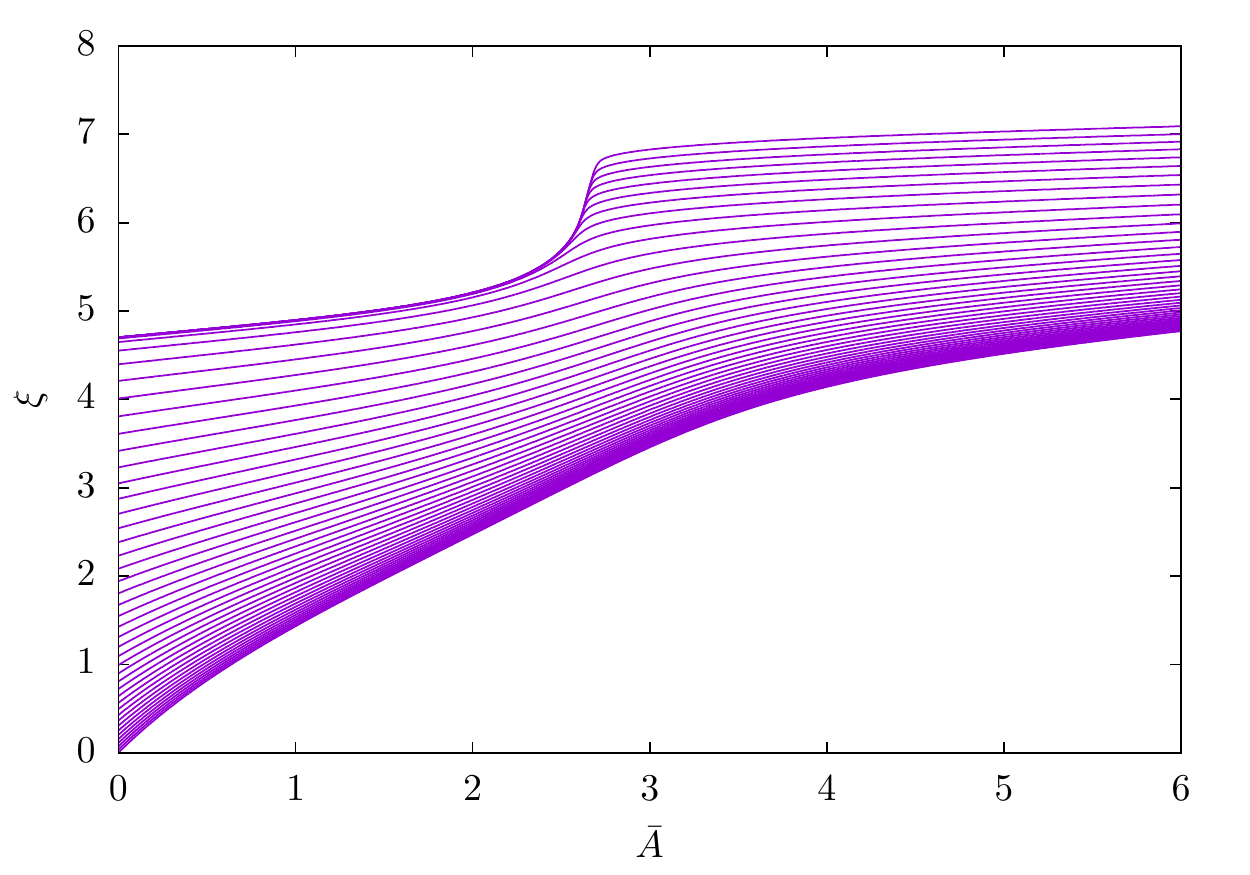}
    \caption{\label{fig:slicing}Plot of null slices in Misner-Sharp coordinates for initial data from which a black hole forms. Slices of constant time in Misner-Sharp coordinates correspond to horizontal lines (constant $\xi$). The way in which null slices avoid the formation of a black hole is evident, and corresponds to an essentially vanishing lapse. Plot data comes from the black hole evolution described in Section \ref{sec:plots}.}
\end{figure}

Let us now show that a horizon never forms in this coordinate system. The condition under which a trapped surface forms is
\begin{align}
  U + \Gamma \equiv D_k R \le 0 \,.
\end{align}
This relationship indicates that despite being evaluated along an outward-traveling null ray, the areal radius $R$ of the ray is decreasing. From Eq. \eqref{eq:HMpsi3} we see that
\begin{align}
  D_k [e^{-\psi} D_k R] = - e^{-\psi} 4 \pi R (\rho + P) \le 0 \,.
\end{align}
Thus, $e^{-\psi} D_k R$ decreases monotonically with increasing comoving radius. As $e^{-\psi} D_k R$ is positive for our observer at the boundary, it must then be positive everywhere. Thus, the condition for a trapped surface to form never eventuates, and a horizon never forms. We see how the null slicing condition avoids a black hole in Figure \ref{fig:slicing}, where we plot an example of null slices in Misner-Sharp coordinates.

Musco \textit{et al.} \cite{Musco2004} point out that as $D_k R$ and $e^\psi$ can only reach zero together as $u \rightarrow \infty$, care must be taken in the numerical evolution equations to ensure that this synchronization occurs, else negative values of $D_k R$ may result. The evolution equation \eqref{eq:HMpsi2} ensures the desired behavior regardless of numerical imprecision, while Eq. \eqref{eq:HMpsi} does not.

The denominator $\Gamma - w (1 + \tilde{Q}) U$ in Eq. \eqref{eq:HMmprime} is of concern. This combination arises from
\begin{align}
  D_k m = 4 \pi \rho R^2 (\Gamma - (w + \tilde{Q}) U)
\end{align}
where it is not guaranteed that $\Gamma - (w + \tilde{Q}) U$ is positive; at sufficiently large radii, the derivative of $m$ along the outgoing null ray can become negative. However, as $\rho$ is always positive, $D_k m$ and $\Gamma - (w + \tilde{Q})U$ must vanish at the same event in spacetime. While a gridpoint is unlikely to exactly land on this point, numerical imprecision is likely to reduce accuracy in $\rho$ in the vicinity of this region. For the background FRW evolution, the location of this crossover is $R = 1/wH$, at a radius a few times the horizon size (depending on $w$). So long as this radius is greater than the computational domain on the initial data for the Hernandez-Misner evolution, any issues related to this turnaround should be absent.

For initial data generated as described below, we can compare the crossover radius to the size of the computational domain for the FRW background. For the FRW background, the null ray's position is given by
\begin{align}
  \bar{A} = \frac{\alpha}{1-\alpha} [e^{(1-\alpha) \xi} - 1] \,.
\end{align}
When the null ray hits the boundary $\bar{A}_\ast$, we have
\begin{align}
  \xi_\ast = \frac{1}{(1-\alpha)} \ln\left(1 + (1 - \alpha) \frac{\bar{A}_\ast}{\alpha}\right) \,.
\end{align}
The crossover radius for the FRW background is
\begin{align}
  \bar{A}_c = \frac{1}{w} e^{(1 - \alpha)\xi}
\end{align}
which at $\xi_\ast$ becomes
\begin{align}
  \bar{A}_c = \frac{1}{w} \left(1 + \frac{1 - \alpha}{\alpha} \bar{A}_\ast \right) \,.
\end{align}
Demanding $\bar{A}_c \gg \bar{A}_\ast$ yields the condition
\begin{align}
  1 \gg - \frac{1+w}{2} \bar{A}_\ast \,.
\end{align}
Thus, for all reasonable $w$, the crossover radius will be well outside the computational domain when the initial data is constructed, and this issue should not arise. We expect this result to still hold when using perturbed initial data, as the crossover radius is so much larger than the outer boundary, which we take to match to FRW (even if only asymptotically).

\subsection{Initial Data}

The biggest issue with this formalism is that it is uncommon to posit initial data along a null slice. As such, it is typical to use the Misner-Sharp formalism to generate initial data, before converting to the Hernandez-Misner formalism.

We suggest generating initial data for the Misner-Sharp formalism, and evolving this data forwards in time. In addition to evolving the physical variables, compute the trajectory of a null ray starting at the origin, and store the values of $U$, $m$ and $R$ along the trajectory. If no black hole holes, then the Hernandez-Misner formalism is not required. If a black hole does form, then the domain can be truncated, excising the black hole, and evolution continued until the null ray reaches the outer boundary of the computational domain.

As no inner boundary condition can be known for the truncated domain, we suggest repeatedly truncating the domain to remove unphysical artifacts. From experimentation, we found that setting a Dirichlet boundary condition of constant $\tilde{\rho}$ at the inner boundary worked well to stabilize the boundary. Note that what occurs at the inner boundary cannot causally affect the evolution at the photon's location, as the speed of sound is less than the speed of the photon.

Once the null ray reaches the outer boundary, the data along this ray can be used as the initial conditions for the Hernandez-Misner formalism. This allows the evolution to be continued, and thus for the final black hole mass to be determined.

In order to record the initial conditions for Hernandez-Misner evolution, a radial null geodesic needs to be tracked. This is described by $du = 0$, which is equivalent to
\begin{align}
  \frac{dA}{dt} = e^\phi \frac{\Gamma}{\partial_A R} \,.
\end{align}
Writing this in terms of $\xi$ and tilded variables, this is
\begin{align}
  \partial_\xi \bar{A} = \alpha e^{\phi} \frac{\bar{\Gamma}}{\partial_{\bar{A}} (\bar{A} \tilde{R})} \,.
\end{align}
The initial condition is that the geodesic begins at the origin, $\bar{A}(\xi = 0) = 0$.

\section{Cosmological Application of the Hernandez-Misner Formalism} \label{sec:hmcosmology}

In the Misner-Sharp formalism, we had much success in factoring the background cosmological evolution out of the evolution variables. This had the benefit of evolving dimensionless quantities that were all typically of $O(1)$, completely stable background evolution, and a simple manner in which to perform linear perturbation theory, which in turn led to an appropriate outer boundary condition. In the Hernandez-Misner formalism, it becomes more challenging to factor out the background evolution, as different values of $A$ are located at different cosmic times $t$. In this section, we develop an approach that accomplishes the appropriate factorization. A similar approach was suggested by Nakama \textit{et al.} \cite{Nakama2013}, although the details of their implementation are somewhat opaque.

We wish to evolve the variables $U$, $m$ and $R$. After each timestep, the auxiliary variables $\Gamma$, $\rho$, $P$, $e^{\lambda/2}$ and $e^\psi$ must also be computed. A further quantity describing the cosmic time $t(u, A)$ can also be evolved. This quantity obeys two relationships, based on Eq. \eqref{eq:dudef}. At fixed value of $A$,
\begin{align} \label{eq:dtdu}
  \frac{dt}{du} = e^{\psi - \phi}
\end{align}
while at fixed $u$,
\begin{align} \label{eq:dtdA}
  \frac{dt}{dA} = e^{\lambda/2 - \phi} \,.
\end{align}

As we did for the Misner-Sharp formalism, let us decompose $R$, $\rho$, $m$ and $U$ as follows.
\begin{subequations} \label{eq:MHdefs}
\begin{align}
  R &= R_b \tilde{R}
\\
  \rho &= \rho_b \tilde{\rho}
\\
  m &= \frac{4\pi}{3} \rho_b R^3 \tilde{m}
\\
  U &= H R \tilde{U}
\\
  P &= \rho_b \tilde{P} = \rho_b \tilde{\rho} (w + \tilde{Q})
\end{align}
\end{subequations}
All of the quantities appearing here are functions of $u$ and $A$, where quantities that are known only explicitly as functions of $t$ and $A$ ($R_b$, $\rho_b$ and $H$) are written as functions of $t(u, A)$ and $A$. In practice, we use the previously defined $\xi$ instead of $t$. We avoid ambiguous overdots from here on, instead expressing time derivatives explicitly.

Let us assume that for a given $u$, we know $\tilde{R}$, $\tilde{U}$, $\tilde{m}$ and $\xi$ as functions of $A$. As previously, we use $A = R_H \bar{A}$, and primes refer to derivatives with respect to $\bar{A}$. We also define our time coordinate as $u = R_H \bar{u}$ in order to make it dimensionless.

The equations of motion in these variables can be computed directly from the Hernandez-Misner equations. This is somewhat cumbersome, and so we relegate details to Appendix \ref{app:derivation}. At each timestep, a number of auxiliary variables need to be computed. The equations for these are as follows.
\begin{subequations}
\begin{align}
  \bar{\Gamma}^2 &= e^{2 (1 - \alpha) \xi} + \tilde{A}^2 \tilde{R}^2 (\tilde{U}^2 - \tilde{m})
\\
  \tilde{\rho} &= \frac{\bar{\Gamma} + \bar{A} \tilde{R} \tilde{U}}{\bar{\Gamma} - (w + \tilde{Q}) \bar{A} \tilde{R} \tilde{U}} \left[
  \tilde{m} + \frac{\bar{A} \tilde{R}}{3} \frac{\tilde{m}' - 2 \tilde{m} \xi'}{\alpha \bar{A} \tilde{R} \xi' + (\bar{A} \tilde{R})'}
  \right] \label{eq:hmrhocomputation}
\\
  e^\phi &= \frac{\alpha \xi' \bar{A} \tilde{R} + (\bar{A} \tilde{R})'}{\alpha \xi' (\bar{\Gamma} + \bar{A} \tilde{R} \tilde{U})}
\\
  e^{\lambda/2} &= \alpha e^{\phi + \xi} \xi'
\\
  & \hspace{-0.8cm} \partial_{\bar{A}} \ln \left(e^{-\psi} (\bar{\Gamma} + \bar{A} \tilde{R} \tilde{U})\right) = - \xi' \bigg[\alpha - 1
\nonumber\\&\qquad \qquad \qquad \qquad
  + \frac{e^\phi \bar{A} \tilde{R} \tilde{\rho}}{\bar{\Gamma} + \bar{A} \tilde{R} \tilde{U}} \frac{1 + w + \tilde{Q}}{1+w}\bigg] \label{eq:hmlapsecomputation}
\end{align}
\end{subequations}
The last of these equations needs to be integrated in order to obtain $e^\psi$, using the boundary condition $e^\psi = e^\phi$ at the outer boundary.

Once the auxiliary variables are known, the evolution equations can be computed.
\begin{subequations} \label{eq:HMcoseqs}
\begin{align}
  \partial_{\bar{u}} \xi &= \frac{1}{\alpha} e^{\psi - \phi - \xi}
\\
  \partial_{\bar{u}} \tilde{R}
  &= e^{\psi - \xi} \tilde{R} (\tilde{U} - e^{- \phi})
\\
  \partial_{\bar{u}} \tilde{m} &= 3 e^{\psi - \xi} \left[ e^{-\phi} \tilde{m} (1 + w) - \tilde{U} (\tilde{P} + \tilde{m}) \right] \label{eq:hmmcomputation}
\\
\partial_{\bar{u}} \tilde{U}
  &= - \frac{e^{\psi - \xi}}{1 - w - \tilde{Q}} \bigg[
  \frac{1}{2} \left(\tilde{m} + 3 \tilde{P} \right) + \tilde{U}^2 - e^{-\phi} \frac{\tilde{U}}{\alpha}
\nonumber\\
  & \qquad + (w + \tilde{Q}) e^{\xi -\lambda/2} \tilde{U}' + \frac{\bar{\Gamma} (w + \tilde{Q})}{\bar{A} \tilde{R} (1 + w + \tilde{Q})} \bigg( 3 \tilde{Q} \tilde{U}
\nonumber \\
  & \qquad + \frac{e^\xi}{w + \tilde{Q}} (e^{-\lambda/2} \tilde{Q}' - e^{-\psi} \partial_{\bar{u}} \tilde{Q})
\nonumber\\
  & \qquad \quad + 3 (1 + w) (\tilde{U} - e^{-\phi}) + e^{\xi-\lambda/2} \frac{\tilde{\rho}'}{\tilde{\rho}} \bigg)
  \bigg] \label{eq:hmucomputation}
\end{align}
\end{subequations}
Methods to evolve the $\tilde{U}$ equation are discussed in Section \ref{sec:code} below.

In the Misner-Sharp formalism, we had the boundary condition $\partial_{\bar{A}} \tilde{X} = 0$ at the origin for all tilded variables. This translates to the coordinate invariant condition $D_r \tilde{X} = 0$ at the origin, and is simply a result of the imposition of spherical symmetry. In the Hernandez-Misner formalism, this condition yields
\begin{align}
  \partial_{\bar{u}} \tilde{X} = e^{\psi - \lambda/2} \partial_{\bar{A}} \tilde{X}
\end{align}
at the origin.

The outer boundary is a little more complicated. Because nonlinearities become important in the black hole formation process, our analysis for the outer boundary conditions in the Misner-Sharp coordinates is typically no longer valid. In particular, in practice, we found that both characteristics were directed inwards, and so no wave-like boundary condition is possible. The most stable outer boundary condition we found was to impose the Dirichlet boundary condition $\tilde{\rho} = \mathrm{const}$, where the constant was determined by the initial conditions. While this tended to inject some energy into the system, there is no effect on the black hole mass, so long as the outer boundary is causally disconnected from the collapsing region.

In order to match results from Misner-Sharp and Hernandez-Misner codes however (on a profile that doesn't lead to black hole formation), it is useful to have the same boundary condition in place. In order to accomplish this, we can simply transform the boundary condition \eqref{eq:outerb} we found in the Misner-Sharp formalism. In particular, derivatives can be transformed using Eq. \eqref{eq:derivtransform} and
\begin{align}
  \left.\partial_{\bar{A}} \right|_{t} = \left.\partial_{\bar{A}}\right|_{\bar{u}} - e^{\lambda/2-\psi} \left.\partial_{\bar{u}}\right|_{\bar{A}}
\end{align}
where we make use of Eq. \eqref{eq:invariantderivs}. The outer boundary condition then becomes
\begin{align}
\alpha e^{\phi + \xi - \psi} \partial_{\bar{u}} \delta_U &=
\left(\frac{c_s}{4} - \frac{1}{2} \frac{c_s^2}{\bar{A}}\right) (\delta_m' - e^{\lambda/2-\psi} \partial_{\bar{u}} \delta_m)
\nonumber\\
&\quad + \frac{c_s \alpha}{2\bar{A}} e^{\phi + \xi - \psi} \partial_{\bar{u}} \delta_m
\nonumber\\
&\quad - c_s (\delta_U' - e^{\lambda/2-\psi} \partial_{\bar{u}} \delta_U) - \frac{1}{4}\delta_m
\end{align}
where primes refer to derivatives in the Hernandez-Misner coordinate system. This can be rearranged to obtain $\partial_{\bar{u}} \delta_U$ as follows.
\begin{align}
\partial_{\bar{u}} \delta_U
&= \frac{1}{\alpha e^{\phi + \xi - \psi} - c_s e^{\lambda/2-\psi}} \bigg[ \frac{c_s \alpha}{2\bar{A}} e^{\phi + \xi - \psi} \partial_{\bar{u}} \delta_m - c_s \delta_U'
\nonumber\\
&\quad + \left(\frac{c_s}{4} - \frac{1}{2} \frac{c_s^2}{\bar{A}}\right) (\delta_m' - e^{\lambda/2-\psi} \partial_{\bar{u}} \delta_m) - \frac{1}{4}\delta_m\bigg]
\end{align}
This can be simplified by using Eq. \eqref{eq:HMlambda2} to obtain
\begin{align}
\partial_{\bar{u}} \delta_U
&= \frac{e^{\psi - \phi - \xi}}{\alpha (1 - c_s \xi')} \bigg[
\frac{c_s \alpha}{2 \bar{A}} \left(1 + c_s \xi' - \frac{\bar{A}}{2} \xi' \right) e^{\phi + \xi - \psi} \partial_{\bar{u}} \delta_m
\nonumber\\
&\quad
+ \left(\frac{c_s}{4} - \frac{1}{2} \frac{c_s^2}{\bar{A}}\right) \delta_m'
- \frac{1}{4}\delta_m
- c_s \delta_U'
\bigg] \,.
\end{align}
Note that the (linearized) speed of sound is still $c_s = e^{\xi/2}/\sqrt{12}$, where $\xi$ is evaluated at the outer boundary. This boundary condition works just as well as it does in the Misner-Sharp formalism.

Obtaining initial data in this formalism is very straightforward. Starting with the Misner-Sharp evolution, trace an outgoing photon geodesic as before, and save the values for $\tilde{m}$, $\tilde{U}$, $\tilde{R}$ and $\xi$. Once the null ray hits the outer boundary, the data can be interpolated to the appropriate gridpoints, the initial value of $\bar{u}$ can be set using $\bar{u} = \alpha e^{\xi(\bar{A}_{max})}$, and the evolution can continue in the new coordinate system.

\subsection{Mass Extraction}

The black hole formation condition in the Hernandez-Misner formalism is the same as in the Misner-Sharp formalism:
\begin{align}
 \frac{2m}{R} \equiv \tilde{R}^2 \tilde{m} \bar{A}^2 e^{2 (\alpha - 1) \xi} \ge 1 \,.
\end{align}
Of course, this condition is never met; when it comes close to being met, the evolution is suppressed by the vanishing of the lapse. We thus need a method by which to extract the black hole mass. The method we describe here is based on that suggested by Baumgarte \textit{et al.} \cite{Baumgarte1995}, and demonstrated on an evolution that started with a Gaussian initial profile (see Section \ref{sec:plots} for full details).

\begin{figure}[t]
    \includegraphics[width=\columnwidth]{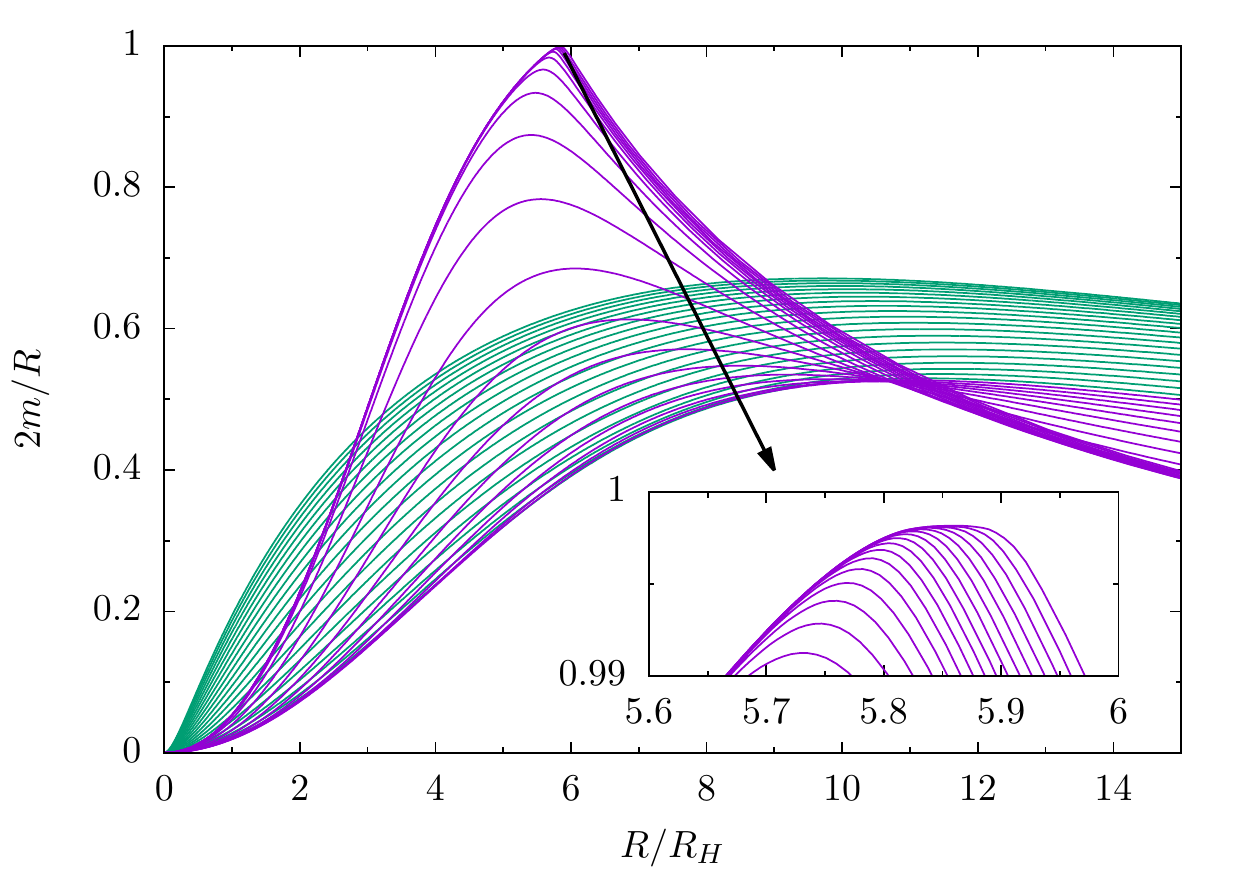}
    \caption{\label{fig:formation}$2m/R$ as a function of areal radius. The black hole formation condition is $2m/R \ge 1$. In the inset, we plot the behavior near the formation condition. The lines in teal are earlier lines, and the lines in purple are taken at later times. We see that the formation condition is almost met at late times. The colors are used simply to help show the evolution.}
\end{figure}

In Figure \ref{fig:formation}, we plot the quantity $2m/R$ as a function of areal radius. We see that as the black hole gets closer to forming, the curve approaches but never reaches the formation condition. As evolution continues, the line continues to approach $2m/R = 1$, but never quite gets there. In our simulations, we found that $2m/R$ often grew to larger than 0.995 before we terminated the evolution.

\begin{figure}[t]
    \includegraphics[width=\columnwidth]{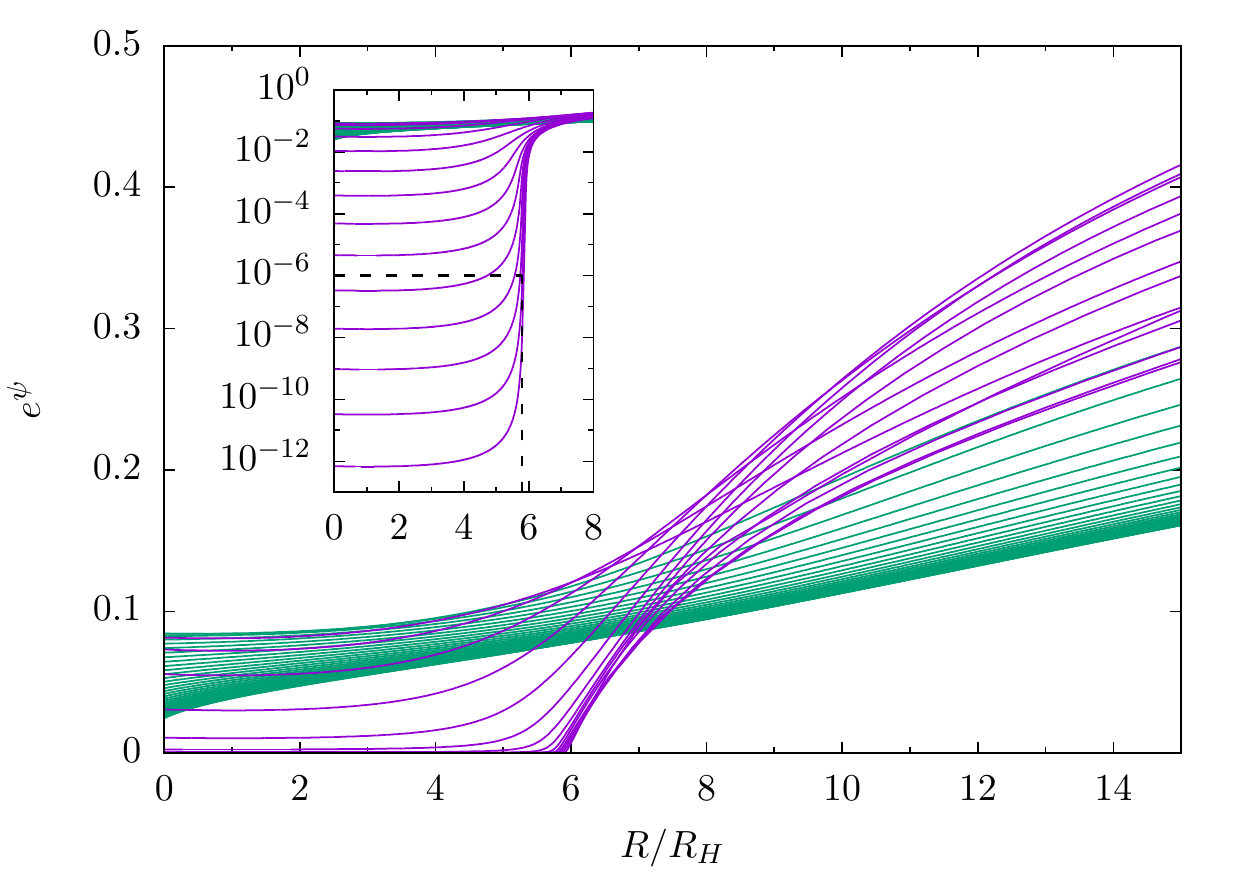}
    \caption{\label{fig:lapse}The lapse $e^\psi$ as a function of areal radius. As with the previous figure, lines in teal are early, while lines in purple are later. The inset contains the same plot, but on a log scale. We see that as the black hole forms, the lapse plummets, essentially freezing evolution. A horizontal dashed black line is drawn at the $10^{-6}$ level, and the corresponding radius $R_0$ for mass extraction is read off from the vertical dashed black line.}
\end{figure}

Figure \ref{fig:lapse} shows a plot of the lapse as a function of areal radius. We see that as the black hole forms, the lapse essentially vanishes at a reasonably well-defined radius. The inset shows that the lapse rapidly approaches zero in a logarithmic sense.

In order to extract the black hole mass, we first find the radius at which the lapse starts to vanish. Programmatically, we chose to wait until the lapse at the origin satisfies $e^\psi \le 10^{-10}$. When this occurs, we find the areal radius $R_0$ at which the lapse $e^\psi = 10^{-6}$. As the lapse is basically falling vertically at this point, it doesn't particularly matter if you choose $10^{-6}$, or a couple of orders of magnitude on either side.

\begin{figure}[t]
    \includegraphics[width=\columnwidth]{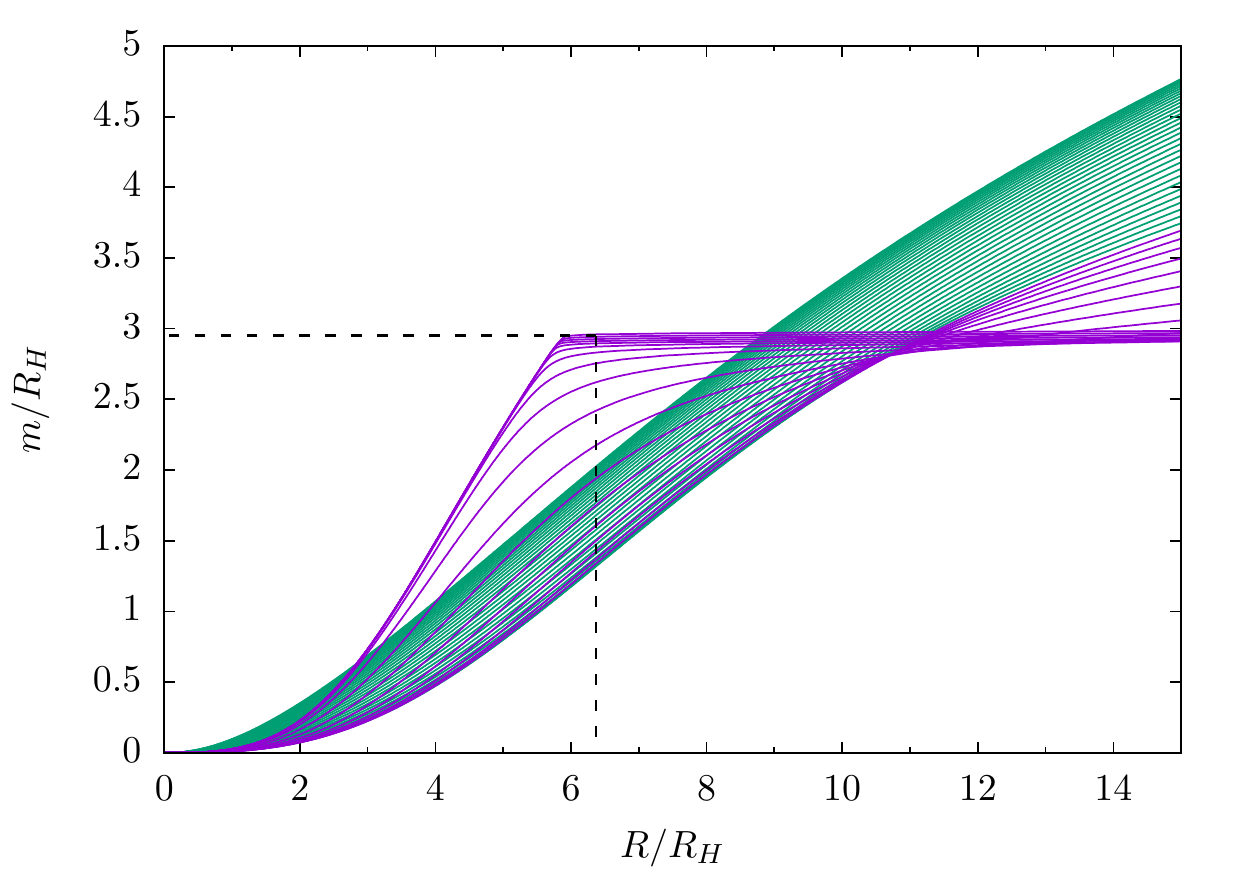}
    \caption{\label{fig:massextraction}Mass enclosed in areal radius $R$ as a function of $R$ (made dimensionless by $R_H$). As with the previous plots, earlier timesteps are in teal, while later timesteps are in purple. We see that when the black hole forms, it sucks in all the nearby matter, and $m$ becomes constant. If you extract the black hole mass at $R = 6 R_H$ or $R = 10 R_H$, you will find essentially the same mass. Off the right of the plot, $m$ starts increasing again. A vertical dashed black line is drawn at $1.1 R_0$ (extracted from the previous figure), and the mass contained in the black hole is read off from the corresponding horizontal dashed black line.}
\end{figure}

This radius $R_0$ is the radius at which we wish to extract the black hole mass. In Figure \ref{fig:massextraction}, we plot the function $m(R)$ as a function of $R$. At $R=R_0$, we can read off the mass at which the lapse starts to vanish. However, note that the black hole has basically consumed all of the nearby matter. If we increase the radius slightly, we won't change the mass of the black hole at all, but we will make sure to include everything in the elbow in the mass function. For this reason, we suggest reading off the mass at $R = 1.1 \, R_0$.

Once the black hole has formed, it very slowly accretes mass, as evidenced by the slowly rising flat line of the black hole mass in Figure \ref{fig:massextraction}. As this process is somewhat slow, it can basically be ignored for the purposes of estimating the final black hole mass. So long as one is consistent in the mass extraction procedure, the amount of accretion after initial formation should just be a small systematic error.

In terms of our variables, mass is written as
\begin{align}
 m = \frac{1}{2} e^{-\xi/2} \tilde{R}^3 \bar{A}^3 \tilde{m} R_H \,.
\end{align}
So, once the radius you wish to read off from is known, that will need to be translated into a value for $\bar{A}$, from which $\xi$, $\tilde{m}$ and $\tilde{R}$ will need to be extracted.

Note that the mass is written in terms of the initial horizon radius scale. In order to compare this to more useful mass scales, we consider the mass within the horizon in an FRW universe (taking $R = R_H(t) = R_H e^\xi$).
\begin{align}
  m_{horizon}(\xi) &= \frac{1}{2} e^{\xi} R_H
\end{align}
It is typical to compare the mass of the black hole with the mass contained in the horizon at the time $\xi_0$ when the perturbation enters the horizon. This time will need to be extracted from the Misner-Sharp evolution. Taking this ratio then eliminates $R_H$.


For the example presented here, we find $R_0 = 5.79 R_H$. This radius corresponds to a mass of $2.89 R_H$, while $1.1 R_0$ corresponds to a mass of $2.95 R_H$. The lines corresponding to these values are shown in Figures \ref{fig:lapse} and \ref{fig:massextraction}. From looking at the Misner-Sharp data, we found that the perturbation entered the horizon at $\xi = 2.47$. The ratio of the black hole mass to the mass enclosed in the horizon at the time of horizon crossing is then
\begin{align}
 \frac{m_{black \ hole}}{m_{horizon \ crossing}} = 0.50.
\end{align}
Taking the black hole mass ten timesteps later in the evolution increased the mass we measured from $2.95 R_H$ to $2.96 R_H$, or approximately $0.3\%$. The fractional mass excess from this evolution was approximately 0.43, just a little above the critical level of 0.4135 estimated by Harada \textit{et al.} \cite{Harada2013}.

If you do not wish to go to the effort of getting the Hernandez-Misner formalism to run, you can always simply estimate the mass of the black hole from the Misner-Sharp formalism: just leave the evolution to run until it dies from an inability to evolve further (due to the formation of a singularity). Then, find the outermost trapped surface, and read off the mass from there.


This method tends to underestimate the black hole mass somewhat, as a fair amount of matter will fall through the horizon soon after its formation. For the example we have provided here, the mass we extracted before the simulation died was $m = 2.25 R_H$, substantially less than the mass in the Hernandez-Misner formalism. We expect that the deficit is strongly dependent upon the initial density profile.

\begin{figure}[t]
    \includegraphics[width=\columnwidth]{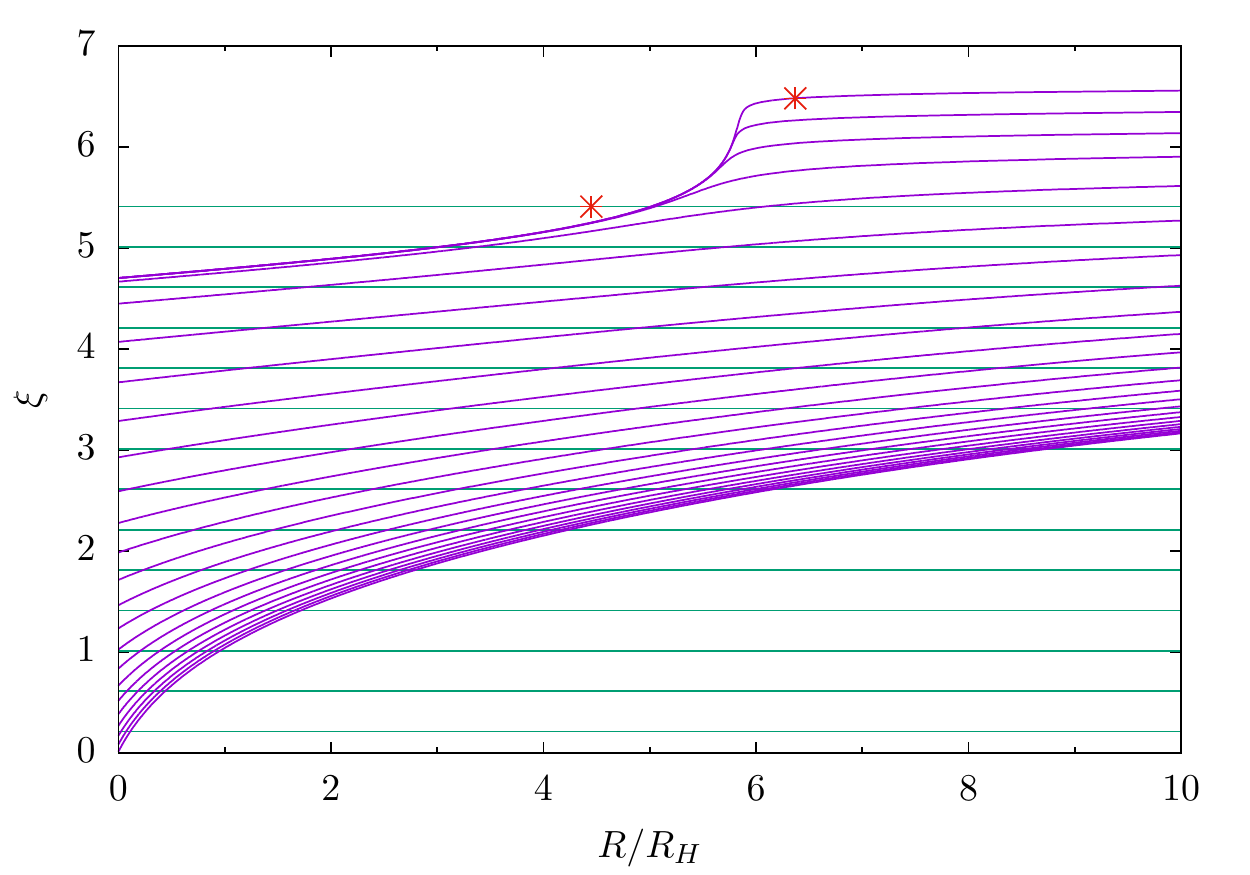}
    \caption{\label{fig:msmass}Time slices in both Misner-Sharp (teal) coordinates and Hernandez-Misner (purple) coordinates, plotted against areal radius. The Misner-Sharp simulation was run until the simulation could not proceed any further due to singularity formation at the origin. A red mark is placed at the outermost trapped surface on the final time slice, from which the mass of the black hole can be estimated. A second red mark is placed on the position at which we extracted the black hole mass from the Hernandez-Misner formalism. The data is the same as was used to generate Figure \ref{fig:slicing} above.}
\end{figure}

In Figure \ref{fig:msmass}, we plot time slices in Misner-Sharp coordinates as well as time slices in Hernandez-Misner coordinates. This figure shows where the Misner-Sharp mass estimate is extracted from, as well as the Hernandez-Misner mass estimate. This figure also shows that the Hernandez-Misner coordinates do an impressive job of hugging the event horizon, and that the Misner-Sharp coordinates can be used to probe a little bit inside the horizon.

\subsection{FRW Evolution}

For code debugging purposes, it is useful to know analytic forms for all the variables in the case of pure FRW evolution. In particular, $\tilde{R} = \tilde{m} = \tilde{\rho} = \tilde{U} = e^\phi = 1$ in FRW. It is straightforward to check that this is preserved under the evolution equations \eqref{eq:HMcoseqs}. Furthermore, in FRW, $e^{\lambda/2} = a$ (the scale factor), and $\bar{\Gamma} = e^{(1 - \alpha) \xi}$. All that remains to be obtained then is $\xi(\bar{A}, \bar{u})$ and $e^{\psi(\bar{A}, \bar{u})}$.

We begin with $\xi$. From Eq. \eqref{eq:HMlambda2}, we have
\begin{align}
  \xi' = \frac{a H R_H}{\alpha} = \frac{e^{(\alpha - 1) \xi}}{\alpha} \,.
\end{align}
This integrates straightforwardly to
\begin{align} \label{eq:xieq}
  e^{(1-\alpha)\xi} = \frac{1 - \alpha}{\alpha} \bar{A} + C
\end{align}
for some constant $C$. Our coordinate system is constructed in such a way that on the outer boundary, $t(\bar{A}_{max}, u) = u$. This implies that
\begin{align}
  e^{\xi(\bar{A}_{max}, \tilde{u})} = \frac{\bar{u}}{\alpha} \,. \label{eq:initial}
\end{align}
Therefore, we have
\begin{align}\label{eq:xifrw}
  e^{(1-\alpha)\xi(\bar{A}, \bar{u})} = \frac{1 - \alpha}{\alpha} (\bar{A} - \bar{A}_{max}) + \left(\frac{\bar{u}}{\alpha}\right)^{1-\alpha} \,.
\end{align}
By evaluating this at $\bar{A} = 0$ and $\bar{A} = \bar{A}_{max}$ and taking the difference of the two results, we obtain
\begin{align}
  e^{(1-\alpha)\xi(\bar{A}_{max}, \bar{u})} - e^{(1-\alpha)\xi(0, \bar{u})}
  =
  \frac{1 - \alpha}{\alpha} \bar{A}_{max} \,. \label{eq:evolutiontime}
\end{align}
In particular, if we look at the original null slice (at $\bar{u} = \bar{u}_0$ such that $\xi(0,\bar{u}_0) = 0$ and Eq. \eqref{eq:initial} holds), this yields
\begin{align}
  e^{(1-\alpha) \xi(\bar{A}_{max}, \bar{u}_0)} = 1 + \frac{1 - \alpha}{\alpha} \bar{A}_{max} \,. \label{eq:initialu}
\end{align}

Looking towards $\psi$, from Eq. \eqref{eq:HMpsi}, we obtain
\begin{align}
  \partial_A \psi &= aH + A a H (\alpha - 1) \partial_A \xi + \frac{1+3w}{2} A H^2 a^2
\end{align}
where we make use of the Friedmann equation and writing $aH$ in terms of $\xi$. Using our knowledge of $\xi'$ from above, this becomes simply
\begin{align}
  \partial_A \psi &= aH \,.
\end{align}
Writing this in terms of $\xi$, we obtain
\begin{align}
  \psi' = e^{(\alpha - 1)\xi}
\end{align}
where the prime denotes a derivative with respect to $\bar{A}$ as usual. We can use Eq. \eqref{eq:xifrw} to integrate this, yielding
\begin{align}
  \psi = \frac{\alpha}{1 - \alpha} \ln \left(
  \bar{A} - \bar{A}_{max} + \frac{\alpha}{1 - \alpha} \left( \frac{\bar{u}}{\alpha}\right)^{1-\alpha}
  \right) + C
\end{align}
for some constant $C$. We know at the outer boundary, $e^\psi = e^\phi = 1$, which sets the constant. We then obtain
\begin{align}
  e^\psi = \left(1 + \frac{1 - \alpha}{\alpha}
  \frac{\bar{A} - \bar{A}_{max}}{\left( \frac{\bar{u}}{\alpha}\right)^{1-\alpha}}
  \right)^{\alpha/(1-\alpha)} \,.
\end{align}
For the initial null ray, we know from Eq. \eqref{eq:initial} that $\bar{u}_0 = \alpha e^{\xi_0}$, which yields
\begin{align}
  e^{\psi(\bar{A}, \bar{u}_0)} = \left(\frac{\alpha + (1-\alpha) \bar{A}}{\alpha + (1 - \alpha) \bar{A}_{max}}
  \right)^{\alpha/(1-\alpha)} \,.
\end{align}
Specializing to $w = 1/3$, these results become simply $u_0 = (1+\bar{A}_{max})^2/2$,
\begin{align}
  e^\psi = 1 + \frac{\bar{A} - \bar{A}_{max}}{\sqrt{2 \bar{u}}}
\end{align}
and
\begin{align}
  e^{\psi(\bar{A}, \bar{u}_0)} = \frac{1 + \bar{A}}{1 + \bar{A}_{max}} \,.
\end{align}

\subsection{Domain Size and Computational Time}

In the Misner-Sharp formalism, doubling the size of the domain (i.e., doubling $\bar{A}_{max}$) while maintaining the same resolution roughly doubles the computational time. In the Hernandez-Misner formalism, this is not the case, as we shall now show.

Consider initial data in the Hernandez-Misner formalism where we specialize to $w=1/3$, $\alpha = 1/2$ for simplicity. From Eq. \eqref{eq:evolutiontime}, we can relate the value of $\bar{u}$ (which corresponds to the value of $\xi$ at the outer boundary) to the value of $\xi$ at the origin, which we denote $\xi(0)$.
\begin{align}
  \bar{u} = \frac{1}{2} \left( e^{\xi(0)/2} + \bar{A}_{max} \right)^2
\end{align}
Let us consider the change in $\bar{u}$ as the value of $\xi$ at the origin progresses from $\xi(0) = 0$ to $\xi(0) = \xi$.
\begin{align}
  \Delta \bar{u} = \frac{1}{2} \left[e^\xi - 1 + 2 \bar{A}_{max} (e^{\xi/2} - 1)\right]
\end{align}
We see that the change in $\bar{u}$ increases linearly with $\bar{A}_{max}$. This implies that if $\bar{A}_{max}$ is doubled, then the computational time to make a timestep equivalent to a timestep in the original domain also doubles, roughly speaking. This is independent of any increase in computational time from increasing the number of gridpoints in the domain.

This problem is magnified when a black hole is forming, as the lapse near the origin becomes increasingly small. We thus see that taking a very large domain is computationally expensive in the Hernandez-Misner formalism, much more so than the Misner-Sharp formalism. This is why we suggest using an appropriate boundary condition instead of steadily increasing the size of the domain in the Misner-Sharp formalism.

\section{Coding Issues} \label{sec:code}

In this section, we detail various issues pertaining to coding the formalism presented here.

\subsection{Evolution Scheme}

For a numerical evolution scheme, we suggest a finite difference method. The reason for this is that a lot of the interesting action tends to occur within a reasonably small radius, where a lot of gridpoints will be necessary, especially for profiles that are close to the threshold of black hole formation.

\begin{figure}[t]
	\includegraphics[width=\columnwidth]{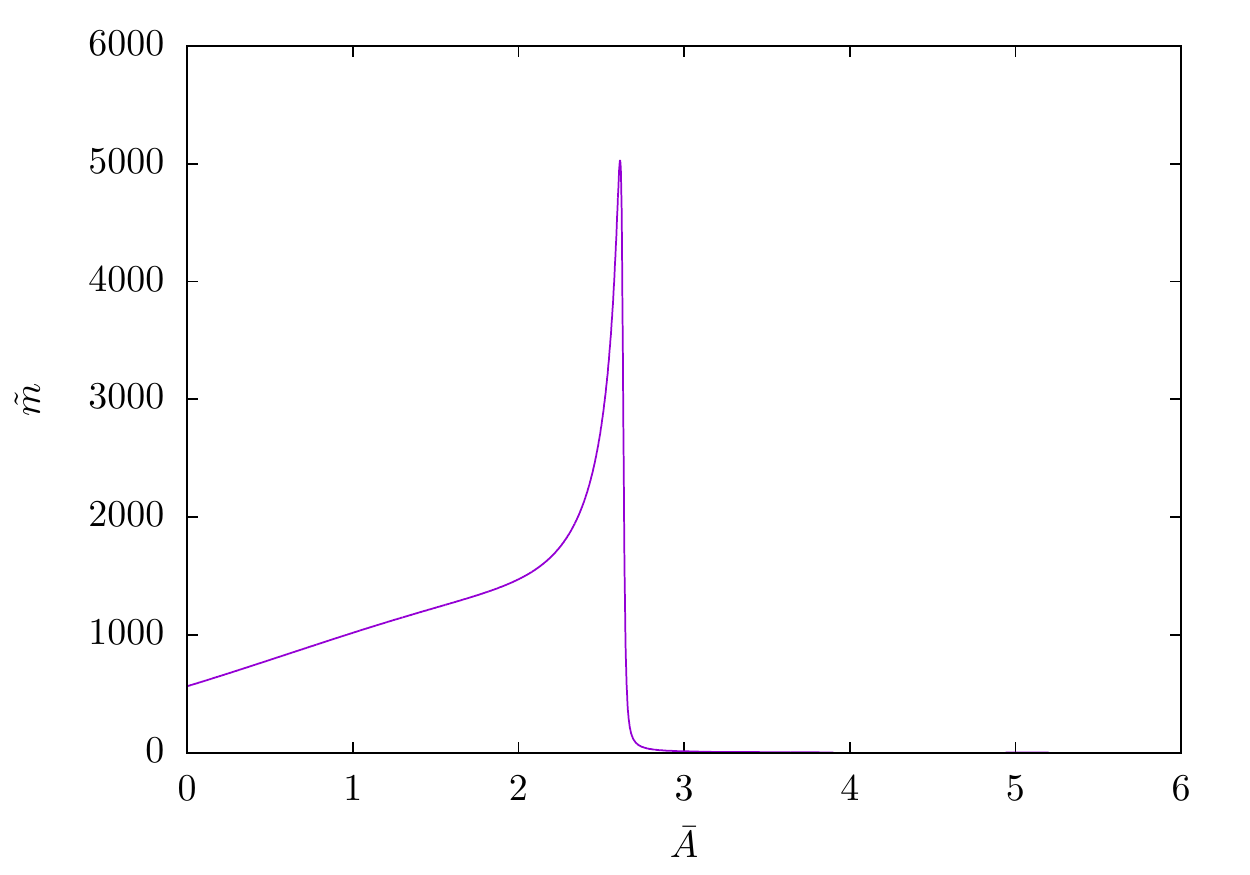}
	\includegraphics[width=\columnwidth]{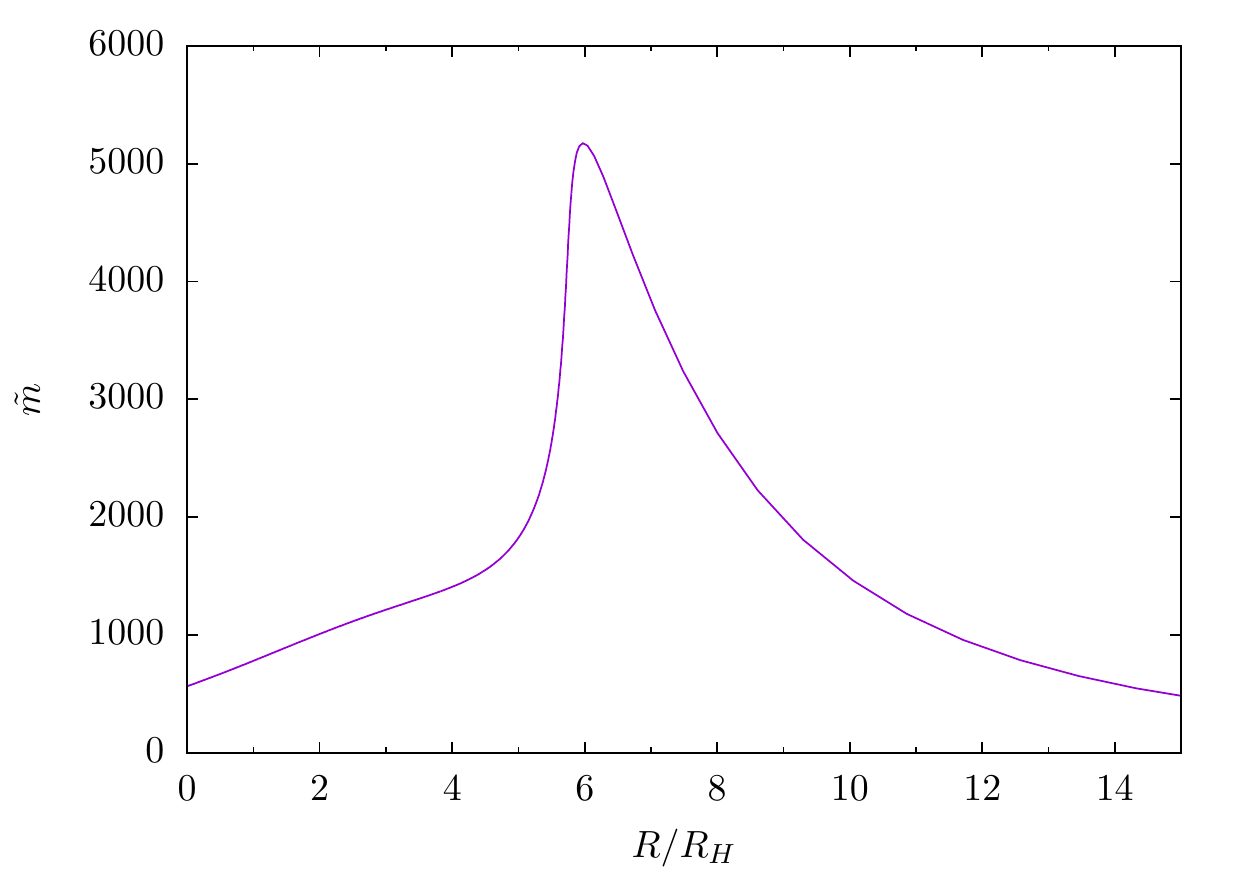}
	\caption{\label{fig:mvsa}$\tilde{m}$ as a function of comoving radius $\bar{A}$ (above) and areal radius $R / R_H$ (below) in the Hernandez-Misner formalism. Data is taken from towards the end of the black hole evolution described in Section \ref{sec:plots}. These plots show the type of behavior one can except to see arising in black hole evolutions in Hernandez-Misner coordinates, demonstrating why finite difference methods with adaptive mesh refinement are necessary. A comparison of the plots shows that the spike in coordinate radius is not a physical shock: at around $\bar{A} \sim 2.8$, the areal radius $R(\bar{A})$ suddenly gets very large very quickly (see Figure \ref{fig:radius}). Thus, moving from $\bar{A}$ of $\sim 2.8$ to $\sim 3$ actually covers a large physical distance, which smooths out this feature somewhat in physical space.}
\end{figure}

\begin{figure}[t]
	\includegraphics[width=\columnwidth]{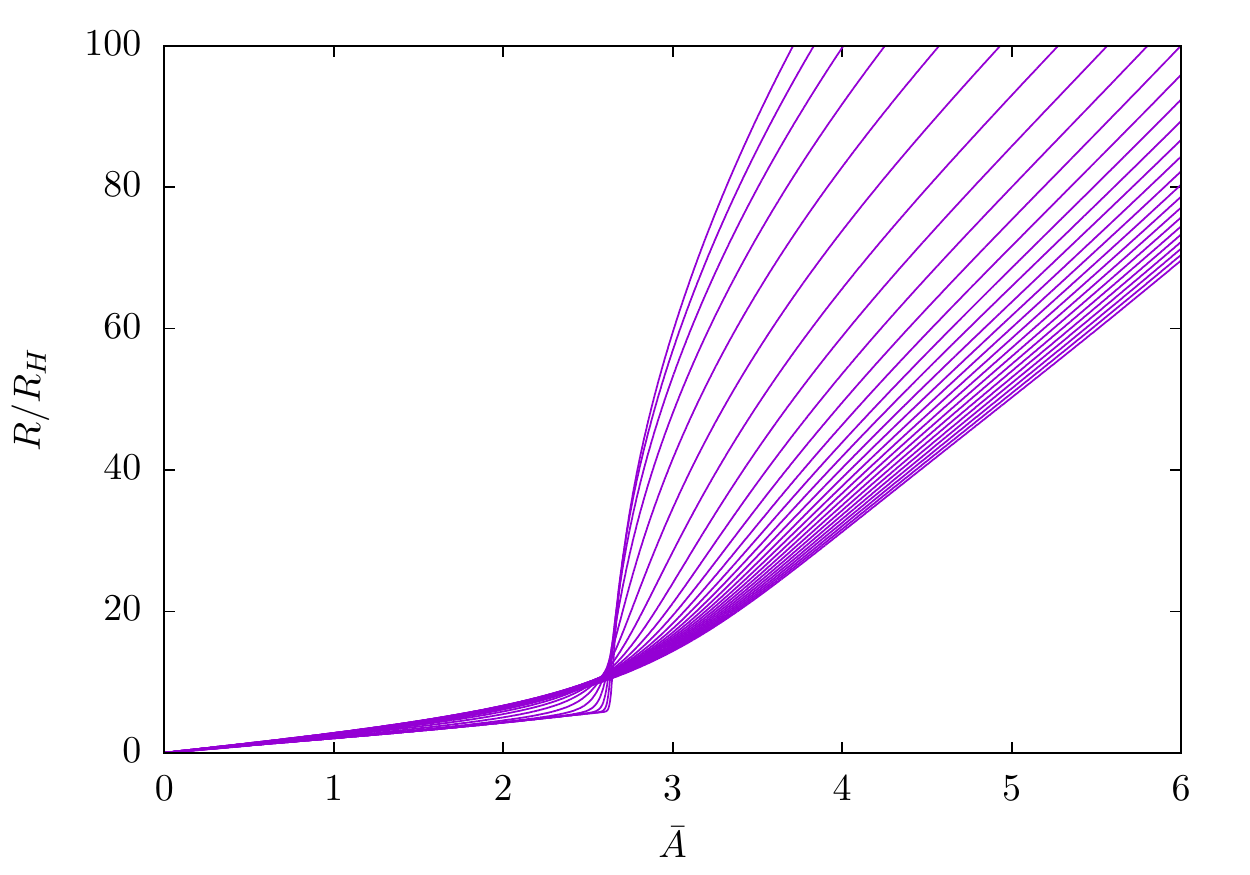}
	\caption{\label{fig:radius}Plot of areal radius $R$ as a function of comoving radius $\bar{A}$ in a black hole evolution (Hernandez-Misner formalism). The smoother lines come from earlier in the evolution, and the elbow develops around the same time that the black hole forms. The relationship between the comoving radius and the areal radius becomes progressively more squished when a black hole forms. This means that without an adaptive mesh refinement scheme, there will be very few gridpoints in $\bar{A}$ for a large change in $R$, which leads to poor sampling and inaccurate derivatives in this region.}
\end{figure}

The Misner-Sharp formalism is reasonably well-behaved. However, in the Hernandez-Misner formalism, we found that strong shocks tended to form in $\tilde{m}$ (see Figure \ref{fig:mvsa}). These shocks were present in coordinate space, but were much less pronounced in physical space. If computing in the Hernandez-Misner formalism, we strongly suggest implementing an adaptive mesh refinement (AMR) scheme in order to evolve the system. An AMR scheme is required to resolve the details of the system appropriately when such shocks occur, as gridpoints in comoving radius become very squashed and then rarified in the vicinity of a black hole (see Figure \ref{fig:radius}).

In both formalisms, we performed time-stepping using an off-the-shelf adaptive time stepping Runge-Kutta (4,5) method. Starting from an initial profile, we could typically extract a black hole mass in $\sim$ 15-20 seconds on a consumer laptop.

For gridpoints, a logarithmic grid spacing structure has been suggested in the literature (e.g., \cite{Musco2012} and citations therein). Putting a lot of gridpoints near the origin and fewer at the outer boundary tends to put gridpoints where they are necessary to capture the dynamics of black hole formation.

In terms of the outer boundary, we suggested above that having $\bar{A}_{max}$ be too large would lead to a very slow evolution. On the other hand, having $\bar{A}_{max}$ be too small may lead to the outer boundary being in causal contact with the collapsing region, in which case an imperfect boundary condition (particularly in Hernandez-Misner evolution) can lead to erroneous results. While the optimal size of $\bar{A}_{max}$ depends on the width of the initial perturbation, we found that a reasonable rule of thumb was to let $\bar{A}_{max}$ be $20\times$ larger than the initial perturbation.

In the Hernandez-Misner formalism, we found that we had positive feedback for high-frequency noise in $\tilde{\rho}$. In particular, when a black hole came close to forming, $\tilde{\rho}$ developed a sawtooth instability, zig-zagging at the Nyquist frequency. We determined that this was due to our sampling scheme, and the instability was being amplified due to the presence of $\tilde{\rho}'$ in the evolution equations. In order to compensate for this, when we calculated $\tilde{\rho}$ in the Hernandez-Misner formalism, we applied a haystack smoothing filter over five gridpoints to $\tilde{\rho}$, which resolved the issue. This stabilization mechanism is a form of artificial viscosity, separate from the viscosity we included in the equations of motion.

\subsection{Timestep Limits}

The Courant-Friedrichs-Lewy (CFL) condition \cite{CFL1928} provides a limit to the size of a timestep that can be taken in a numerical evolution. Here, we provide the limit in both the Misner-Sharp and Hernandez-Misner formalisms \cite{Baumgarte1995}.

In Misner-Sharp coordinates, the CFL condition requires
\begin{align}
c_s = \sqrt{w} < \frac{e^{\lambda/2} \Delta A}{e^{\phi} \Delta t}
\end{align}
where $c_s$ is the speed of sound, $\Delta A$ is the grid spacing, and $\Delta t$ is the amount of time being stepped forwards, and we have used the speed of sound of a perfect fluid. This reduces to the condition
\begin{align}
\frac{\Delta t}{t_0} < \alpha \frac{e^{\lambda/2} \Delta \bar{A}}{e^{\phi} \sqrt{w}} \,.
\end{align}
Writing this in terms of $\xi$, we have
\begin{subequations}
	\begin{align}
	\Delta \xi &= \ln \left(1 + \frac{\Delta t}{t} \right)
	\\
	&< \ln \left(1 + e^{\lambda/2 - \phi -\xi} \alpha \frac{\Delta \bar{A}}{\sqrt{w}} \right) \,.
	\end{align}
\end{subequations}

In the Hernandez-Misner formalism, we can use the discrete form of Eq. \eqref{eq:dudef} to obtain
\begin{align}
e^\phi \Delta t = e^\psi \Delta u + e^{\lambda/2} \Delta A \,.
\end{align}
Eliminating $\Delta t$ and solving for $\Delta u$, we then obtain
\begin{align}
\Delta \bar{u} < \left(\frac{1}{\sqrt{w}} - 1\right) e^{\lambda/2 - \psi} \Delta \bar{A} \,.
\end{align}

The CFL condition is a necessary but insufficient condition on stability. Further conditions on timestepping in the Misner-Sharp and Hernandez-Misner formalisms have been used in the literature \cite{Baumgarte1995, Jedamzik1999}, including limiting density changes to $\Delta \rho / \rho < 0.02$, radius changes to $\Delta R / R < 0.005$, and changes in the quantity $1 - 2m(R)/R$ to less than 0.1. Sometimes, these conditions can be much more stringent than the CFL condition. For our purposes, the adaptive time-stepper in our integration library proved sufficient, and we only demanded the CFL condition be met.

\subsection{Artificial Viscosity}

In our equations of motion, we included artificial viscosity from the very beginning, although we never gave it a form. We finally come to address this issue.

Artificial viscosity is an additional pressure that is added into the system in order to mitigate the formation of shocks. The original implementations of the Misner-Sharp and Hernandez-Misner formalism included artificial viscosity in their equations of motion, and subsequent investigations of supernovae collapse also utilized it. Papers in the literature using these formalisms for primordial collapse have sometimes implemented it, but documentation has been somewhat scarce.

When discussing shocks, there are two distinct effects. Generally speaking, a shock is a discontinuity in a curve, but can also simply describe very steep-sloping data. The second effect is a shockwave, which describes a moving shockfront or discontinuity. When shockwaves form, the characteristics tend to get distorted in such a manner as to guarantee that an actual discontinuity forms, which then propagates. This leads to all manner of difficulties when evaluating spatial derivatives.

From our investigations, artificial viscosity, as it applies to the system that we have been investigating, is useful for smoothing out shockwaves, but much less effective at smoothing static ``shocks'', which can simply be addressed through increased local resolution. We found that outgoing shockwaves tended to form in the Misner-Sharp formalism when an overdensity failed to collapse into a black hole, and artificial viscosity was useful for smoothing these over. However, we never saw the formation of a shockwave in the Hernandez-Misner formalism; as a black hole forms, the vanishing lapse tends to ensure that everything freezes. We did see shocks forming, but artificial viscosity did nothing to assist with these.

When shocks form in the Hernandez-Misner formalism, they tend to be a coordinate effect (see Figure \ref{fig:mvsa}, for example). In our experience, these shocks can and should be suitably described through the implementation of an AMR scheme.

Thus, our recommendation is to implement artificial viscosity in the Misner-Sharp formalism if you desire to describe outgoing shockwaves. We recommend ignoring artificial viscosity in the Hernandez-Misner formalism, noting that a few other authors have done likewise \cite{Musco2012, Nakama2013}.

\subsubsection{Misner-Sharp Formalism}

The origin of the form of dissipation that we describe goes back to Neumann and Richtmyer \cite{Neumann1950}, which has been used in many evolutions of this kind \cite{MayWhite1966, MayWhite1967, Miller1989, Baumgarte1995, Jedamzik1999, Musco2004}, although documentation is somewhat sparse. The basic idea is to identify where sharp discontinuities may occur, and artificially increase the pressure there in order to smear out the discontinuity over a few gridpoints.

We investigated two triggers for identifying when to turn on artificial viscosity. The first condition, used by May and White \cite{MayWhite1967}, is to trigger on $\partial_t {\rho} > 0$. The second condition, used by Baumgarte \textit{et al.} \cite{Baumgarte1995} (BST), instead triggers on $\partial_A U < 0$. Both of these triggers seem reasonable: in the first instance, if the density is increasing, then something is collapsing. In the second instance, the idea is to identify shockwaves (moving in either direction) as requiring a negative velocity gradient.

The suggested form of the viscosity term is
\begin{align}
P_{vis} = \left\{
\begin{array}{ll}
\kappa \rho (\Delta A \partial_A U)^2 \quad & \text{when triggered} \\
0 & \text{otherwise}
\end{array}
\right.
\end{align}
where $\kappa$ is a dimensionless constant that controls the strength of the viscosity, $\Delta A$ is the grid spacing, and we take derivatives at constant $t$.

In our description, we can construct $\tilde{Q}$ as
\begin{align}
  \tilde{Q} = \kappa (\Delta \bar{A})^2 e^{2 (\alpha - 1) \xi} \left[\partial_{\bar{A}} \left( \bar{A} \tilde{R} \tilde{U}\right)\right]^2 \,.
\end{align}
The two triggering conditions become
\begin{subequations}
\begin{align}
  (\bar{A} \tilde{R})' \tilde{U} &< - \frac{\bar{A} \tilde{R} \tilde{U}'}{3} \quad \text{(May and White)}
\\
  (\bar{A} \tilde{R})' \tilde{U} &< - \bar{A} \tilde{R} \tilde{U}' \quad \text{(BST)}
\end{align}
\end{subequations}
where we use Eq. \eqref{eq:rhodot} for the first condition, and note that $(\bar{A} \tilde{R})' > 0$ by Eq. \eqref{eq:condition1}. Despite rather different origins, these conditions are very similar. Indeed, the two conditions are equivalent to taking the divergence of $U$ in spherical polar coordinates ($\partial_{\bar{A}} (R^2 U) < 0$) or in a linear sense ($\partial_{\bar{A}} U < 0$). When the velocities are converging instead of diverging, the viscosity triggers.

\begin{figure}[t]
	\includegraphics[width=\columnwidth]{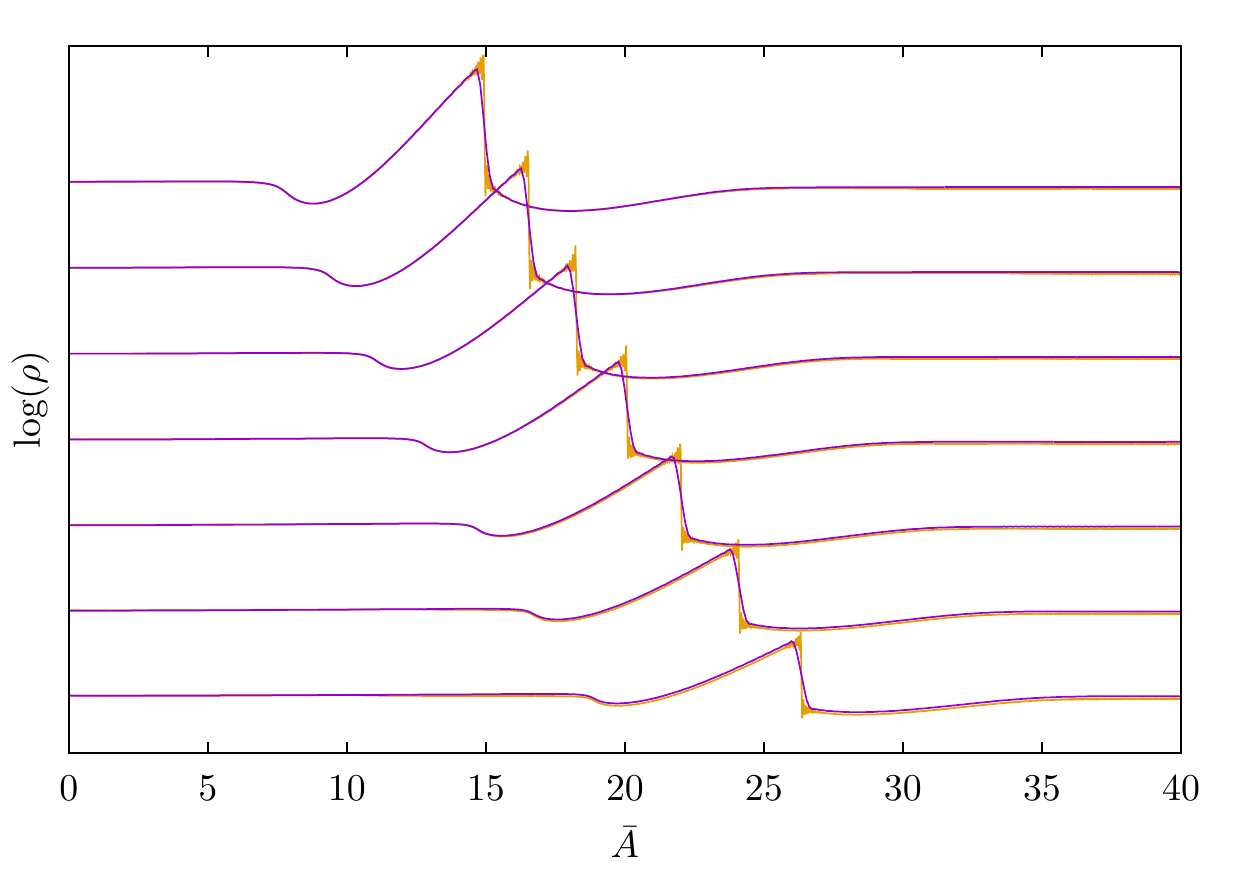}
	\caption{\label{fig:viscosity}Density profiles from two Misner-Sharp evolutions from the same initial data. These evolutions did not form a black hole, but sent out a spherical wave of matter, which will tend to form a shockwave. In the purple evolution, artificial viscosity was turned on with $\kappa = 2$, while in the yellow evolution, artificial viscosity was turned off. The manner in which artificial viscosity has smoothed the shock is evident.}
\end{figure}

We found that the BST trigger tended to work better than the May and White trigger, and we used it to smooth outgoing shockwaves in our Misner-Sharp evolutions. In particular, the smoothing provided by the artificial viscosity greatly increased the performance of our boundary condition. The effects of the artificial viscosity are evident in Figure \ref{fig:viscosity}, where we show data from evolutions with the same initial data, but with $\kappa = 2$ and $\kappa = 0$. In general, we found that $\kappa \simeq 2$ is a reasonable value of the coefficient.

\subsubsection{Hernandez-Misner Formalism}

\begin{figure}[t]
	\includegraphics[width=\columnwidth]{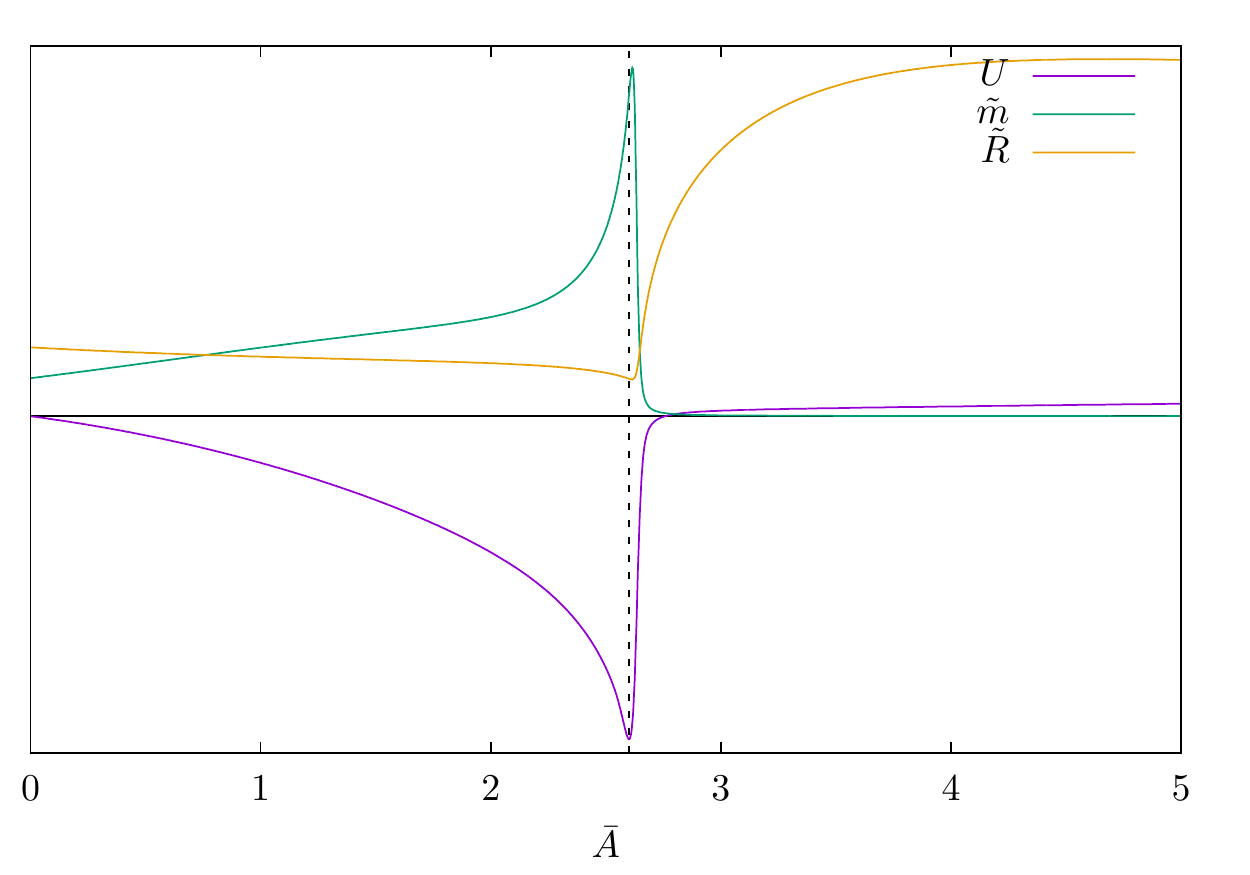}
	\caption{\label{fig:shock}This plot shows three curves from late in the evolution of a black hole formation in the Hernandez-Misner formalism, plotted as a function of comoving radius.. The purple curve is $U$, which can be seen to have a negative derivative everywhere to the left of the dashed vertical line ($U$ having a negative gradient is the trigger for artificial viscosity). The teal curve is $\tilde{m}$, which is the most shock-prone variable we evolve. We see that the shock is occurring outside the region in which artificial viscosity kicks in (to the left of the dashed vertical line), and so it does little to help smooth the shock. Furthermore, the lapse has essentially vanished to the left of the dashed curve, and so very little evolution is occurring anyway. The yellow curve represents $\tilde{R}$, which increases sharply around the point of the shock, causing a small distance in coordinate radius to correspond to a large distance in areal radius, which is the real cause of the shock. This leads to a dilution of gridpoints in physical space in this region, which can be addressed through the use of AMR. These curves have all been rescaled, so the vertical axis is unimportant. The horizontal black line shows zero on all three curves.}
\end{figure}

In the Hernandez-Misner formalism, we found that artificial viscosity did not help much at all. The times when we thought that artificial viscosity may be helpful were when stationary shocks formed in the evolution variables. However, we discovered that artificial viscosity wasn't triggering in the vicinity of these shocks. An example demonstrating this issue is described in Figure \ref{fig:shock}. As the shocks that we saw forming were best addressed through the use of adaptive mesh refinement, we decided against the use of artificial viscosity in this formalism (setting $\tilde{Q} = 0$). However, for completeness' sake, we now describe the implementation of artificial viscosity in this formalism.

It turns out that transforming the artificial viscosity of the Misner-Sharp formalism into the null slices makes for a circular headache, due to $\tilde{Q}$ depending on spatial derivatives in Misner-Sharp that correspond to both spatial and temporal derivatives in Hernandez-Misner. Instead, it is much easier to construct new conditions in the Hernandez-Misner coordinates. In particular, Baumgarte \textit{et al.} \cite{Baumgarte1995} suggest triggering on
\begin{align}
  \partial_A U < 0
\end{align}
in Hernandez-Misner coordinates, which for us is equivalent to
\begin{align}
  \partial_{\bar{A}} \left(e^{(\alpha - 1) \xi} \bar{A} \tilde{R} \tilde{U} \right) < 0 \,.
\end{align}
Baumgarte \textit{et al.}'s suggested form of the artificial viscosity is the same as in the Misner-Sharp formalism, but taking the derivative of $U$ in the Hernandez-Misner coordinates. In our variables, this becomes
\begin{align}
  \tilde{Q} = \kappa (\Delta \bar{A})^2 \left[\partial_{\bar{A}} \left( e^{(\alpha - 1) \xi} \bar{A} \tilde{R} \tilde{U}\right)\right]^2 \,.
\end{align}
This can be computed at every timestep. We again suggest $\kappa \simeq 2$ as a coefficient.

Artificial viscosity appears in Eqs. \eqref{eq:hmrhocomputation}, \eqref{eq:hmlapsecomputation}, \eqref{eq:hmmcomputation} and \eqref{eq:hmucomputation} in the Hernandez-Misner formalism. Two derivatives of artificial viscosity are required: $\partial_{\bar{A}} \tilde{Q}$, and $\partial_{\bar{u}} \tilde{Q}$. The first of these is straightforward to evaluate. However, $\tilde{Q}$ tends to trigger at a single gridpoint initially, which leads to somewhat unpleasant spatial derivatives. We suggest smoothing $\tilde{Q}$ over a few gridpoints in order to assist with this derivative. 

The temporal derivative is somewhat harder. Baumgarte \textit{et al.} suggest taking a one-sided first-order difference $(\tilde{Q}^{n} - \tilde{Q}^{n-1}) / \Delta \bar{u}$ to estimate this quantity. In most instances, this will be zero. The code will only be reduced to first order accuracy when shocks form and the triggering condition is met.

\section{Implementation Examples} \label{sec:plots}

In this section, we discuss animations from simulations of the formalism that we have described in this paper\footnote{These animations may be found in a playlist on youtube at \url{http://www.youtube.com/playlist?list=PLqRPTTDr7yKTD9HExr7Fgk_RVrg5RwTau}.}. We have included still frames from the animations in this paper, but we believe that the animations provide a much better picture of the evolution than these frames can describe. For the benefit of those only interested in understanding the animations, we include a brief summary of the formalisms.

\subsection{Density Animations}

\begin{figure}[t]
	\includegraphics[width=\columnwidth]{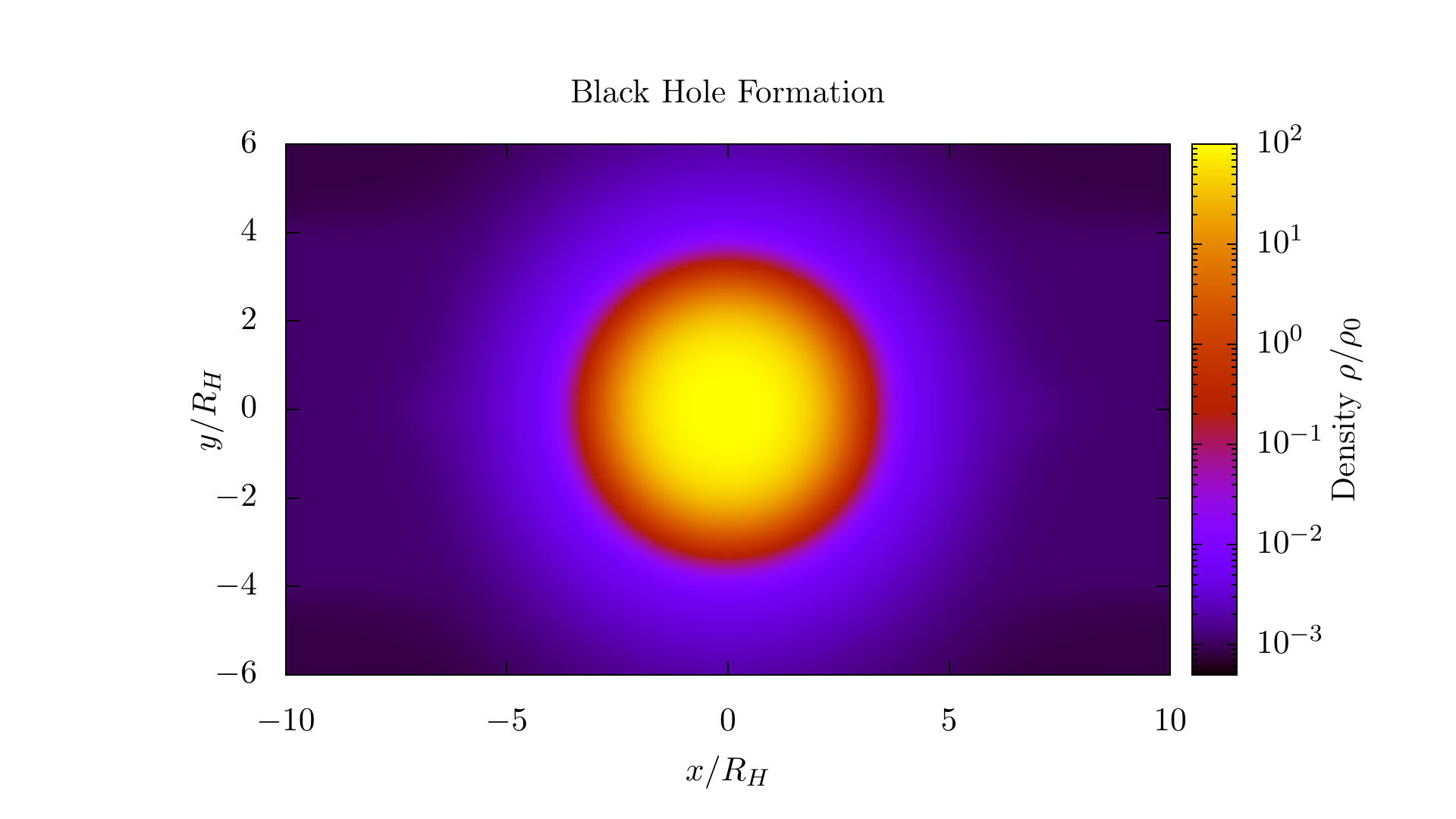}
	\caption{\label{fig:logmagic}Still frame taken from an animation describing the formation of a black hole in spherical symmetry.}
\end{figure}

We start our animations with a 2D density map that evolves in time to form a black hole (``Black Hole Formation''). This animation requires no understanding of our formalism to interpret, and gives a good idea of how a black hole can form in the early universe. The expansion of the universe and the corresponding dilution of energy is evident as the map evolves, before the density collapses to form a black hole. The animation runs through twice; the first cycle shows the density using a linear scale, while the second cycle uses a logarithmic scale. A still frame from the end of the logarithmic version is shown in Figure \ref{fig:logmagic}. 

\subsection{Misner-Sharp Formalism}

The rest of our animations are animated plots of our evolution variables. In order to understand them, it helps to know a little about the formalism we have used.

We simulate the early universe using a constant equation of state perfect fluid. For the purpose of these simulations, we take $w = 1/3$, corresponding to a radiation fluid. We use $A$ as a comoving radius, and define a dimensionless comoving radius $\bar{A} = A / R_H$, where $R_H$ is the horizon radius at the time of the initial data. The Misner-Sharp formalism uses a cosmic time variable defined by appropriate gauge conditions, and we evolve forwards using dimensionless time $\xi$ defined by $t = t_0 e^\xi$.

The Misner-Sharp formalism evolves three quantities forwards in time: the areal radius $R(A,t)$, which is a gauge variable, the fluid coordinate velocity $U(A,t)$, and the mass
\begin{align}
  m(A,t) = 4 \pi \int_0^A \rho R^2 \partial_A R dA
\end{align}
which can be thought of as the mass enclosed in the comoving radius $A$. The density $\rho$ can be computed from $m$ and $R$. For a number of reasons, it makes more sense to evolve quantities normalized by the values of the background cosmology, and so we define dimensionless tilded quantities $\tilde{R}$, $\tilde{U}$, $\tilde{m}$ and $\tilde{\rho}$ as
\begin{subequations}
\begin{align}
  R &= a A \tilde{R} = R_b \tilde{R}
\\
  U &= H R \tilde{U}
\\
  m &= \frac{4 \pi}{3} \rho_b R^3 \tilde{m}
\\
  \rho &= \rho_b \tilde{\rho}
\end{align}
\end{subequations}
where $\rho_b$ is the background (asymptotic) energy density. When evolving FRW, all tilded quantities are unity. It turns out that the initial background density $\rho_0$ scales out of all the equations, and that all that is needed to specify initial data that consists only of a growing mode is an initial function for $\tilde{m} = \tilde{m}_0$, from which initial data for $\tilde{R}_0$ and $\tilde{U}_0$ can be computed.

For the next six animations, we demonstrate the evolution of two Gaussian initial profiles for $\tilde{m}_0$, taken a little over and under the threshold for black hole formation. In particular, we take
\begin{align}
  \tilde{m}_0 = 1 + \kappa e^{-A^2 / 2 \sigma^2}
\end{align}
with $\sigma = 2 R_H$ and $\kappa_1 = 0.173$, $\kappa_2 = 0.175$\footnote{Our first animation is of the energy density from the evolution using $\kappa_2$.}. We take our domain to be $A_{max} = 20 R_H$. In the animations, the teal curve represents $\kappa_1$, which doesn't form a black hole, and the purple curve represents $\kappa_2$, which does.

After a black hole has formed, the singularity at its center rapidly forms, we can no longer evolve the simulation, and the purple curve in the animations vanishes. All of the plots displayed here are still frames from the animations, taken shortly after black hole formation.

\begin{figure}[t]
	\includegraphics[width=\columnwidth]{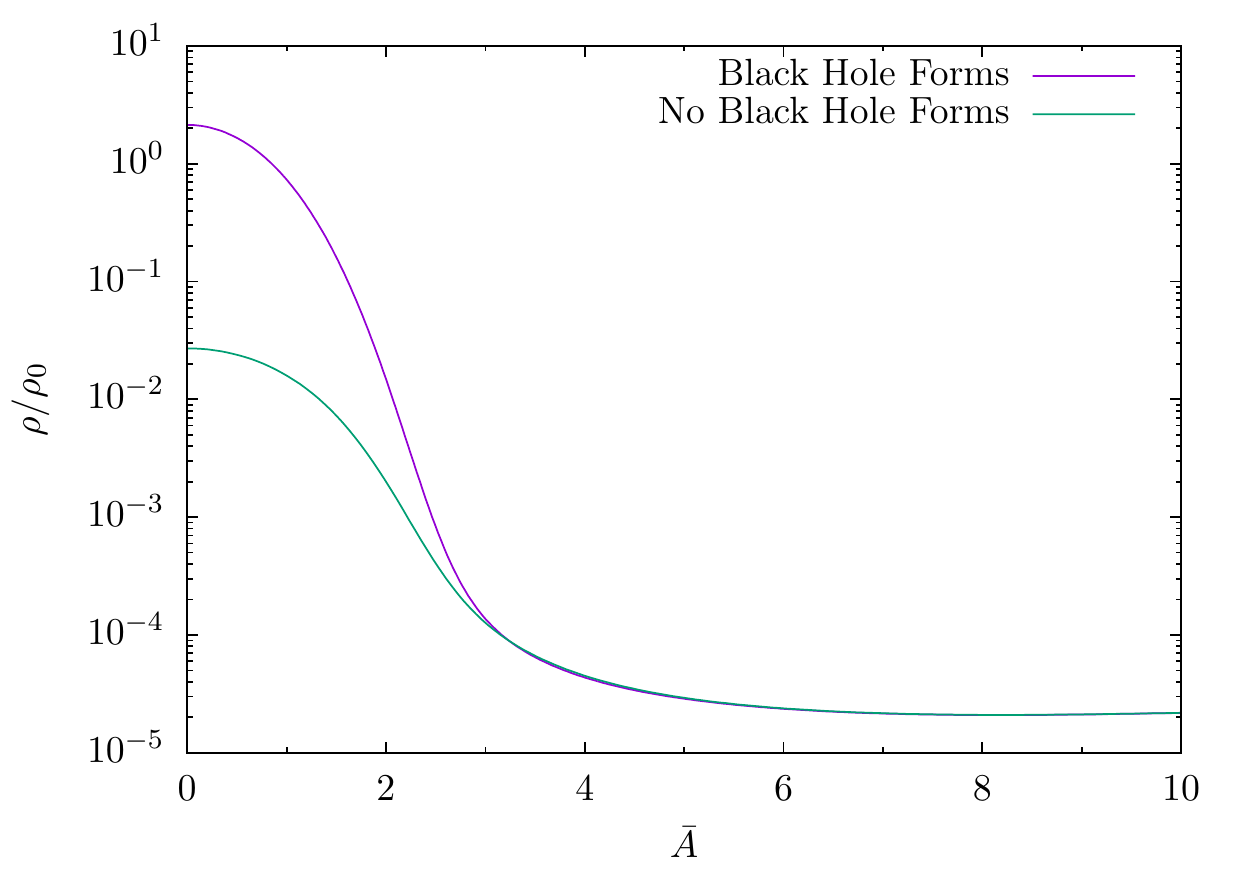}
	\caption{\label{fig:rho}Density profile $\rho/\rho_0$ as a function of comoving radius just after black hole formation.}
\end{figure}

The animation ``Density Profile'' plots the density $\rho/\rho_0$ as a function of comoving radius $\bar{A}$. The initial part of the animation looks incredibly flat, and shows the dilution of energy density due to the expansion of the universe. After some time, mass begins to accumulate near the origin and the black hole formation process begins. The density near the forming black hole remains many times higher than the background density, which continues to decay. As the purple curve forms a black hole, the density at the origin grows without bound. Our numerical simulation breaks down just before the singularity forms, and the purple curve vanishes at this point. When the teal curve fails to form a black hole, the extra energy is blasted out in a spherical wave. Note that the asymptotic (background) energy density drops at a constant rate, which occurs because we use a logarithmic time coordinate. The density profiles just after black hole formation are shown in Figure \ref{fig:rho}.

\begin{figure}[t]
	\includegraphics[width=\columnwidth]{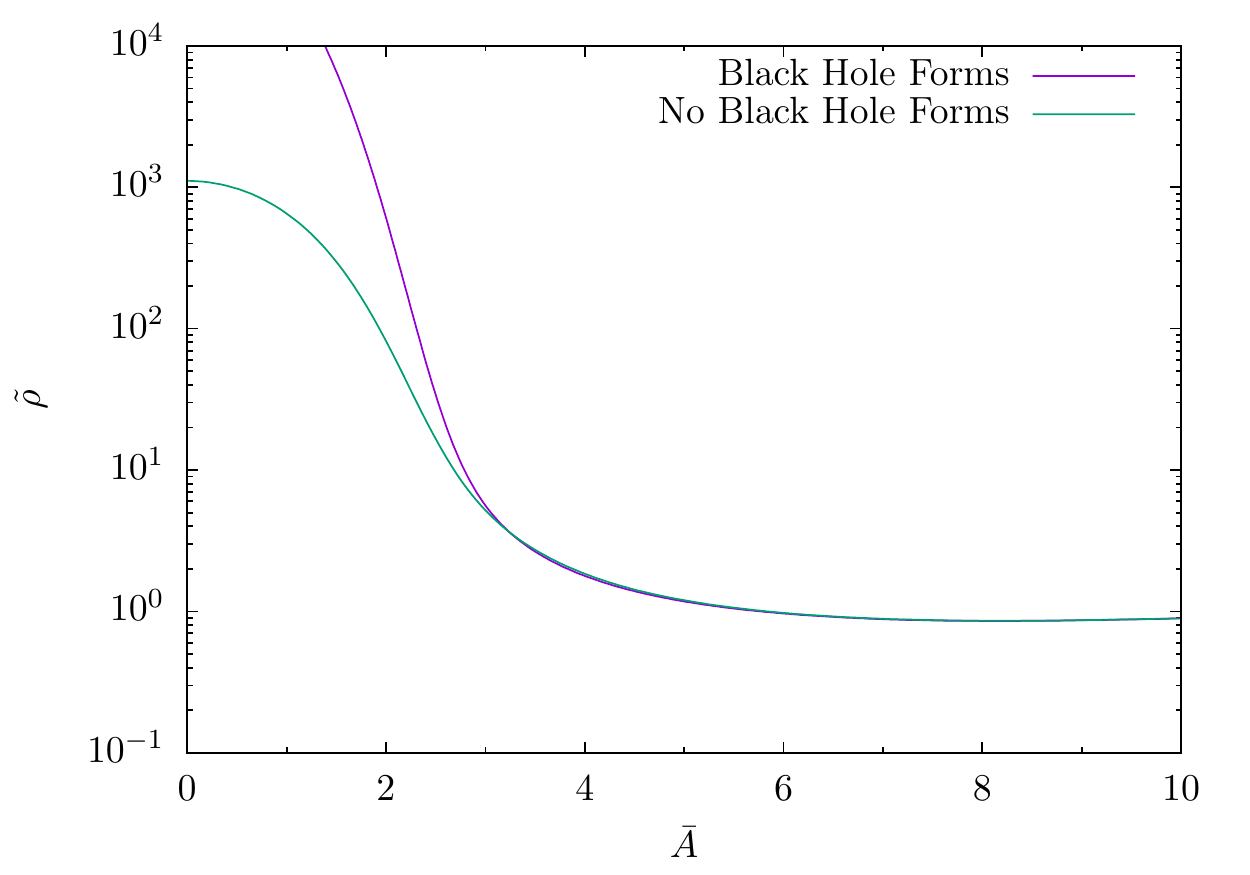}
	\includegraphics[width=\columnwidth]{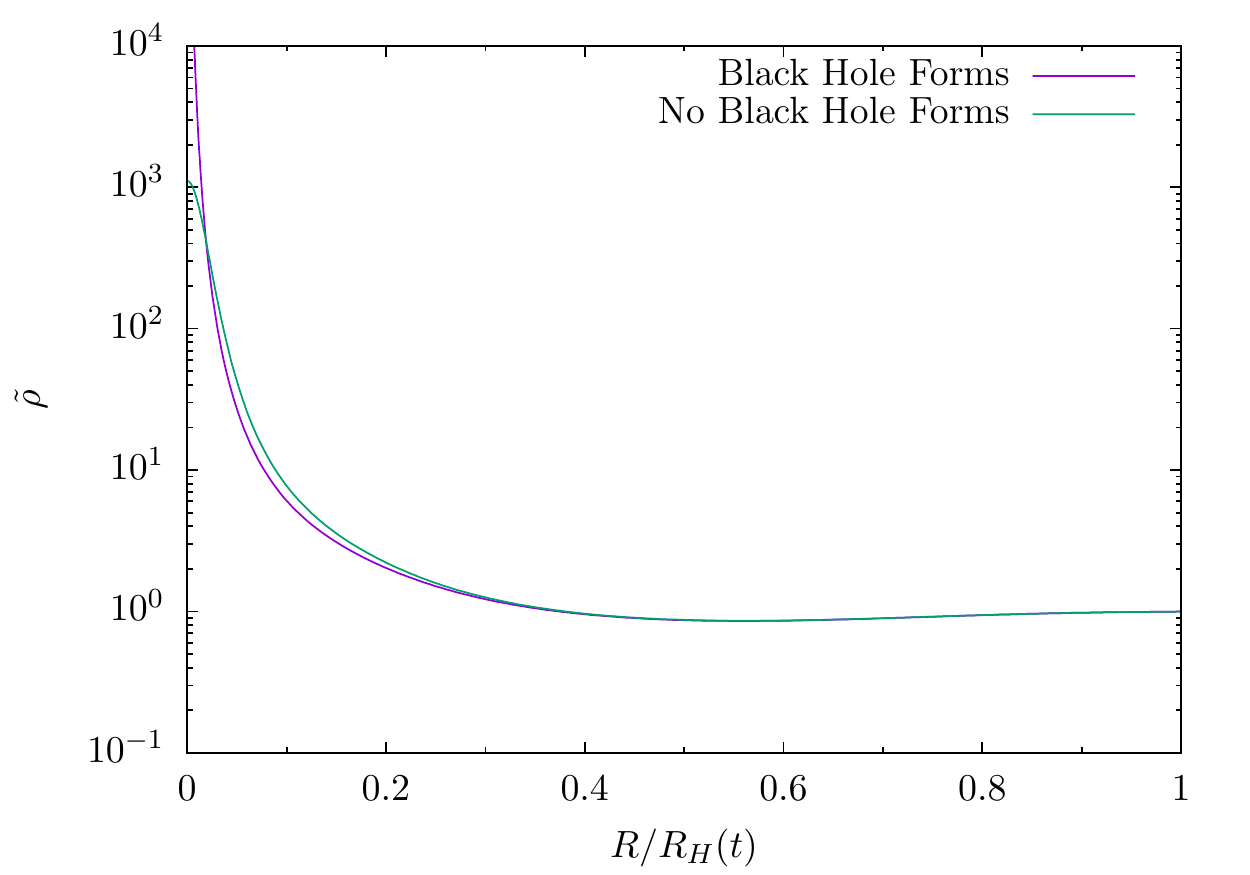}
	\caption{\label{fig:tilderho}Density profile $\tilde{\rho}$ as a function of comoving radius (above) and areal radius (below) just after black hole formation. The below curve has its $x$-axis scaled such that the horizon scale is at 1.}
\end{figure}

The next animation ``Density Perturbation (Comoving Radius)'' shows the density profile relative to the background density, which scales out the dilution due to the expansion of the universe. We see that the purple curve continues to increase in relative density until the black hole actually forms, while the teal curve reaches a peak of around a thousand times the background density before coming crashing down and releasing an outwards-moving density wave. The pressure when the teal curve peaks pushes the material out so rapidly that there is a strong rarefaction at the origin. The outgoing wave eventually forms a shockwave, which travels much faster than the speed of sound in the fluid. A still frame is shown in Figure \ref{fig:tilderho}.

The fourth animation ``Density Perturbation (Physical Radius)'' again shows the density profile relative to the background density, but this time plots it as a function of physical radius normalized by the horizon radius at the time (computed based on the background cosmology). Thus, $R/R_H = 1$ shows what is happening at the horizon scale at that time. Since the horizon grows with time, the horizontal scale corresponds to larger and larger physical distances as time progresses. We can clearly see when the perturbation has entered the horizon (when $\tilde{\rho}$ drops below unity at $R/R_H = 1$). Note that at the very end, the horizon scale grows larger than our domain of evolution, and so the outermost point we have data for recedes towards the left. It is interesting to compare the two plots in Figure \ref{fig:tilderho}; by looking at features in these plots, we can see the difference between comoving radius and physical radius. In particular, the inner radius has been squashed somewhat, while the region where the teal curve lies above the purple curve has been extended.

\begin{figure}[t]
    \includegraphics[width=\columnwidth]{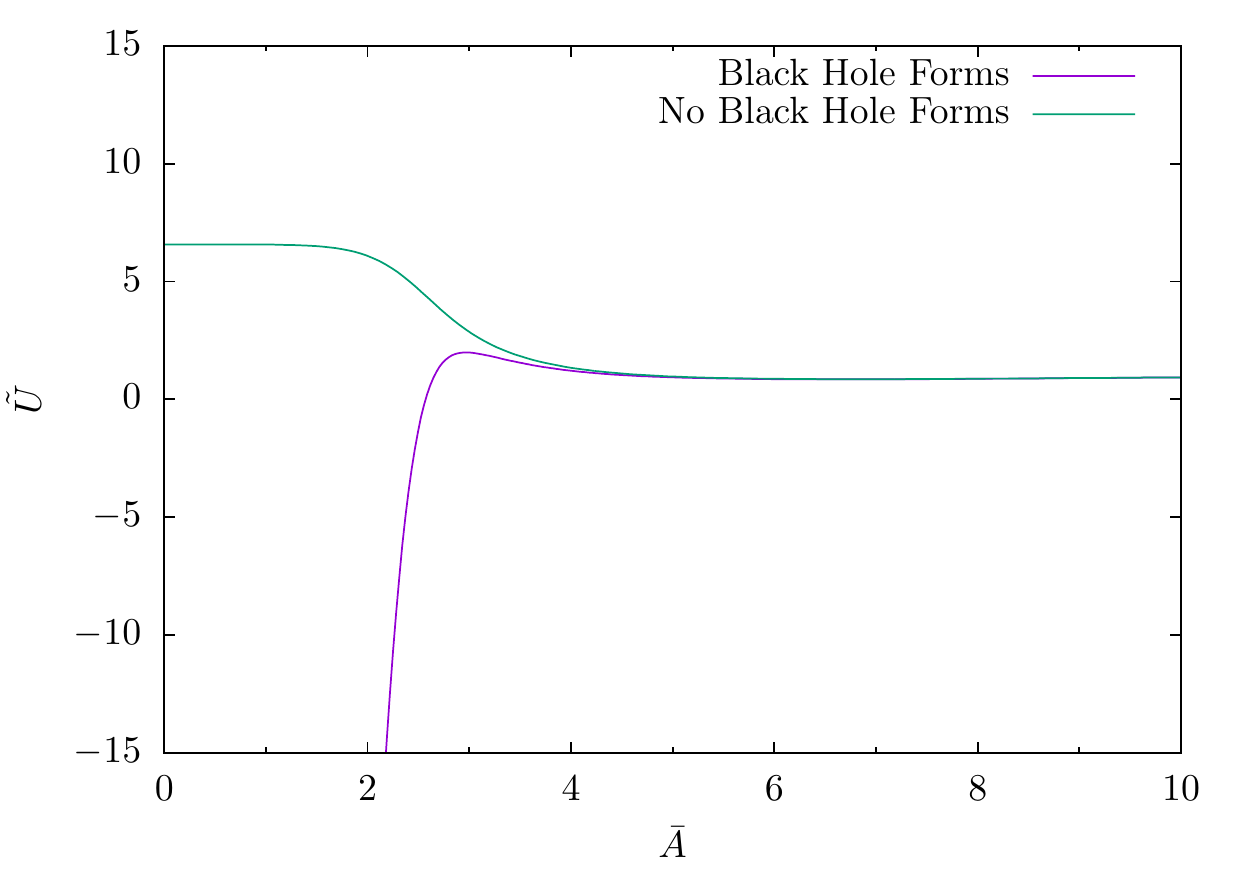}
    \caption{\label{fig:velocity}Velocity perturbation as a function of comoving radius just after black hole formation.}
\end{figure}

The fifth video ``Velocity Perturbation'' shows the normalized velocity perturbation as a function of comoving radius. Normalized velocities less than one are expanding slower than the Hubble rate, while negative velocities are moving in the opposite direction of the Hubble flow, attracted into the growing overdensity. Initially, the velocity is essentially unity everywhere, indicating that matter is moving with the Hubble flow. We can clearly see when matter begins to fall towards the central overdensity. Eventually, the pressure of the overdensity dominates over the gravitational pull for the teal curve, and the velocity rises very sharply, explosively sending out a wave of matter (or photons, as the case may be). For the purple curve, the velocity continues to grow negative as matter continues to fall into the overdensity, until a black hole is formed. It is interesting to note a small apparent outflow ($\tilde{U} > 1$) from the black hole-forming region. This turns out to be a coordinate velocity effect; as the relationship between physical velocity and coordinate velocity has become rather stretched in this vicinity, matter is still actually flowing inwards. The normalized velocity for the teal curve returns very rapidly to unity, despite being largely uninfluenced by the background cosmology. We believe that this is at least partly due to our gauge condition. A still frame can be seen in Figure \ref{fig:velocity}.

\begin{figure}[t]
    \includegraphics[width=\columnwidth]{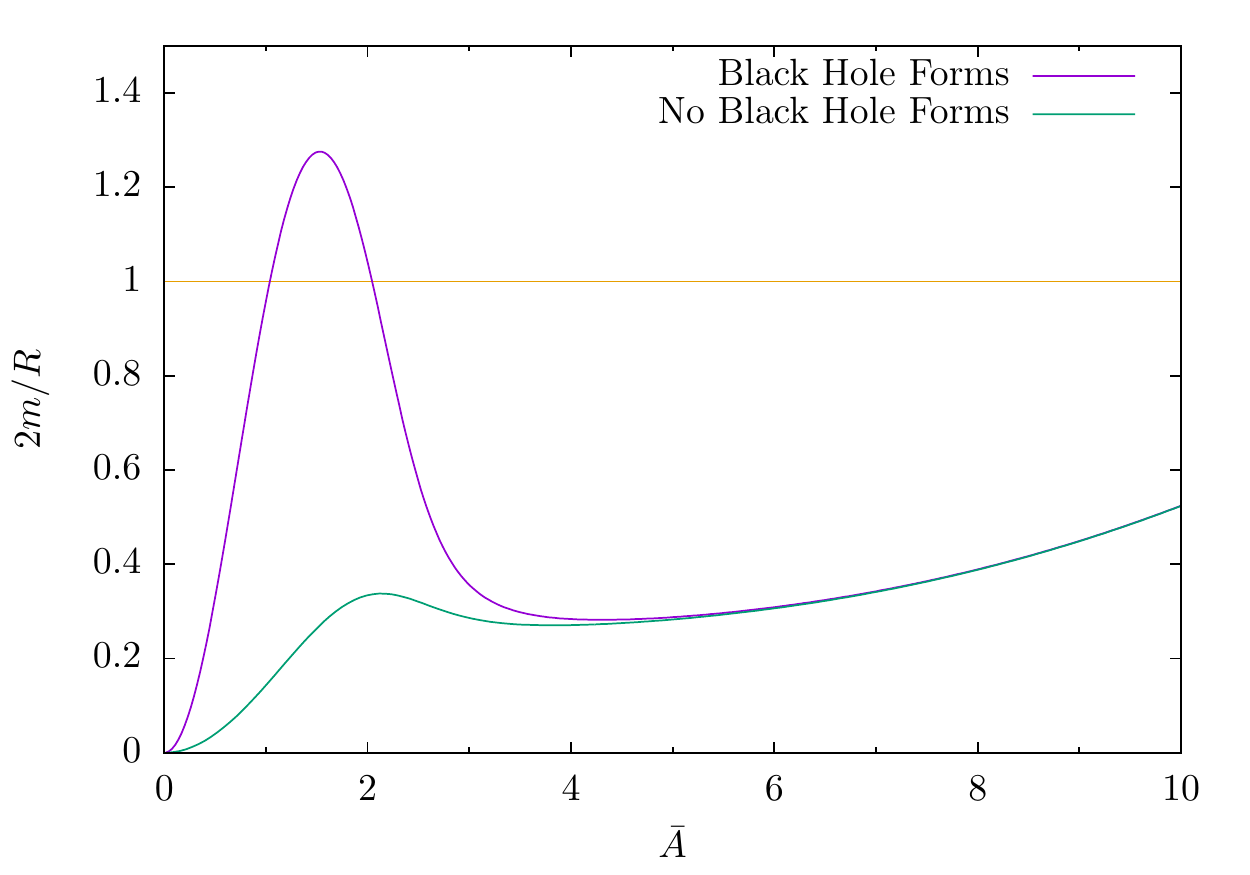}
    \includegraphics[width=\columnwidth]{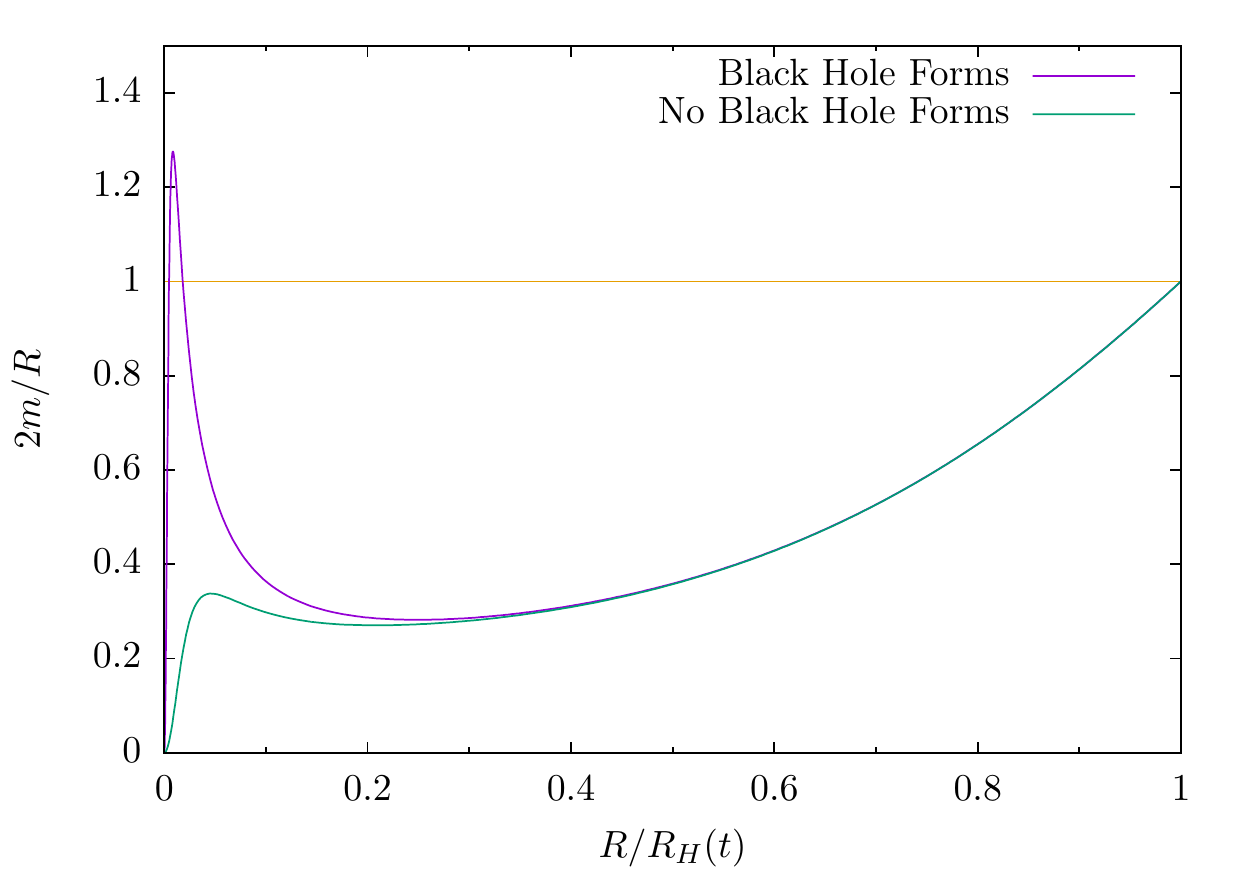}
    \caption{\label{fig:bhdetect}Above: Plot of the black hole detection condition as a function of comoving radius. Below: The same plot, but as a function of normalized areal radius (such that 1 is the horizon scale at that time). A horizon has formed when a curve lies above the yellow line and $\tilde{U}<0$ (cf. Figure \ref{fig:velocity}).}
\end{figure}

Animations 6 and 7 (``Black Hole Detection (Comoving Radius)'' and ``Black Hole Detection (Physical Radius)'') demonstrate the evolution of the quantity $2m/R$ as a function of radius. The black hole formation condition is $2m/R = 1$ (with $G = c = 1$; Eq. \eqref{eq:bhcondition}), which signifies the formation of an apparent horizon (which must be contained within an event horizon). We see that after the purple curve crosses the yellow line, a singularity rapidly forms, and we cannot evolve it any further (the purple curve vanishes). There is a second condition for a black hole to form, namely that $\tilde{U}$ must be negative at that point, which is why the curves can cross the yellow line at large radii without issue (the velocity there is roughly the Hubble flow, or $\tilde{U} \sim 1$). Still frames are shown in Figure \ref{fig:bhdetect}. Again, the difference between comoving coordinates and physical radius can be seen in the comparison between these two plots.

\subsection{Hernandez-Misner Formalism}

The Misner-Sharp formalism cannot evolve much past the formation of a black hole, due to the formation of a singularity. The Hernandez-Misner formalism aims to overcome this shortcoming by using a null time coordinate $u$. Data on a slice of constant $u$ represents what an observer at the outer boundary of the computational domain sees at that instant, retarded by the distance that the light needs to travel to get to them. As no null rays ever leave an event horizon, a black hole never forms from the perspective of this observer. In particular, the lapse near the formation of a black hole approaches zero, which indicates that evolution is essentially frozen. Even though a black hole never forms, one can extract information about the black hole by looking at where the lapse becomes nonzero again.

The Hernandez-Misner formalism evolves the same variables as the Misner-Sharp formalism, just using a different coordinate system. The initial data for these variables has to come from a Misner-Sharp evolution, as constructing initial data on a null slice is somewhat unusual. A strong test of one's code is to check to see that the evolution of a system is the same in the two different coordinate systems. This comparison is demonstrated in the eighth animation, ``Two Coordinate Systems''.

\begin{figure}[t]
    \includegraphics[width=\columnwidth]{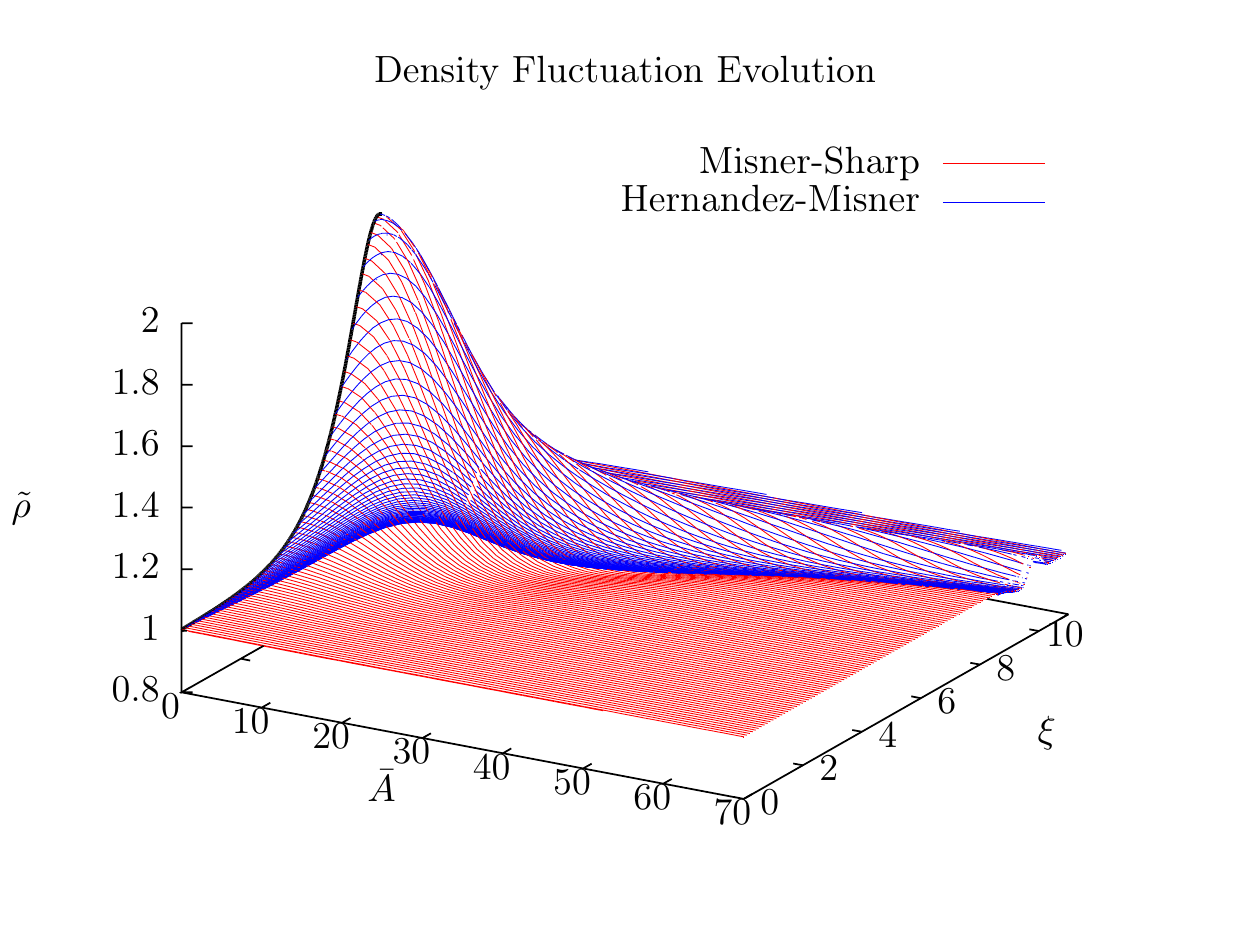}
    \caption{\label{fig:3d}A plot of the density perturbation $\tilde{\rho}$ as a function of $\bar{A}$ and $\xi$ in both the Misner-Sharp (red) and Misner-Hernandez (blue) coordinate systems.}
\end{figure}

In this animation, we show a three-dimensional plot with comoving radius $\bar{A}$ on the $x$-axis, logarithmic time $\xi$ on the $y$-axis, and density perturbation $\tilde{\rho}$ on the $z$-axis for an evolution that did not form a black hole. The red lines show timesteps from the Misner-Sharp evolution, where each time step is at constant $\xi$. The blue lines show timesteps from the Hernandez-Misner evolution, where each timestep is a null ray in the spacetime. The excellent agreement of the two surfaces shows that the evolution is in agreement between the two coordinate systems, including at the boundaries. A still frame from this animation is shown in Figure \ref{fig:3d}.

The rest of our animations are of the same purple black-hole-forming data that we showed in the Misner-Sharp animations above.

\begin{figure}[t]
	\includegraphics[width=\columnwidth]{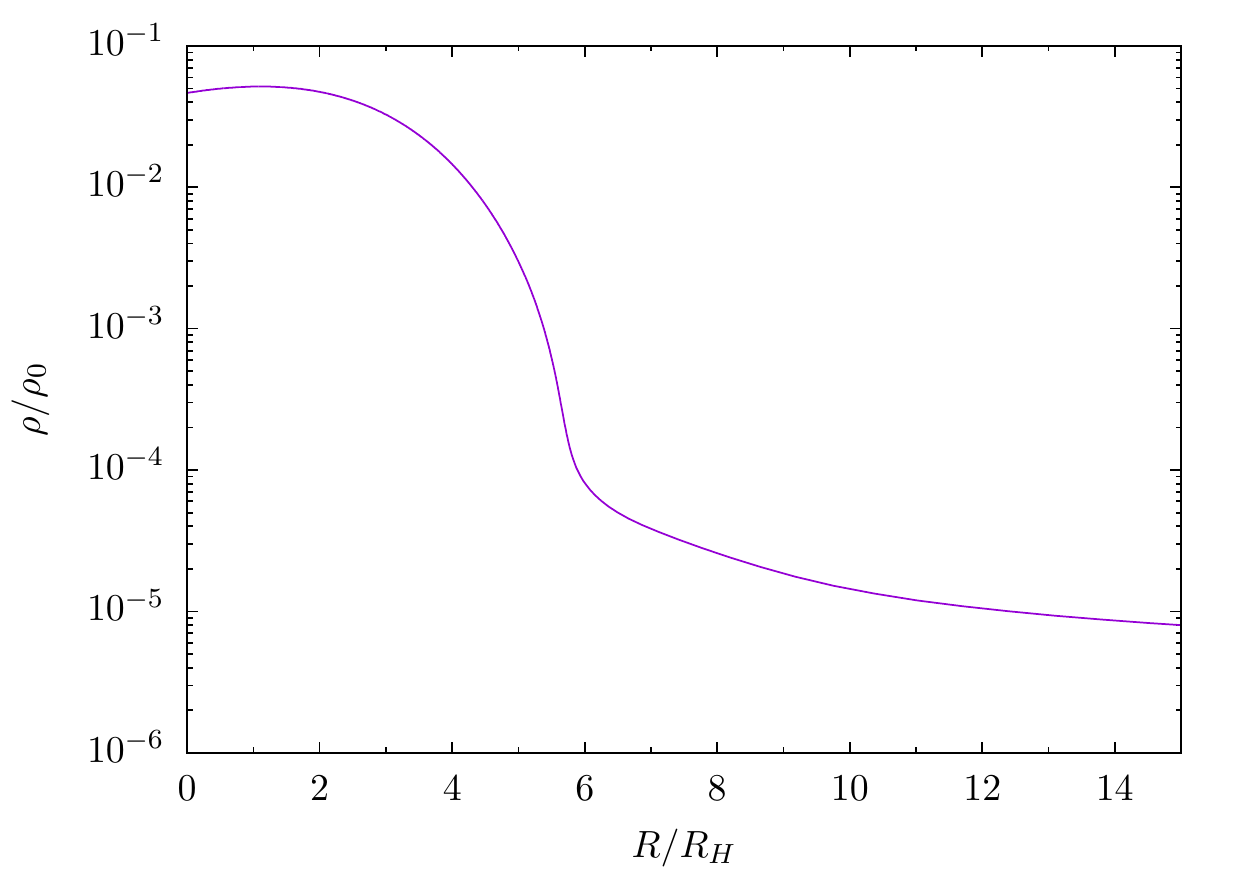}
	\caption{\label{fig:hrho}A plot of the physical density $\rho$ as a function of areal radius. The large bump to the left has essentially been frozen in place. Together, the energy contained in this frozen bump constitutes the mass of the black hole.}
\end{figure}

Our next animation, ``Density Profile (Null Slicing)'', is of the density profile in the Hernandez-Misner coordinates, plotting the physical density $\rho$ as a function of areal radius. We see that the density profile forms a large bump near the origin, as shown in Figure \ref{fig:hrho}. However, this bump is very much frozen in time by the vanishing lapse. In the Misner-Sharp coordinates, the density grows without bound at the origin; here, because we do not pierce the event horizon, it merely reaches a large density and stays fixed. As the evolution progresses, we see the energy density outside the black hole continues to fall off with cosmic expansion, though some matter accretes onto the black hole and freezes. Note that due to the null slicing, points on the curve at larger radius correspond to later times, and thus more expansion has occurred, decreasing the energy density.

\begin{figure}[t]
	\includegraphics[width=\columnwidth]{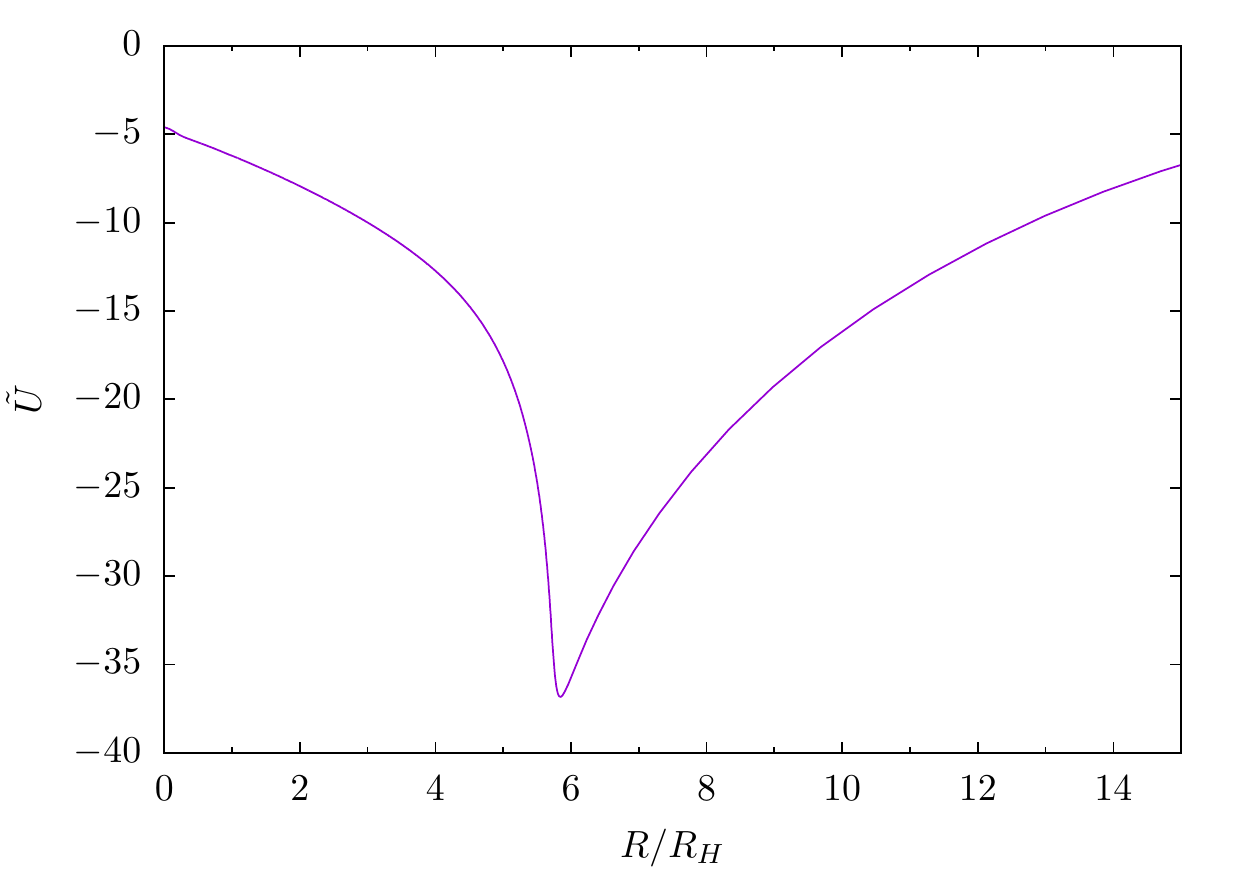}
	\caption{\label{fig:hmu}This plot shows a frame from an animation of $\tilde{U}$ as a function of areal radius, late in the formation of the black hole. As $\tilde{U}$ is always negative, matter is sharply falling into the black hole.}
\end{figure}

The animation ``Velocity Perturbation (Null Slicing)'' of the normalized velocity perturbation $\tilde{U}$ as a function of areal radius shows the formation of a huge negative spike when the black hole forms. Figure \ref{fig:hmu} shows a frame from late in the animation. We see that all matter is falling sharply inwards (recall that $\tilde{U} = 1$ is the Hubble flow). The animation also exhibits the freezing of the lapse. Starting from the origin, a trail of ``frozen'' velocity profile creeps outwards.

Our final animation, ``Lapse in the Null Slicing'', is that of the lapse as a function of areal radius. We do not include a still frame plot here, as the animation is simply an animated version of Figure \ref{fig:lapse}. The animation clearly shows the lapse rapidly vanishing, which freezes the evolution of all variables.

\section{Conclusions} \label{sec:conclusions}

We have presented a comprehensive formalism for investigating primordial black hole formation in full general relativity. The basic idea is to construct an initial density profile and evolve it forwards in time using a standard time-slicing until either a black hole forms or the overdensity disperses. If a black hole forms, a second formalism that avoids the formation of a singularity is used to continue the evolution and compute the mass of the resulting black hole.

The foundation of the approach is the Misner-Sharp formalism, which we have adapted to cosmological applications. We paid special attention to the outer boundary, where we constructed a non-reflecting boundary condition. For extracting information about black holes that form, we turned to the Hernandez-Misner formalism. Here, we adapted the system of equations to use variables more suited to cosmological and numerical evolution.

The issue of initial conditions was explored in detail, where we identified an expansion in two small parameters, and explicitly constructed initial conditions consisting only of the linear cosmological growing mode to second order, based on a single function $\delta_{m0}$. The regime of validity of the expansion was investigated, including both perturbative convergence and physical constraints. We related these initial conditions to a variety of approaches in the literature, including Carr's fractional mass excess, Polnarev's initial curvature perturbation, Bardeen gauge invariant variables, and physical perturbation profiles at the end of inflation.

The approach that has been developed here builds upon numerous other approaches in the literature. Most importantly, various ideas from the literature have been combined to form a clear and comprehensive picture. Numerically, there are numerous benefits to the enhancements we have incorporated, including completely stable cosmological background evolution, increased computational efficiency, and increased numerical accuracy. Theoretically, we have investigated initial conditions in unprecedented detail, constructed a rigorous connection to inflation, introduced appropriate cosmological boundary conditions, and clarified the role of the initial curvature perturbation within this formalism.

In addition to describing the formalism, we have constructed a number of animations to describe what happens in primordial black hole formation and near misses. We hope that these animations will help spur interest in the subject, and provide a visual interpretation of what actually happens in primordial black hole formation.

We plan to use the formalism developed here to investigate properties of initial profiles that give rise to black holes and to quantify the resulting black hole masses that are produced. In the long term, we hope to run Monte Carlo simulations from inflationary initial profiles to identify the number density and mass spectrum of primordial black holes formed from different models of inflation.

\acknowledgments

We thank Mark Hertzberg, Evangelos Sfakianakis, Alan Guth, Curran Muhlberger, Scott Hughes, Geoffrey Lovelace and Gerard Pascual-Lopez for helpful discussions. This work is supported in part by the U.S. Department of Energy under grant Contract Number DE-SC00012567.

\appendix
\section{Stress-Energy Tensor} \label{app:stressenergytensor}

In this appendix, we derive the stress-energy tensor for a spherically symmetric fluid, based on a statistical description of the system. This reproduces standard derivations, and is included for completeness.

The description of a particle in a generic space-time is given by its position and momentum $\{ x^{\alpha} (\lambda) , p^{\alpha} (\lambda) \}$, where an affine parameter $\lambda$ is employed. The momentum is related to the position through
\begin{align}
p^{\alpha} \equiv \frac{d x^{\alpha}}{d \lambda} \left[ = m \frac{d x^{\alpha}}{d \tau} \right]
\end{align}
where the second equivalence is only defined for massive particles of mass $m$. A particle's momentum obeys the mass shell equation,
\begin{align}
g^{\alpha \beta} p_{\alpha} p_{\beta} = - m^2 \,. \label{eq:mshell}
\end{align}
Note that given the spatial components of a particle's four-momentum, this constraint determines the four-momentum time component, and thus only three of the four components of momentum are independent.

Analogously to the definition in Special Relativity, we definite the {\it number density of particles in phase space} $\mathcal{N}$ to be the function which describes the number of particles per unit volume in both position and momentum space. When integrated over any space-like hypersurface $\Sigma$ and set of momenta $\mathcal{R}$, this yields the total number of particles in $\Sigma$ and ${\cal R}$.
\begin{align}
N = \int\limits_{\Sigma} d^3 \Sigma_{\alpha}(x^{\beta}) \int\limits_{\mathcal{R}} d \mu (p_{\beta}) p^{\alpha} \mathcal{N} (x^{\beta},p_{\beta}) \label{eq:Ndef}
\end{align}
Here, $p^{\alpha} d\Sigma_{\alpha}(x^{\beta})$ and $d \mu (p_{\beta})$ are the invariant volume measures of $\Sigma$ and $\mathcal{R}$ respectively. The 3-dimensional volume element $d\Sigma_{\alpha}(x^{\beta})$ takes the form,
\begin{align}
d \Sigma_{\alpha}(x^{\beta}) = \epsilon_{\mu \nu \sigma \alpha} \frac{\partial x^{\mu}}{\partial a} \frac{\partial x^{\nu}}{\partial b} \frac{\partial x^{\sigma}}{\partial c} da db dc
\end{align}
where $\epsilon_{\mu \nu \sigma \alpha}$ is the antisymmetric tensor with $\epsilon_{0123}=+1$, and we have parameterized $\Sigma$ by $(a,b,c)$ so that $x^{\alpha}=x^{\alpha}(a,b,c)$. Since $\Sigma$ is a 3-dimensional hypersurface in a 4-dimensional manifold, its volume element is a vector pointing along the direction normal to it. In order to construct a scalar from $d \Sigma_{\alpha}(x^{\beta})$ we must take its inner product with another four-vector. The only other true four-vector available in a statistical picture is the four-momentum $p^{\alpha}$, which we use to construct the invariant volume. 

The invariant momentum volume element is
\begin{align}
d \mu (p_{\beta}) = 2 \delta \left( p^{\beta} p_{\beta} +m^2 \right) \Theta(p^0) \frac{d^4p}{\sqrt{-g}} \,.
\end{align}
The delta function guarantees the momenta obey the constraint \ref{eq:mshell}, and the step-function $\Theta$ picks out the positive energy solution. The factor of two arises from the fact that momentum is squared in the delta function. Note that the infinitesimals $dp$ refer to momenta with lowered index (this arises because $p_\alpha$ is the conjugate momentum to $x^\alpha$).

Note that in Minkowski spacetime on a constant time slice, we have $d \Sigma_{\alpha}(x^{\beta})=(1,\vec{0}) d^3 x$, $p^{\alpha}=(E,\vec{p})$, and $d \mu (p^{\beta}) = \frac{1}{E} d^3 p$, so \ref{eq:Ndef} reduces to
\begin{align}
N = \int\limits_{\Sigma} d^3x \int\limits_{\mathcal{R}} d^3p \ \mathcal{N}(t,\vec{x},\vec{p})
\end{align}
as expected, where we have integrated over the delta function in momentum space. A number density per volume in space can be obtained by simply not integrating over the spatial hypersurface.

We can take different moments of the number density $\mathcal{N}$ by multiplying it by $p^{\alpha}$ and integrating over momentum space. In particular, we can define the {\it stress-energy tensor} by the integral
\begin{align}
T^{\alpha \beta}(x^{\nu}) \equiv \int\limits d \mu (p_{\nu}) p^{\alpha} p^{\beta} \mathcal{N}(x^{\nu},p_{\nu}) \,. \label{eq:Tdef}
\end{align}
Ignoring normalization, this is the only rank two tensor which we can construct from the invariants $\mathcal{N}$, $p^{\mu}$, and $d \mu(p_{\nu})$ that is linear in $\mathcal{N}$. One can be readily check that this object agrees with the usual stress-energy tensor by transforming to a local Lorentz frame.

Let us now construct the stress-energy tensor in spherical coordinates. In order to compute the stress-energy tensor, we need to integrate over all possible momenta at a given position in space $x^i = (r, \theta, \phi)$. To facilitate this, we construct an orthonormal Cartesian basis at $x^i$, oriented such that the $x$ direction aligns with $\hat{r}$, the $y$ direction aligns with $\hat{\theta}$, and the $z$ direction aligns with $\hat{\phi}$. We can then describe momenta at that point in space through the components $(p^x, p^y, p^z)$ in the orthonormal basis. For the metric described in Eq. \eqref{eq:MSmetric}, these components are related to the momenta $p^\mu = (p^0, p^r, p^\theta, p^\phi)$ by
\begin{subequations}
\begin{align}
  p^r &= \frac{p^x}{e^{\lambda / 2}}
\\
  p^\theta &= \frac{p^y}{R}
\\
  p^\phi &= \frac{p^z}{R \sin{\theta}} \,.
\end{align}
\end{subequations}

We now write the components of the momentum in the orthonormal basis in spherical polar coordinates defined by
\begin{align}
 (p^x, p^y, p^z) = (p \cos \Theta, p \sin \Theta \cos \Phi, p \sin \Theta \sin \Phi) \,. \label{eq:app:porthonormal}
\end{align}
Here, $\Theta$ denotes the angle away from the radial ($x$) axis, while $\Phi$ is the azimuthal angle about the radial direction, and $p^2 = (p^x)^2 + (p^y)^2 + (p^z)^2$.

We can now construct $p_r$, $p_\theta$ and $p_\phi$ in terms of $p$, $\Theta$ and $\Phi$. Begin by noting that $p_x = p^x$, thanks to the orthonormal basis. We can then construct the following relationships.
\begin{subequations}
\begin{align}
  p_r &= e^{\lambda \slash 2} p_x = e^{\lambda \slash 2} p \cos \Theta
\\
  p_\theta &= R p_y = R p \sin \Theta \cos \Phi
\\
  p_\phi &= R \sin \theta p_z = R p \sin \Theta \sin \Phi
\end{align}
\end{subequations}

The volume element $d^4 p / \sqrt{-g}$ on momentum space then becomes the following.
\begin{align}
  \frac{d^4p}{\sqrt{-g}} &= \frac{1}{e^{\phi} e^{\lambda / 2} R^2 \sin \theta} dp_0 \, dp_r \, dp_\theta \, dp_\phi \nonumber
\\
  &= e^{- \phi} dp_0 \, dp_x \, dp_y \, dp_z \nonumber
\\
  &= e^{- \phi} dp_0 \, p^2 \, \sin \Theta \, dp \, d\Theta \, d\Phi
\end{align}
The range of $p$ is from 0 to $\infty$, while $0 \le \Theta \le \pi$ and $0 \le \Phi < 2 \pi$. The complete integration measure on momentum space then becomes
\begin{align}
  d \mu (p^\nu) = 2 \delta \left( - \frac{p_0^2}{e^{2 \phi}} + p^2 \right) \Theta(p^0) \frac{dp_0}{e^\phi} \, p^2 \, \sin \Theta \, dp \, d\Theta \, d\Phi
\label{eq:intmes}
\end{align}
where we take $m=0$.

We now calculate the stress-energy tensor. Inserting the integration measure \eqref{eq:intmes} into Eq. \eqref{eq:Tdef} and integrating over the delta function, we arrive at
\begin{align}
  T^{\alpha \beta} = \int p^3 \, \sin \Theta \, dp \, d\Theta \, d\Phi \, n^\alpha n^\beta \cn \label{eq:app:stresspartway}
\end{align}
where we have defined 
\begin{align}
  n^\alpha &= p^\alpha / p 
  = \left(e^{-\phi}, \frac{\cos \Theta}{e^{\lambda/2}}, \frac{\sin \Theta \cos \Phi}{R}, \frac{\sin \Theta \sin \Phi}{R \sin \theta} \right) \,.
\end{align}

In order for the distribution to be spherically symmetric, we require it to be a function of
\begin{align}
  \cn(x^\alpha, p_\alpha) = \cn(t, r, p, \Theta)
\end{align}
only, where $\cos \Theta$ is related to the rotationally invariant dot product $x^i p_i$. Note that $p = \sqrt{g^{ij} p_i p_j}$ is also rotationally invariant.

Evaluating the integrals in Eq. \eqref{eq:app:stresspartway} over $\Theta$ and $\Phi$ component by component, we find
\begin{subequations}
\begin{align}
  T^{tt} &= 2 \pi e^{-2 \phi} \int f \, d\zeta
\\
  T^{tr} &= 2 \pi e^{- \phi - \lambda \slash 2} \int \zeta f \, d\zeta
\\
  T^{rr} &= 2 \pi e^{- \lambda} \int \zeta^2 f \, d\zeta
\\
  T^{\theta \theta} &= \frac{\pi}{R^2} \int (1 - \zeta^2) f \, d\zeta
\\
  T^{\phi \phi} &= \frac{\pi}{R^2 \sin^2 \theta} \int (1 - \zeta^2) f \, d\mu
\end{align}
\end{subequations}
with all other components vanishing, where we use $\zeta = \cos \Theta$ and define
\begin{align}
  f(t, r, \zeta) = \int p^3 \, \cn \, dp \,.
\end{align}
Each of these integrals involves different moments of the function $f$ with respect to the variable $\zeta$.

If we assume that the particles in the spacetime in question are in thermal equilibrium, then at a given point in space, particles have equal probability to be going in any direction, and so $f$ cannot depend on $\zeta$. Under this assumption, the off-diagonal component vanishes, and if we set
\begin{subequations}
\begin{align}
u^{\alpha} &= (e^{-\phi},0,0,0) 
\\
\rho &= 2 \pi \int f \, d\zeta 
\\
P &= \frac{\rho}{3}
\end{align}
\end{subequations}
then we can write $T^{\alpha \beta}$ as
\begin{align}
T^{\alpha \beta} = P g^{\alpha \beta} + u^{\alpha} u^{\beta} (\rho + P)
\end{align}
which is a perfect fluid in comoving coordinates, and the desired result. Note that this also yields that the equation of state of radiation is $w = 1/3$.

One might ask how non-interacting radiation (photons and neutrinos) can yield a fluid description. The key assumption that facilitated this was that of thermal equilibrium. So long as there are other species present in the universe to scatter radiation efficiently enough to maintain thermal equilibrium, then the radiation bath behaves like a fluid. When neutrinos decouple from the bath and begin to freely stream, then the thermal equilibrium assumption will no longer hold, and the stress-energy tensor will develop an off-diagonal $T^{tr}$ component. At this stage, the photon bath may still be described by a perfect fluid, but the evolution of the neutrino species will have to be described by the full Boltzmann equation. Such a description is beyond the scope of the formalism described in this paper.

\section{Method of Characteristics} \label{app:char}

Here we use the method of characteristics on the linearized Misner-Sharp equations to find inward and outward moving components of the solutions. We want to find the inward moving component so that we can create a nonreflecting boundary at the outer edge of our computational domain. We find the characteristic curves along which the solutions propagate, but the equations do not appear to permit a direct solution. For this appendix, we restrict our analysis to the case where $w=1/3$. We also define $c_s^2 \equiv e^\xi/12$. Using these, we write the linear evolution equation, Eq. \eqref{eq:fullm} as
\begin{align}
	\partial_\xi^2{\delta_m} - \frac{1}{2}\partial_\xi{\delta_m} - \frac{1}{2}\delta_m - c_s^2\left(\frac{4\delta_m'}{\bar{A}}+\delta_m''\right) = 0.
\end{align}

To use the method of characteristics, we need a first order linear equation, so we create a vector with the derivatives of $\delta_m$ and write the matrix equation:
\begin{align}
	\partial_\xi{\vec{f}} + M \cdot \vec{f}' = \vec{k}, \label{eq:mat}
\end{align}
where
\begin{align}
	\vec{f} =& \begin{pmatrix}
	\delta_m \\
	\partial_\xi\delta_m \\
	\delta_m'
	\end{pmatrix}\\
	M =& \begin{pmatrix}
	0 & 0 & 0 \\
	0 & 0 & -c_s^2 \\
	0 & -1 & 0
	\end{pmatrix}\\
	k =&\begin{pmatrix}
	f_2 \\
	\frac{1}{2}\left(f_1+f_2\right) + \frac{4 c_s^2}{\bar{A}}f_3 \\
	0
	\end{pmatrix}.
\end{align}

To study the solutions of this equation, it is helpful for us to find the eigenvalues and eigenvectors of $M$. In this basis, we will still have a coupled system, but this will allow us to write down three equations where each has derivatives of only one the scalar functions appearing. The eigenvalues and eigenvectors are
\begin{subequations}
\begin{align}
\lambda_1&= 0 & \vec{v}_1&=\begin{pmatrix}1\\0\\0\end{pmatrix}\\
\lambda_2 &= -c_s& \vec{v}_2&=\begin{pmatrix}0\\c_s\\1\end{pmatrix}\\
\lambda_3 &= +c_s& \vec{v}_3&=\begin{pmatrix}0\\-c_s\\1\end{pmatrix}.
\end{align}
\end{subequations}

We introduce $u_i$ such that $\vec{f} = u_1 \vec{v}_1 + u_2 \vec{v}_2 + u_3 \vec{v}_3$, from which we can identify
\begin{subequations}
\begin{align}
  \delta_m &= u_1
\\
  \partial_\xi \delta_m &= c_s (u_2 - u_3)
\\
  \delta_m' &= u_2 + u_3 \,.
\end{align}
\end{subequations}
Substituting $\vec{f}$ into the equation of motion, we find
\begin{align}
\sum_i \vec{v}_i \left(\partial_\xi({u_i}) + \lambda_i u_i' \right) &= \vec{k} - \sum_i u_i \partial_\xi{\vec{v}_i} \,.
\end{align}
Note that because $c_s$ depends on $\xi$, $\partial_\xi{\vec{f}}$ has terms coming from the time dependence of the eigenvectors. Extracting the coefficient of each eigenvector from this equation, we obtain
\begin{subequations} \label{eq:charwave}
\begin{align}
\partial_\xi {u_1}& = c_s \left(u_2-u_3\right)\\
\partial_\xi {u_2}-c_s u_2'& = \frac{1}{4c_s}u_1 + \frac{2c_s}{\bar{A}}\left(u_2 + u_3\right) \label{eq:inwave}\\
\partial_\xi {u_3}+c_s u_3'& = -\frac{1}{4c_s}u_1 - \frac{2c_s}{\bar{A}}\left(u_2 + u_3\right).
\end{align}
\end{subequations}

With the equations in this form, we may apply the method of characteristics. This method takes a first order PDE and computes curves called characteristics along which the equation becomes an ODE. We introduce $\lambda$ to parameterize these curves.
\begin{subequations}\label{eq:charpara}
\begin{align}
\partial_\lambda \xi_1 &= 1,
&
\partial_\lambda \bar{A}_1 &= 0,
&
\partial_\lambda u_1 &= c_s(u_2 - u_3)
\\
\partial_\lambda \xi_2 &= 1,
&
\partial_\lambda \bar{A}_2 &= -c ,
&
\partial_\lambda u_2 &= \frac{u_1}{4c_s} + \frac{2c_s}{\bar{A}}\left(u_2 + u_3\right)
\\
\partial_\lambda \xi_3 &= 1,
&
\partial_\lambda \bar{A}_3 &= c,
&
\partial_\lambda u_3 &= -\frac{u_1}{4c_s} - \frac{2c_s}{\bar{A}}\left(u_2 + u_3\right)
\end{align}
\end{subequations}
Here, $\bar{A}_i$, $\xi_i$ and $u_i$ are only functions of the parameter $\lambda$.

We can straightforwardly integrate the equations for $\xi_i$ and $\bar{A}_i$, eliminating $\lambda$.
\begin{subequations}
\begin{align}
\bar{A}_1(\xi_1) &= \bar{A}_1(0) \\ \bar{A}_2(\xi_2) &= \bar{A}_2(0) - 2 c_s \\ \bar{A}_3(\xi_3) &= \bar{A}_3(0) + 2 c_s
\end{align}
\end{subequations}
$\bar{A}_i(0)$ are arbitrary constants giving the value of $\bar{A}$ on the slice $\xi = 0$. Thus we can see that the second family of solutions correspond to inward traveling waves, so if we wish to set a nonreflecting boundary, it is $u_2$ for which we need a boundary condition.

Unfortunately, the equations cannot be solved explicitly. The condition we wish to satisfy is $u_i = 0$ for $\bar{A}>\bar{A}_{max}$ on the initial slice, $\xi = 0$ (assuming FRW initial conditions outside $\bar{A}_{max}$). As time passes and $u_3$ tracks outwards, $u_i=0$ does not remain true beyond the boundary because $u_3$ sources the evolution of the other variables in Eqs. \eqref{eq:charpara}. Since beyond the boundary the values of $u_1$ and $u_2$ are initially zero and sourced only by $u_3$, all of the information about their evolution should be encoded in the history of $u_3$ at the boundary $\bar{A}=\bar{A}_{max}$. However, we could not find a way to use this knowledge to our advantage. Instead, we take the characteristic equations \eqref{eq:charwave} and pursue a local boundary condition, to reasonable success.

\section{Connection to Cosmological Perturbation Theory} \label{app:perturbation}

In this appendix, we connect the formalism developed in this paper to standard cosmological perturbation theory.

\subsection{Standard Cosmological Perturbation Theory}

We consider the perturbed FRW metric
\begin{align} \label{eq:whollyperturbed}
  ds^2 = {}& - (1 + 2 \phi) dt^2 + 2 \tilde{\nabla}_i B \, dx^i dt
\nonumber\\ & \;
  + a^2 \left[ (1 + 2 C) \tilde{g}_{ij} + 2 \tilde{\nabla}_i \tilde{\nabla}_j E \right] dx^i dx^j
\end{align}
where $\tilde{g}_{ij}$ is the background flat, time-independent three-dimensional metric on space, and $\tilde{\nabla}_i$ are covariant derivatives associated with this metric (we raise and lower indices on $\tilde{\nabla}_i$ with $\tilde{g}_{ij}$). We include only scalar perturbations, as that is sufficient for our purposes. Consider the transformation
\begin{align}
  x^\mu \rightarrow x^\mu + \xi^\mu (t, x^i)
\end{align}
with
\begin{align}
  \xi^0 = T, \qquad \xi^i = L^i = \tilde{\nabla}^i L \,.
\end{align}
Under this transformation,
\begin{align}
  \delta g_{\mu \nu} \rightarrow \delta g_{\mu \nu} - {\cal L}_\xi \bar{g}_{\mu \nu}
\end{align}
where $\bar{g}_{\mu \nu}$ is the background metric, and the Lie derivative is given by
\begin{align}
  {\cal L}_\xi \bar{g}_{\mu \nu} = \xi^\lambda \partial_\lambda \bar{g}_{\mu \nu} + \partial_\mu \xi^\lambda \bar{g}_{\lambda \nu} + \partial_\nu \xi^\lambda \bar{g}_{\mu \lambda} \,.
\end{align}
Evaluating the Lie derivative yields the following, where dots represent time derivatives.
\begin{align}
  {\cal L}_\xi \bar{g}_{tt} &= - 2 \dot{T}
\\
  {\cal L}_\xi \bar{g}_{ti} &= \tilde{\nabla}_i (a^2 \dot{L} - T)
\\
  {\cal L}_\xi \bar{g}_{ij} &= 2 a^2 (T H \tilde{g}_{ij} + \tilde{\nabla}_i \tilde{\nabla}_j L)
\end{align}
The metric scalars in Eq. \eqref{eq:whollyperturbed} then transform in the following manner.
\begin{align}
  \phi &\rightarrow \phi - \dot{T}
\\
  B &\rightarrow B + T - a^2 \dot{L}
\\
  C &\rightarrow C - H T
\\
  E &\rightarrow E - L
\end{align}
The Bardeen gauge invariant potentials \cite{Bardeen1980} are the following combinations of these potentials.
\begin{align} \label{eq:bardeen}
  \Phi &= \phi + \partial_t(B - a^2 \dot{E})
\\
  \Psi &= - C - H (B - a^2 \dot{E})
\end{align}

We should also consider how the stress-energy tensor transforms under a gauge transformation. Let us split the stress-energy tensor into a background and perturbed component as
\begin{align}
 \tensor{T}{^\mu_\nu} = \tensor{\bar{T}}{^\mu_\nu} + \delta \tensor{T}{^\mu_\nu}
\end{align}
where
\begin{align}
 \tensor{\bar{T}}{^\mu_\nu} = (\rho_b + P_b) u_b^\mu u_{b\nu} + P_b \delta^\mu_\nu
\end{align}
with $u_b^\mu$ being the background value of $u^\mu$. Let us write the perturbed stress-energy tensor as
\begin{align}
 \delta \tensor{T}{^\mu_\nu} = {}& (\delta \rho + \delta P) u_b^\mu u_{b\nu} + (\rho_b + P_b) (\delta u^\mu u_{b \nu} + u^\mu_b \delta u_\nu)
\nonumber\\
 & \qquad + \delta P \delta^\mu_\nu + \tensor{\Pi}{^\mu_\nu}
\end{align}
where we again work only with scalar perturbations. We decompose $u^\mu$ as
\begin{align}
 u^\mu &= [1 - \phi, \tilde{\nabla}^i v]
\\
 u_\mu &= [- (1 + \phi), \tilde{\nabla}_i (a^2 v + B)]
\end{align}
and use
\begin{align}
 \tensor{\Pi}{^0_\nu} &= \tensor{\Pi}{^\mu_0} = 0
\\
 \tensor{\Pi}{^i_j} &= \left(\tilde{\nabla}^i \tilde{\nabla}_j - \frac{\delta^i_j}{3} \tilde{\nabla}^2\right) \Pi \,.
\end{align}
The components of the stress-energy tensor are then
\begin{align}
 \tensor{T}{^0_0} &= - (\rho_b + \delta \rho)
\\
 \tensor{T}{^i_0} &= - (\rho_b + P_b) \tilde{\nabla}^i v
\\
 \tensor{T}{^0_j} &= (\rho_b + P_b) \tilde{\nabla}_i (a^2 v + B)
\\
 \tensor{T}{^i_j} &= (P_b + \delta P) \delta^i_j + \tensor{\Pi}{^i_j} \,.
\end{align}

Under a gauge transformation, the stress-energy tensor transforms as
\begin{align}
 \delta \tensor{T}{^\mu_\nu} \rightarrow \delta \tensor{T}{^\mu_\nu} - {\cal L}_\xi \tensor{\bar{T}}{^\mu_\nu}
\end{align}
where the Lie derivative is given by
\begin{align}
 {\cal L}_\xi \tensor{\bar{T}}{^\mu_\nu} = \xi^\lambda \partial_\lambda \tensor{\bar{T}}{^\mu_\nu} + \partial_\nu \xi^\lambda \tensor{\bar{T}}{^\mu_\lambda} - \partial_\lambda \xi^\mu \tensor{\bar{T}}{^\lambda_\nu} \,.
\end{align}
Evaluating the Lie derivatives for $\xi^\mu$ defined above yields the following.
\begin{align}
 {\cal L}_\xi \tensor{\bar{T}}{^0_0} &= - T \dot{\rho}_b
\\
 {\cal L}_\xi \tensor{\bar{T}}{^i_0} &= (\rho_b + P_b) \tilde{\nabla}^i \dot{L}
\\
 {\cal L}_\xi \tensor{\bar{T}}{^0_j} &= - (\rho_b + P_b) \tilde{\nabla}_j T
\\
 {\cal L}_\xi \tensor{\bar{T}}{^i_j} &= T \dot{P}_b \delta^i_j
\end{align}
The scalars in the stress-energy tensor then transform in the following manner.
\begin{align}
 \delta \rho &\rightarrow \delta \rho - T \dot{\rho}_b
\\
 \delta P &\rightarrow \delta P - T \dot{P}_b
\\
 v &\rightarrow v + \dot{L}
\\
 \Pi &\rightarrow \Pi
\end{align}
A variety of gauge invariant quantities can be constructed for the matter perturbations. We focus on two of them. The first is the comoving density contrast $\Delta$, defined by
\begin{align}
 \rho_b \Delta = \delta \rho + \dot{\rho}_b (a^2 v + B) \,.
\end{align}
The second, which we shall refer to as the gauge invariant velocity perturbation (as we are unaware of any standard name), is given by
\begin{align}
 V = v + \dot{E} \,.
\end{align}

\subsection{Connection to Misner-Sharp Formalism}

We now cast the Misner-Sharp metric \eqref{eq:MSmetric} into the above language. Writing
\begin{align}
  R &= a A \tilde{R} = a A (1 + \delta_R)
\\
  e^{\lambda/2} &= \frac{\partial_A R}{\Gamma} = a \frac{1 + \delta_R + A \partial_A \delta_R}{1 + \delta_\Gamma}
\end{align}
where
\begin{align}
  \delta_\Gamma = e^{2 (\alpha - 1 ) \xi} \bar{A}^2 \left(\delta_U - \frac{1}{2} \delta_m\right) \label{eq:app:deltag}
\end{align}
and expanding the metric to first order in $\delta_R$, $\delta_\Gamma$ and $\phi$, we obtain the following.
\begin{align} \label{eq:pertmetric}
  ds^2 &= - (1 + 2 \phi) dt^2 + a^2 (1 + 2 \delta_R) (dA^2 + A^2 d\Omega^2)
\nonumber\\&\qquad
  + a^2 (2 A \partial_A \delta_R - 2 \delta_\Gamma) dA^2
\end{align}
This metric contains three scalar perturbations, as expected in a comoving gauge. No vector or tensor components are present.

Looking back at the metric in Eq. \eqref{eq:whollyperturbed}, let us take the metric $\tilde{g}_{ij}$ to be the metric in spherical polar coordinates with radius $A$. We evaluate the quantity
\begin{align}
  \tilde{\nabla}_i \tilde{\nabla}_j E dx^i dx^j = \partial_A^2 E dA^2 + A \partial_A E d\Omega^2 \,.
\end{align}
Letting $\phi$, $C$ and $E$ become functions of $A$ and $t$ alone, we can identify the coefficients in Eqs. \eqref{eq:whollyperturbed} and \eqref{eq:pertmetric} to obtain
\begin{align}
  \delta_R &= C + \frac{\partial_A E}{A}
\\
  \delta_\Gamma &= A \partial_A C \,.
\end{align}
The quantity $\phi$ is the same in both descriptions, and we have $B = 0$.

We now relate our quantities to the Bardeen potentials \eqref{eq:bardeen}. In order to do so, we need to compute $\dot{E}$, which in turn requires $\dot{C}$. We thus start with $C$.
\begin{align}
  C = - \int^\infty_{\bar{A}} \frac{\delta_\Gamma}{\bar{A}'} d\bar{A}'
\end{align}
The limits of integration are set by the requirement that all perturbations vanish at infinity. In order to compute $\dot{C}$, we first compute
\begin{align}
  \partial_\xi C = - \int^\infty_{\bar{A}} \frac{\partial_\xi \delta_\Gamma}{\bar{A}'} d\bar{A}'
\end{align}
where
\begin{align}
  \partial_\xi \delta_\Gamma &= 2 (\alpha - 1) \delta_\Gamma + e^{2 (\alpha - 1 ) \xi} \bar{A}^2 \left(\partial_\xi \delta_U - \frac{1}{2} \partial_\xi \delta_m\right)
\nonumber \\
  &= - \bar{A} \frac{3 w\alpha^2}{2} \partial_{\bar{A}} \delta_\rho
\end{align}
and we have used Eqs. \eqref{eq:reducedlinear} and $\alpha (1 + 3w) = 2 (1 - \alpha)$ to simplify this expression. We then have
\begin{align}
  \partial_\xi C = -\frac{3 w\alpha^2}{2} \delta_\rho \,.
\end{align}
Converting from $\xi$ to $t$, we obtain
\begin{align}
  \dot{C} = - \frac{3 w\alpha}{2} H \delta_\rho \,.
\end{align}

Next, we look towards $E$. Taking our expression for $\delta_R$ and integrating as we did for $C$, we obtain
\begin{align}
  E = R_H^2 \int^\infty_{\bar{A}} \bar{A}'(C - \delta_R) d\bar{A}' \,.
\end{align}
We then compute $\partial_\xi E$ as a step towards $\dot{E}$, using Eqs. \eqref{eq:deltaphieq} and \eqref{eq:deltareq} to simplify the result.
\begin{align}
  \partial_\xi E &= - R_H^2 \int^\infty_{\bar{A}} \alpha \bar{A}' \delta_U d\bar{A}'
\\
  \dot{E} &= - H R_H^2 \int^\infty_{\bar{A}} \bar{A}' \delta_U d\bar{A}'
\end{align}
The first gauge invariant variable is then
\begin{align}
  \Psi = - \frac{1}{2} e^{2 (\alpha - 1) \xi} \int^\infty_{\bar{A}} \bar{A}' \delta_m d\bar{A}'
\end{align}
where we use the definition of $\delta_\Gamma$ \eqref{eq:app:deltag}.

For the second gauge invariant variable, we need $\partial_t (a^2 \dot{E})$. As before, it is simplest to calculate $\partial_\xi (a^2 \dot{E})$ and then convert the result.
\begin{align}
  \partial_\xi (a^2 \dot{E}) &= \frac{\alpha}{2} a^2 H R_H^2 \int^\infty_{\bar{A}} \bar{A}' \, \delta_m d \bar{A}' - \frac{3 w\alpha^2}{2 H} \delta_\rho
\\
  \partial_t (a^2 \dot{E}) &= \frac{1}{2} e^{2(\alpha - 1) \xi} \int^\infty_{\bar{A}} \bar{A}' \, \delta_m d \bar{A}' - \frac{3 w \alpha}{2} \delta_\rho
\end{align}
Here, we've used Eq. \eqref{eq:deltaueqprelim} to substitute $\partial_\xi \delta_U$. The second gauge invariant variable is then
\begin{align}
  \Phi = - \frac{1}{2} e^{2 (\alpha - 1)\xi} \int^\infty_{\bar{A}} \bar{A}' \, \delta_m d \bar{A}' = \Psi
\end{align}
where we substitute $\phi$ from Eq. \eqref{eq:deltaphieq}. We could have anticipated this result; a perfect fluid has no anisotropic shear stress, and so the two Newtonian potentials, and thus the Bardeen gauge invariant variables, are the same. Note that as our linear solution for $\delta_m$ grows as $e^{2 (1 - \alpha) \xi}$, $\Phi$ and $\Psi$ are invariant to first order in our expansion.

The gauge invariant density contrast is simply $\Delta = \delta_\rho$, thanks to the comoving gauge and diagonal metric. The gauge invariant velocity perturbation is
\begin{align}
  V = \dot{E} = - R_H e^{-\xi} \int^\infty_{\bar{A}} \bar{A}' \delta_U d\bar{A}' \,.
\end{align}

\section{Derivation of Hernandez-Misner Results} \label{app:derivation}

In this appendix, we include some derivations relating to the Hernandez-Misner results.

\subsection{Hernandez-Misner}

In the Hernandez-Misner formalism, the equation of motion for $U$ needs to be taken carefully. We begin with Eq. \eqref{eq:UdotMH}. Substituting for the derivatives in Eq. \eqref{eq:invariantderivs}, we obtain
\begin{align}
e^{- \psi} \partial_u {U} &= - \left(\frac{\Gamma}{ \rho + P} (D_r P - D_t P)+ \frac{m + 4 \pi R^3 P}{R^2} \right) \,.
\end{align}
We can write
\begin{align}
D_t P = D_t [\rho (w + \tilde{Q})] = D_t \rho (w + \tilde{Q}) + \rho D_t \tilde{Q} \,.
\end{align}
Inserting this back into the equation of motion, we obtain the following.
\begin{align}
e^{- \psi} \partial_u {U} &= - \frac{\Gamma}{ \rho + P} (D_r P - D_t \rho (w + \tilde{Q}) - \rho D_t \tilde{Q})
\nonumber\\&\qquad
- \frac{m + 4 \pi R^3 P}{R^2}
\end{align}
We then insert $D_t \rho = e^{-\phi} \partial_t \rho$ using Eq. \eqref{eq:rhodot}, being careful with the Misner-Sharp spatial derivatives of $U$ and $R$.
\begin{align}
e^{- \psi} \partial_u {U} &=
- \frac{\Gamma}{\rho + P} e^{- \lambda / 2}P'
- \frac{m + 4 \pi R^3 P}{R^2}
+ \frac{\Gamma \rho D_t \tilde{Q}}{\rho + P}
\nonumber\\&\quad
- (w + \tilde{Q}) \left(2 \frac{U\Gamma}{R} + e^{-\lambda/2} U' - e^{-\psi} \partial_u U \right)
\end{align}
Primes here refer to derivatives with respect to $A$ in Hernandez-Misner coordinates. This can be solved for $\partial_u U$ as follows.
\begin{align}
\partial_u {U}
&= - \frac{e^\psi}{1 - w - \tilde{Q}} \bigg[
\Gamma e^{- \lambda / 2} \frac{P'}{\rho + P}
+ \frac{m + 4 \pi R^3 P}{R^2}
\nonumber\\&\quad
- \frac{\Gamma \rho}{\rho + P} e^{-\psi} \partial_u \tilde{Q}
+ (w + \tilde{Q}) \left(2 \frac{U\Gamma}{R} + e^{-\lambda/2} U' \right)\bigg]
\end{align}

\subsection{Cosmological Hernandez-Misner}

We now derive the cosmological equations of motion for the Hernandez-Misner formalism. In particular, we use the definitions \eqref{eq:MHdefs} in the equations of motion \eqref{eq:HM} and \eqref{eq:HMpsi2}. We assume that at any given time $u = R_H \bar{u}$, we know $\tilde{U}$, $\tilde{m}$, $\tilde{R}$ and $\xi$ as functions of $\bar{A}$.

The first step is to compute $\Gamma$. As previously, it is useful to work with $\bar{\Gamma} = \Gamma / H a R_H$.
\begin{align}
  \bar{\Gamma}^2 = e^{2 (1 - \alpha) \xi} + \bar{A}^2 \tilde{R}^2 (\tilde{U}^2 - \tilde{m})
\end{align}
Next, we construct $\tilde{\rho}$ from Eq. \eqref{eq:HMmprime}.
\begin{align}
  \tilde{\rho} = \frac{m'}{4 \pi \rho_b R^2 R'} \frac{\bar{\Gamma} + \bar{A} \tilde{R} \tilde{U}}{\bar{\Gamma} - (w + \tilde{Q}) \bar{A} \tilde{R} \tilde{U}}
\end{align}
To expand this further, we need
\begin{align}
  \frac{R'}{R} = \alpha \xi' + \frac{1}{\bar{A}} + \frac{\tilde{R}'}{\tilde{R}}
\end{align}
and
\begin{align}
  m' = 4 \pi \rho_b R^3 \tilde{m} \left[
    \left( \alpha - \frac{2}{3} \right) \xi'
  + \frac{1}{\bar{A}}
  + \frac{\tilde{R}'}{\tilde{R}}
  + \frac{1}{3} \frac{\tilde{m}'}{\tilde{m}}
  \right] \,.
\end{align}
The end result is
\begin{align}
  \tilde{\rho} = \frac{\bar{\Gamma} + \bar{A} \tilde{R} \tilde{U}}{\bar{\Gamma} - (w + \tilde{Q}) \bar{A} \tilde{R} \tilde{U}} \left[
  \tilde{m} + \frac{\bar{A} \tilde{R}}{3} \frac{\tilde{m}' - 2 \tilde{m} \xi'}{\alpha \bar{A} \tilde{R} \xi' + (\bar{A} \tilde{R})'}
  \right] \,.
\end{align}
Knowing $\tilde{\rho}$, we can compute $\tilde{P} = \tilde{\rho} (w + \tilde{Q})$ (given a form for $\tilde{Q}$).

We now turn our attention to $e^{\lambda/2}$. From Eq. \eqref{eq:dtdA}, we find
\begin{align}
  e^{\lambda/2} = \alpha e^{\phi + \xi} \xi' \,. \label{eq:HMlambda2}
\end{align}
From Eq. \eqref{eq:HMlambda}, we have
\begin{align} \label{eq:othereq}
  e^{-\lambda/2} = e^{-\xi} \frac{\bar{\Gamma} + \bar{A} \tilde{R} (\tilde{U} - e^{-\phi})}{(\bar{A} \tilde{R})'} \,.
\end{align}
As we do not yet know $e^\phi$, we can use these equations to eliminate the unknown $e^{\lambda/2}$ and solve for $e^\phi$, and then compute $e^{-\lambda/2}$ using Eq. \eqref{eq:HMlambda2}.
\begin{align}
e^{\phi} = \frac{\alpha \xi' \bar{A} \tilde{R} + (\bar{A} \tilde{R})'}{\alpha \xi' (\bar{\Gamma} + \bar{A} \tilde{R} \tilde{U})}
\end{align}

The final auxiliary variable we need to compute is $e^\psi$. From \eqref{eq:HMpsi2}, we have
\begin{align}
&\partial_{\bar{A}} [e^{-\psi} (\bar{\Gamma} + \bar{A} \tilde{R} \tilde{U})]
\nonumber\\
& \quad = - e^{-\psi} \xi' \left[e^\phi \bar{A} \tilde{R} \tilde{\rho} \frac{1 + w + \tilde{Q}}{1+w}
+ (\alpha - 1) (\bar{\Gamma} + \bar{A} \tilde{R} \tilde{U}) \right]
\end{align}
subject to the boundary condition that $e^\psi = e^\phi$ at the outer boundary. To integrate this equation, let $\chi = e^{-\psi} (\bar{\Gamma} + \bar{A} \tilde{R} \tilde{U})$, and rewrite it as
\begin{align}
  \partial_{\bar{A}} (\ln \chi) = - \xi' \left[\frac{e^\phi \bar{A} \tilde{R} \tilde{\rho}}{\bar{\Gamma} + \bar{A} \tilde{R} \tilde{U}} \frac{1 + w + \tilde{Q}}{1+w} + \alpha - 1\right]
\end{align}
which can be integrated straightforwardly.

The next step is to compute the evolution equations for $\tilde{R}$, $\tilde{U}$, $\tilde{m}$ and $\xi$. For $\xi$, $\tilde{m}$ and $\tilde{R}$, the results can be found by changing coordinates in the Misner-Sharp evolution equations in these coordinates, Eqs. \eqref{eq:scaledMS}. In particular,
\begin{align}
D_t \tilde{X} = e^{-\phi} \partial_t \tilde{X} = e^{-\psi} \partial_u \tilde{X} \,.
\end{align}
Writing this in terms of derivatives with respect to $\bar{u}$ and $\chi$, we have
\begin{align} \label{eq:derivtransform}
\partial_\xi \tilde{X} = \alpha e^{\phi + \xi -\psi} \partial_{\bar{u}} \tilde{X} \,.
\end{align}
Applying this to $\xi$, $\tilde{R}$ and $\tilde{m}$, we obtain the following.
\begin{align}
  \partial_{\bar{u}} \xi &= \frac{1}{\alpha} e^{\psi - \phi - \xi} \,. \label{eq:xidot}
\\
  \partial_{\bar{u}} \tilde{R}
  &= e^{\psi - \xi} \tilde{R} (\tilde{U} - e^{- \phi})
\\
  \partial_{\bar{u}} \tilde{m} &= 3 e^{\psi - \xi} \left[ e^{-\phi} \tilde{m} (1 + w) - \tilde{U} (\tilde{P} + \tilde{m}) \right]
\end{align}
Note that when the lapse $e^\psi \rightarrow 0$ near horizon formation, all quantities are essentially frozen; we find the same behavior in the $\tilde{U}$ evolution equation too.

To obtain the equation of motion for $\tilde{U}$ requires a bit more work. Start with Eq.\eqref{eq:newu}, transform the $\partial_\xi$ derivative as above, and write the spatial derivatives in terms of the invariant derivative operators.
\begin{align} \label{eq:partialu}
  e^{\xi - \psi} \partial_{\bar{u}} \tilde{U}
  &= e^{-\phi} \frac{\tilde{U}}{\alpha} -
  \bar{\Gamma}^2 \frac{D_r \tilde{P}}{\bar{A} \tilde{R} D_r (\bar{A} \tilde{R}) (\tilde{\rho} + \tilde{P})}
\nonumber \\
  & \qquad \qquad \qquad - \frac{1}{2} \left(2 \tilde{U}^2 + \tilde{m} + 3 \tilde{P} \right)
\end{align}
Now, we can write
\begin{align}
  D_r (\bar{A} \tilde{R}) &= \frac{1}{R_H} \left( e^{-\lambda/2} \partial_{\bar{A}} - e^{-\psi} \partial_{\bar{u}} \right) (\bar{A} \tilde{R})
\\
  &= H \bar{\Gamma}
\end{align}
where we make use of the result for $\partial_{\bar{u}} \tilde{R}$ above and Eq. \eqref{eq:othereq}.
The other derivative we need to evaluate is
\begin{align}
  D_r \tilde{P} = (w + \tilde{Q}) D_r \tilde{\rho} + \tilde{\rho} D_r \tilde{Q} \,.
\end{align}
The first term here can be written
\begin{align}
  D_r \tilde{\rho} = \frac{1}{R_H} e^{-\lambda/2} \left.\partial_{\bar{A}} \tilde{\rho}\right|_{\bar{u}} - D_t \tilde{\rho} \,.
\end{align}
To evaluate the final term here, we turn to Eq. \eqref{eq:rhodot} in the form
\begin{align}
  D_t \rho &= - (\rho + P) \left(2 \frac{U}{R} + \frac{D_r U}{D_r R} \right) \,.
\end{align}
Writing this in terms of tilded quantities, we have
\begin{align} \label{eq:dtrhotwiddle}
  D_t \tilde{\rho} &= 3 H e^{-\phi} (1 + w) \tilde{\rho} - H (\tilde{\rho} + \tilde{P}) \left(3 \tilde{U} + \frac{\tilde{R} \bar{A} D_r \tilde{U}}{D_r (\bar{A} \tilde{R})} \right)
\end{align}
where we use the continuity equation in the form $D_t \rho_b = - e^{-\phi} 3 H \rho_b (1+w)$. Combining all these results, we can write $D_r \tilde{P}$ as \begin{align}
  D_r \tilde{P} = {}& (w + \tilde{Q}) \Bigg[ \frac{e^{-\lambda/2}}{R_H} \tilde{\rho}' + \frac{\tilde{\rho} D_r \tilde{Q}}{w + \tilde{Q}}
\nonumber\\
  & \quad + (\tilde{\rho} + \tilde{P}) \frac{\tilde{R} \bar{A}}{R_H \bar{\Gamma}} (e^{-\lambda/2} \tilde{U}' - e^{-\psi} \partial_{\bar{u}} \tilde{U})
\nonumber\\
  & \quad + 3 H \tilde{\rho} \left[(1 + w) (\tilde{U} - e^{-\phi}) + \tilde{Q} \tilde{U} \right]
  \Bigg]
\end{align}
where primes once again refer to derivatives in the Hernandez-Misner coordinate system.

We can now combine these results in Eq. \eqref{eq:partialu}.
\begin{align}
  e^{\xi - \psi} \partial_{\bar{u}} \tilde{U}
  ={}& e^{-\phi} \frac{\tilde{U}}{\alpha} - \frac{1}{2} \left(2 \tilde{U}^2 + \tilde{m} + 3 \tilde{P} \right)
\nonumber \\  
  & - \frac{\bar{\Gamma} (w + \tilde{Q})}{\bar{A} \tilde{R} H (\tilde{\rho} + \tilde{P})} \Bigg[ \frac{e^{-\lambda/2}}{R_H} \tilde{\rho}' + \frac{\tilde{\rho} D_r \tilde{Q}}{w + \tilde{Q}}
\nonumber\\
  & \quad + 3 H \tilde{\rho} \left[(1 + w) (\tilde{U} - e^{-\phi}) + \tilde{Q} \tilde{U} \right]
  \Bigg]
\nonumber\\
  & - (w + \tilde{Q}) (e^{\xi -\lambda/2} \tilde{U}' - e^{\xi -\psi} \partial_{\bar{u}} \tilde{U})
\end{align}
We can finally solve this for $\partial_{\bar{u}} \tilde{U}$.
\begin{align}
  \partial_{\bar{u}} \tilde{U}
  ={}& - \frac{e^{\psi - \xi}}{1 - w - \tilde{Q}} \bigg[
  \frac{1}{2} \left(\tilde{m} + 3 \tilde{P} \right) + \tilde{U}^2 - e^{-\phi} \frac{\tilde{U}}{\alpha}
\nonumber\\
  & \qquad + (w + \tilde{Q}) e^{\xi -\lambda/2} \tilde{U}'
\nonumber \\  
  & \qquad + \frac{\bar{\Gamma} (w + \tilde{Q})}{\bar{A} \tilde{R} (1 + w + \tilde{Q})} \bigg( e^{\xi-\lambda/2} \frac{\tilde{\rho}'}{\tilde{\rho}} + \frac{D_r \tilde{Q}}{H (w + \tilde{Q})}
\nonumber\\
  & \qquad \quad + 3 (1 + w) (\tilde{U} - e^{-\phi}) + 3 \tilde{Q} \tilde{U} \bigg)
  \bigg]
\end{align}

This equation looks like it diverges as $\bar{A} \rightarrow 0$. In order for it not to diverge, we require
\begin{align} \label{eq:limiteq}
  \lim_{\bar{A} \rightarrow 0} \left[e^{\xi-\lambda/2} \frac{\tilde{\rho}'}{\tilde{\rho}} + 3 (1 + w) (\tilde{U} - e^{-\phi}) + 3 \tilde{Q} \tilde{U}\right] = 0 \,.
\end{align}
This is a sufficient condition, as this quantity is just being divided by $\bar{A}$. (Note that $D_R \tilde{Q} = 0$ at the origin by virtue of being an even function in Misner-Sharp coordinates.) In order to show this, we write $\tilde{\rho}'$ in terms of $D_r$ and $D_t$ once again.
\begin{align}
  \tilde{\rho}' = R_H e^{\lambda/2} (D_t + D_r) \tilde{\rho}
\end{align}
As $\tilde{\rho}$ is an even function in Misner-Sharp coordinates, $D_r \tilde{\rho}$ vanishes at the origin, and we can write
\begin{align}
  \tilde{\rho}' (\bar{A} = 0) = 3 e^{\lambda/2 - \xi} \left[e^{-\phi} (1 + w) \tilde{\rho} - (\tilde{\rho} + \tilde{P}) \tilde{U}\right]
\end{align}
using Eq. \eqref{eq:dtrhotwiddle} from above. Inserting this in Eq. \eqref{eq:limiteq}, it is straightforward to show that the limit vanishes. We thankfully do not need to evaluate $\partial_{\bar{u}} \tilde{U}$ at the origin, as there is a boundary condition there, but it is good to know that the equation of motion is not divergent as one approaches the origin.

\bibliographystyle{utphys}
\bibliography{pbhrefs}

\end{document}